\pdfoutput=1
\documentclass[11pt]{article}
\usepackage[letterpaper,margin=1in]{geometry}
\usepackage[T1]{fontenc}
\usepackage[utf8]{inputenc}
\usepackage{lmodern}
\usepackage{amsmath,amssymb,amsthm,mathtools}
\usepackage{enumitem}
\usepackage{booktabs,array}
\usepackage{xcolor}
\usepackage{microtype}
\usepackage{tikz}
\usepackage[colorlinks=true,linkcolor=blue!50!black,citecolor=green!35!black,urlcolor=blue!50!black]{hyperref}
\usepackage[capitalise,nameinlink]{cleveref}\crefname{figure}{Figure}{Figures}

\usepackage{aliascnt}
\newcommand{\sharedthm}[2]{\newaliascnt{#1}{theorem}\newtheorem{#1}[#1]{#2}\aliascntresetthe{#1}%
  \crefname{#1}{#2}{#2s}\Crefname{#1}{#2}{#2s}}
\theoremstyle{plain}
\newtheorem{theorem}{Theorem}[section]
\sharedthm{lemma}{Lemma}
\sharedthm{corollary}{Corollary}
\sharedthm{proposition}{Proposition}
\sharedthm{claim}{Claim}
\sharedthm{conjecture}{Conjecture}
\sharedthm{question}{Question}
\theoremstyle{definition}
\sharedthm{definition}{Definition}
\sharedthm{construction}{Construction}
\theoremstyle{remark}
\sharedthm{remark}{Remark}
\sharedthm{observation}{Observation}
\newcommand{\restatedtitle}{}
\newtheorem*{restatedthm*}{\restatedtitle}
\newenvironment{restated}[1]{\renewcommand{\restatedtitle}{#1 (restated)}\begin{restatedthm*}}{\end{restatedthm*}}

\newcommand{\N}{\mathbb{N}}
\newcommand{\E}{\mathbb{E}}
\newcommand{\Ot}{\widetilde{O}}
\newcommand{\Tmpl}{\operatorname{Tmpl}}
\newcommand{\Jst}{J^{*}}
\newcommand{\tow}{\operatorname{tow}}
\newcommand{\lgs}{\log^{*}}
\newcommand{\chidet}{\chi^{\mathrm{det}}}
\newcommand{\chirand}{\chi^{\mathrm{rand}}}
\newcommand{\chiol}{\chi_{\mathrm{ol}}}
\newcommand{\col}{\operatorname{col}}
\newcommand{\Res}{\operatorname{Res}}
\newcommand{\width}{\operatorname{width}}
\newcommand{\Sat}{\operatorname{Sat}}
\newcommand{\serve}{\operatorname{serve}}
\newcommand{\idx}{\operatorname{idx}}
\newcommand{\Alg}{\mathsf{Alg}}
\newcommand{\cA}{\mathcal{A}}
\newcommand{\OB}{\mathrm{OB}}
\newcommand{\down}{{\downarrow}}

\usepackage[numbers,sort&compress]{natbib}
\usetikzlibrary{positioning,arrows.meta,calc,fit,shapes.geometric,decorations.pathreplacing,backgrounds,plotmarks}

\definecolor{cA}{RGB}{166,206,227}   %
\definecolor{cB}{RGB}{253,191,111}   %
\definecolor{cC}{RGB}{178,223,138}   %
\definecolor{cD}{RGB}{251,154,153}   %
\tikzset{
  rv/.style={circle,draw,inner sep=0pt,minimum size=9pt},
  iv/.style={circle,draw,thick,inner sep=0pt,minimum size=12pt},
  mv/.style={circle,draw,inner sep=0pt,minimum size=10pt},
  resv/.style={line width=1.4pt},
  blk/.style={draw=gray!80,rounded corners=2pt},
  oe/.style={black!55,line width=0.35pt},
  me/.style={black!60,line width=0.45pt},
  iarc/.style={-{Stealth[length=4pt]},thin},
  gl2/.style={thick,black!50,mark options={fill=black!50}},
  gl3/.style={thick,blue!70!black,mark options={fill=blue!70!black}},
  gl4/.style={thick,red!70!black,mark options={fill=red!70!black}},
  ours/.style={very thick,blue!70!black},
  li/.style={thick,red!70!black,mark options={fill=red!70!black}},
  exkyy/.style={thick,black!60,mark options={fill=black!60}},
  exapp/.style={thick,orange!80!black,mark options={fill=orange!80!black}},
  exthmA/.style={thick,blue!70!black,mark options={fill=blue!70!black}},
  exrand/.style={thick,green!45!black,dashed,mark options={fill=green!45!black,solid}},
  hasse/.style={circle,draw,inner sep=1pt,minimum size=13pt,font=\scriptsize},
  hmini/.style={circle,draw,inner sep=0pt,minimum size=8.5pt,font=\tiny},
  hin/.style={fill=blue!25},
  hout/.style={fill=white},
  fbox/.style={draw,rounded corners=3pt,align=center,font=\footnotesize,inner sep=4pt},
  farr/.style={-{Stealth[length=5pt]},thick},
}

\newcommand{\colourkey}{%
\begin{tikzpicture}[baseline=-3pt]
\node[rv,fill=cA] at (0,0) {\tiny 1}; \node[rv,fill=cB] at (0.4,0) {\tiny 2};
\node[rv,fill=cC] at (0.8,0) {\tiny 3}; \node[rv,fill=cD] at (1.2,0) {\tiny 4};
\end{tikzpicture}}

\newcommand{\hasseNode}[4]{%
  \node[hasse,hout] (#1) at (#2,#3) {#4};}
\newcommand{\hasseEdges}{%
  \draw (h1)--(h12) (h1)--(h13) (h2)--(h12) (h2)--(h23) (h3)--(h13) (h3)--(h23)
        (h12)--(h123) (h13)--(h123) (h23)--(h123);}
\def\hasseLabel h#1\relax{#1}%
\newcommand{\hasseP}[2][1]{%
\begin{tikzpicture}[x=#1 cm,y=#1 cm,baseline=(h1.base)]
  \foreach \n/\x/\y/\l in {h1/-0.75/0/1,h2/0/0/2,h3/0.75/0/3,h12/-0.75/0.85/12,h13/0/0.85/13,h23/0.75/0.85/23,h123/0/1.7/123}
    {\node[hmini,hout] (\n) at (\x,\y) {\l};}
  \begin{scope}[on background layer]\hasseEdges\end{scope}
  \foreach \e in {#2} {\node[hmini,hin] at (\e) {\phantom{\expandafter\hasseLabel\e\relax}};}
  \foreach \n/\l in {h1/1,h2/2,h3/3,h12/12,h13/13,h23/23,h123/123} {\node[font=\tiny] at (\n) {\l};}
\end{tikzpicture}}

\newcommand{\figPosetOne}{%
\begin{tikzpicture}[x=1.15cm,y=1.1cm]
  \hasseNode{h1}{-1}{0}{$\{1\}$} \hasseNode{h2}{0}{0}{$\{2\}$} \hasseNode{h3}{1}{0}{$\{3\}$}
  \hasseNode{h12}{-1}{1}{$\{1,2\}$} \hasseNode{h13}{0}{1}{$\{1,3\}$} \hasseNode{h23}{1}{1}{$\{2,3\}$}
  \hasseNode{h123}{0}{2}{$[3]$}
  \begin{scope}[on background layer]\hasseEdges\end{scope}
\end{tikzpicture}}

\newcommand{\figDownsets}{%
\begin{tabular}{@{}c@{\quad}c@{\quad}c@{\quad}c@{\quad}c@{\quad}c@{}}
\hasseP[0.55]{h1} & \hasseP[0.55]{h1,h2,h12} & \hasseP[0.55]{h1,h3,h13} & \hasseP[0.55]{h2,h3,h23} &
\hasseP[0.55]{h1,h2,h3} & \hasseP[0.55]{h1,h2,h3,h12,h13,h23,h123}\\[2pt]
\scriptsize ${\downarrow}\{\{1\}\}$ & \scriptsize ${\downarrow}\{\{1,2\}\}$ & \scriptsize ${\downarrow}\{\{1,3\}\}$ &
\scriptsize ${\downarrow}\{\{2,3\}\}$ & \scriptsize $\mathcal M=\{\{1\},\{2\},\{3\}\}$ & \scriptsize $P_1(3)$
\end{tabular}}

\newcommand{\figRuleLeft}{%
\begin{tikzpicture}
  \node[iv] (w) at (0,0) {\scriptsize $w$};
  \node[iv] (v) at (2.4,0) {\scriptsize $v$};
  \draw[iarc] (w) -- (v);
\end{tikzpicture}}
\newcommand{\figRuleRight}{%
\begin{tikzpicture}[baseline=(current bounding box.center)]
    \foreach \i in {0,...,3} {\node[rv,fill=white] (s\i) at (0.42*\i,0) {};}
    \node[rv,fill=white,resv] (s4) at (1.68,0) {};
    \foreach \i in {0,...,4} {\node[rv,fill=white] (t\i) at (0.42*\i+0.6,-1.4) {};}
    \draw[blk] ($(s0.north west)+(-3pt,3pt)$) rectangle ($(s4.south east)+(3pt,-3pt)$);
    \draw[blk] ($(t0.north west)+(-3pt,3pt)$) rectangle ($(t4.south east)+(3pt,-3pt)$);
    \foreach \i in {0,...,4} {\draw[oe] (s4) -- (t\i);}
    \node[font=\scriptsize,left] at ($(s0)+(-0.2,0)$) {$\sigma(w)$};
    \node[font=\scriptsize,left] at ($(t0)+(-0.2,0)$) {$\sigma(v)$};
    \node[font=\scriptsize,right,align=left] at ($(s4)+(0.2,0)$) {$x=x_{w,v}$: $f(x)=\{2\}\in\psi(w)\setminus\psi(v)$};
    \node[font=\scriptsize,right,align=left] at ($(t4)+(0.2,0)$) {$u\in\sigma(v)$: $f(u)\in\psi(v)$, e.g.\ $\{1\}$};
    \node[font=\scriptsize,align=left,anchor=north west] at (-0.9,-1.85)
      {$\psi(v)$ is down-closed and $f(x)\notin\psi(v)$,\\
       hence $f(x)\not\le f(u)$ for each edge $x\to u$.};
\end{tikzpicture}}
\newcommand{\figRule}{%
\begin{tabular}{@{}c@{\hspace{1.2cm}}c@{}}
\begin{tabular}[c]{@{}c@{}}
\figRuleLeft\\[2pt]
\begin{tabular}{@{}c@{\hspace{0.9cm}}c@{}}
\hasseP[0.55]{h1,h2,h12} & \hasseP[0.55]{h1,h3,h13}\\
\scriptsize $\psi(w)={\downarrow}\{\{1,2\}\}$ & \scriptsize $\psi(v)={\downarrow}\{\{1,3\}\}$
\end{tabular}
\end{tabular}
&
\figRuleRight
\end{tabular}}

\newcommand{\figSchematic}{%
\begin{tikzpicture}[x=1cm,y=1cm]
  \node[font=\footnotesize,align=center] at (1.0,2.65) {level $\ell+1$};
  \node[iv,fill=cA] (W) at (0.2,0.9) {\scriptsize $w$};
  \node[iv,fill=cB] (V) at (1.8,1.5) {\scriptsize $v$};
  \node[iv,fill=cC] (V2) at (1.8,0.3) {\scriptsize $v'$};
  \draw[iarc] (W) -- (V);
  \draw[iarc] (W) -- (V2);
  \node[font=\scriptsize,align=center,anchor=north] at (1.0,-0.2) {fills: induced\\ colors $\col(\cdot)$};
  \draw[-{Stealth[length=6pt]},very thick,black!50] (2.5,0.9) -- node[above,font=\scriptsize]{blow-up} (3.5,0.9);
  \foreach \i/\c/\r in {0/cA/1,1/cB/0,2/cA/1,3/cC/0,4/cA/1,5/cA/0,6/cB/0} {
    \ifnum\r=1 \node[rv,fill=\c,resv] (s\i) at (4.6+0.42*\i,1.5) {};
    \else \node[rv,fill=\c] (s\i) at (4.6+0.42*\i,1.5) {}; \fi}
  \draw[blk] ([xshift=-3pt,yshift=3pt]s0.north west) rectangle ([xshift=3pt,yshift=-3pt]s6.south east);
  \node[font=\scriptsize,anchor=east] at ([xshift=-5pt]s0.west) {$\sigma(w)$};
  \node[font=\scriptsize,anchor=south] at ([yshift=3pt]s0.north) {$x_{w,v}$};
  \node[font=\scriptsize,anchor=south] at ([yshift=3pt]s4.north) {$x_{w,v'}$};
  \node[font=\scriptsize,anchor=west,align=left] at (7.6,1.5)
     {$R=(K-1)(D-1)+1=7$ vertices; reserve $\Res(w)$:\\ $D=3$ vertices of the induced color $\col(w)=1$ (thick)};
  \foreach \i/\c/\r in {0/cB/1,1/cC/0,2/cB/1,3/cB/1,4/cC/0,5/cB/0,6/cC/0} {
    \ifnum\r=1 \node[rv,fill=\c,resv] (t\i) at (4.0+0.42*\i,-0.4) {};
    \else \node[rv,fill=\c] (t\i) at (4.0+0.42*\i,-0.4) {}; \fi}
  \foreach \i/\c/\r in {0/cC/1,1/cC/1,2/cB/0,3/cC/1,4/cB/0,5/cC/0,6/cC/0} {
    \ifnum\r=1 \node[rv,fill=\c,resv] (u\i) at (8.0+0.42*\i,-0.4) {};
    \else \node[rv,fill=\c] (u\i) at (8.0+0.42*\i,-0.4) {}; \fi}
  \draw[blk] ([xshift=-3pt,yshift=3pt]t0.north west) rectangle ([xshift=3pt,yshift=-3pt]t6.south east);
  \draw[blk] ([xshift=-3pt,yshift=3pt]u0.north west) rectangle ([xshift=3pt,yshift=-3pt]u6.south east);
  \begin{scope}[on background layer]
    \foreach \i in {0,...,6} {\draw[blue!55!black,line width=0.4pt] (s0) -- (t\i);}
    \foreach \i in {0,...,6} {\draw[red!60!black,line width=0.4pt] (s4) -- (u\i);}
  \end{scope}
  \node[font=\scriptsize,anchor=north] at ([yshift=-5pt]t3.south) {$\sigma(v)$: $\col(v)=2$};
  \node[font=\scriptsize,anchor=north] at ([yshift=-5pt]u3.south) {$\sigma(v')$: $\col(v')=3$};
  \node[font=\scriptsize,anchor=west,align=left] at ([xshift=6pt]u6.east)
     {every vertex sees\\ color $1=\col(w)$};
  \node[font=\footnotesize,align=center] at (5.9,2.65) {level $\ell$};
  \node[anchor=west] at (11.2,2.25) {\colourkey};
\end{tikzpicture}}

\numberwithin{equation}{section}

\title{An $n^{1/O(\log^*n)}$ Lower Bound for Online Coloring\\ of $k$-Colorable Graphs}
\author{Sajad Faghfoor Maghrebi\\
  {\small University of Toronto}\\
  {\small\texttt{smaghrebi@cs.toronto.edu}}}
\date{}

\begin{document}
\maketitle

\begin{abstract}
In online coloring, the vertices of a graph arrive one at a time with their edges to earlier vertices, and each vertex
must be colored irrevocably on arrival. For graphs promised to be $k$-colorable, with $k\ge 3$ fixed, the best
deterministic algorithms use $n^{1-\Theta(1/k^2)}$ colors, while the best lower bound was Vishwanathan's $\Omega(\log^{k-1}n)$ from
1990 until Li recently proved $2^{\Omega((\log n)^{1/3}(\log\log n)^{2/3})}$. For every fixed $k\ge 3$ and all
sufficiently large $n$, we prove a lower bound of $n^{1/(\log^*n+3)}$ colors for deterministic online algorithms on
$n$-vertex $k$-colorable graphs, and the same lower bound in expectation for randomized online algorithms. Both bounds hold
on a single triangle-free $3$-colorable graph: even an algorithm that is shown this graph in advance, but is not told
which of its vertices each arriving vertex is, can be forced to use this many colors by a suitable arrival order. The hard instances are iterated
blow-ups of a clique, in which every vertex becomes a block of independent vertices and every edge a private server in
the earlier block; each blow-up keeps the forced colors and replaces the chromatic number $\chi$ by about $\log\chi$, so about $\log^*n$ blow-ups bring it down to $k$. We also show that deterministic lower bounds for different
values of $k\ge 3$ are polynomially equivalent, and that no adversary presenting such iterated blow-ups can force more than
$n^{1/(\log^*n-O(1))}$ colors. All numbered results are formalized in
Lean~4.\footnote{The formalization is available at \url{https://github.com/sfmqrb/lean-verifications-k-colorable}.}
\end{abstract}

\thispagestyle{empty}
\newpage
\setcounter{page}{1}

\section{Introduction}\label{sec:intro}

In online graph coloring, the vertices of an unknown graph arrive one at a time, each with its edges to the vertices
that arrived before it. An online algorithm must color every vertex on arrival, irrevocably and differently from its
earlier neighbors, and its cost is the number of colors it uses. Even bipartite graphs are hard in this model:
on $n$ vertices $\Theta(\log n)$ colors are both necessary and sufficient~\citep{GL1988,LST1989}; cf.~\citep{Bean1976}. We study graphs promised to be $k$-colorable for a fixed $k\ge 3$, where the gap has been far wider.

For $k\ge 3$ the best algorithms use polynomially many colors. Kierstead's deterministic algorithm uses
$O(n^{2/3}\log^{1/3}n)$ colors for $k=3$ and $O(n^{1-1/k!})$ in general~\citep[Thms.~1.6 and~1.5]{Kie1998}, and
Kawarabayashi, Yoneda and Yoneda recently improved the general bound to $\Ot(n^{1-2/(k(k-1))})$, and to
$\Ot(n^{1-6/(3k(k-1)-2)})$ for $k\ge 4$~\citep[Thm.~3.17 and Cor.~4.15]{KYY2025}. Randomized algorithms of Vishwanathan
and of Halld\'orsson use $O(n^{1-1/(k-1)}(\log n)^{1/(k-1)})$ colors in expectation~\citep[Thm.~2.2]{Vis1990},
\citep[Thm.~4.1]{Hal1997}, and Li reduced this to $O(\sqrt n)$ for $k=3$~\citep[Thm.~1.2]{Li2026}.

The lower bounds, in contrast, stayed polylogarithmic for three decades. Vishwanathan proved in 1990 that $\Omega(\log^{k-1}n)$ colors are
necessary, even for randomized algorithms~\citep[Thm.~3.1]{Vis1990}, and this stood until Li recently
proved a bound of $2^{\Omega((\log n)^{1/3}(\log\log n)^{2/3})}$ for $k=3$, for deterministic algorithms and for
randomized ones~\citep[Thms.~1.1 and~A.1]{Li2026}. Kierstead asked whether $O(\sqrt n)$ colors suffice for
$k=3$~\citep[Problem~12]{Kie1998survey}, and posed as a challenge to prove that between $\Omega(n^{\varepsilon})$ and
$O(n^{1/2})$ colors are needed~\citep[Problem~5.1]{Kie1998}. Before this work, the number of colors needed for $3$-colorable graphs was
thus known only to lie between $2^{(\log n)^{1/3+o(1)}}$ and $O(\sqrt n)$ for randomized algorithms, and up to
$O(n^{2/3}\log^{1/3}n)$ for deterministic ones.

\paragraph{Why lower bounds are hard.} A lower bound is an adversary that must do two things at once: force many
colors, and keep its graph $k$-colorable. A clique on $K$ vertices forces $K$ colors, but it is not $k$-colorable
once $K>k$. Three known facts show how little room the adversary has. First, a strong adversary must keep its own
coloring undecided, as Kawarabayashi, Yoneda and Yoneda also observe~\citep[\S4]{KYY2026comp}. The reason is the
transparent model of Halld\'orsson and Szegedy~\citep{HS1994}, where the adversary reveals its own $k$-coloring of
each vertex right after the algorithm has colored it. In that model, $2^k-1$ colors always
suffice~\citep[\S4.5]{HS1994}. Second, the adversary must control the arrival order, since under a uniformly random
order a randomized algorithm that knows $n$ and $k$ needs only polylogarithmically many
colors~\citep[Thm.~1]{YY2026random}. Third, a polynomial lower bound must use short odd cycles, since graphs of
sufficiently large odd girth can be colored online with $O(n^{\varepsilon})$ colors for every fixed
$\varepsilon>0$~\citep[Thm.~1]{YY2026oddgirth}.

\subsection{Our results}

Let $\chidet_k(n)$ be the least number of colors that a deterministic online algorithm needs, in the worst case, on
ordered $k$-colorable graphs with $n$ vertices, and let $\chirand_k(n)$ be the analogue for randomized algorithms,
measured in expectation against an oblivious adversary (\cref{def:chi}); thus $\chirand_k(n)\le\chidet_k(n)$. Let $\lgs
n$ be the number of times $\log_2$ must be applied to $n$ to reach a number at most $1$. Our main result is that $\chidet_k(n)$
and $\chirand_k(n)$ are both at least $n^{1/(\lgs n+O(1))}$ for every fixed $k\ge 3$; whether either is at least
$n^{\varepsilon}$ for a fixed $\varepsilon>0$ remains open.

\begin{theorem}[main theorem]\label{thm:main-logstar}
For every $n\ge 3$ and $L=\lgs n$,
\[
  \chirand_3(n)\;\ge\;\frac{n^{1/(L+2)}}{8(1+\ln 4n)}\qquad\text{and}\qquad
  \chirand_k(n)\;\ge\;\frac{n^{1/L}}{8(1+\ln 4n)}\quad\text{for every }k\ge 4.
\]
Moreover $\chidet_3(n)\ge\chirand_3(n)\ge n^{1/(\lgs n+3)}$ for all $n\ge 2^{406}$, and $\chidet_k(n)\ge\chirand_k(n)\ge
n^{1/(\lgs n+1)}$ for every $k\ge 4$ and all $n\ge 2^{201}$.
\end{theorem}

Roughly speaking, \cref{thm:main-logstar} says that online coloring of $3$-colorable graphs needs $n^{1/(\lgs
n+O(1))}$ colors, against deterministic and randomized algorithms alike. This bound is still $n^{o(1)}$, far from the $n^{1-\Theta(1/k^2)}$ colors of the best deterministic algorithms (\cref{tab:bounds}),
but no adversary of the kind we use can do better, up to the additive constant in the denominator of the exponent
(\cref{thm:main-ceiling}). The thresholds $2^{406}$ and $2^{201}$ are the sizes from which our proof absorbs the factor
$8(1+\ln 4n)$ into one more unit in the denominator of the exponent. Below these thresholds the first bound still holds, and at small sizes (for $3$-colorable graphs up to about
$n=2^{50}$, see \cref{sec:logstar-proof}) a simpler deterministic construction,
which also underlies \cref{thm:main-logstar}, gives better and fully explicit bounds.

\begin{theorem}[explicit deterministic bound]\label{thm:main-det}
For every $n\ge 3$ and $L=\lgs n$,
\[
  \chidet_3(n)\;\ge\;\big\lfloor n^{2/((L+2)(L+3))}\big\rfloor\qquad\text{and}\qquad
  \chidet_k(n)\;\ge\;\big\lfloor n^{2/(L(L+1))}\big\rfloor\quad\text{for every }k\ge 4.
\]
In particular $\chidet_k(n)\ge n^{(2-o(1))/(\lgs n)^2}$ for every fixed $k\ge 3$. For all $3\le n\le 2^{65536}$,
$\chidet_3(n)\ge\lfloor n^{1/28}\rfloor$ and $\chidet_k(n)\ge\lfloor n^{1/15}\rfloor$ for every $k\ge 4$.
\end{theorem}

\Cref{thm:main-det} alone raises the deterministic lower bound from $2^{(\log n)^{1/3+o(1)}}$ to $2^{\log
n/O((\lgs n)^2)}$, and its construction is small: one level forces $7$ colors on $3$-colorable graphs with $187$
vertices, and two levels force $18$ colors with about $2.1\cdot10^7$ vertices (\cref{tab:sizes}). A version that holds
with probability $1-\varepsilon$ against randomized algorithms, at the cost of a factor $O(\log n)$, is in
\cref{sec:rand}.

\paragraph{High-level idea of the construction.} The lower bounds of \cref{thm:main-logstar,thm:main-det} come from one
construction. Since a clique on $K$ vertices forces $K$ colors but is not $k$-colorable once $K>k$, their adversaries
keep a clique in the background and never present it. They present instead a graph obtained from the clique by applying
one operation, the \emph{reservation blow-up}, once or several times.

Let $H$ be an ordered graph, that is, a graph together with the arrival order of its vertices. A reservation
blow-up of $H$ replaces every vertex $v$ by a nonempty \emph{block} $\sigma(v)$ of new vertices with no edges among them, and
the blocks arrive one after the other, in the order of the vertices of $H$. For every arc $w\to v$ of $H$, that is,
every edge from an earlier vertex $w$ to a later vertex $v$, it picks a vertex of $\sigma(w)$, the \emph{server}
of the arc, and joins it to all of $\sigma(v)$. Different arcs have different servers, and the blow-up has no other
edges.

A blow-up has two effects, one for each task of the adversary. First, a blow-up need not reduce the number of
colors that the adversary forces. Every vertex of
$\sigma(v)$ is joined to one server in the block of each earlier neighbor of $v$, so it cannot receive the color of
any of these servers. If the blocks are large enough, the servers can be chosen, after their colors are known, so that
an adversary that forces $K$ colors still forces $K$ colors when it presents a blow-up of its graph instead. Second, a blow-up lowers a large
chromatic number to about its logarithm. It is
an online version of one step of the arc-graph operator, whose effect on the chromatic number is
classical~\citep{HE1972,PR1981}.

\paragraph{Towers and the choice of servers.} A \emph{full tower} with $j$ levels is obtained from an ordered clique by
$j$ successive blow-ups in which an arc may have several servers: every arc has one or more, and no vertex serves two
arcs. A \emph{tower} is an initial
segment of a full tower, possibly followed by isolated vertices (\cref{def:tower}). A full tower
over a clique on $K$ vertices is $k$-colorable once $j$ is about $\lgs K$ (Corollaries~\ref{cor:towers} and~\ref{cor:levels}). What remains is to choose the servers so
that many colors are still forced, and we do this in two ways. \Cref{thm:main-det} chooses each server after its block
has been colored, from a color that occurs often in the block. This forces all $K$ colors, but every block must then
contain many vertices of one color, one for each later neighbor of the vertex that the block replaces, and the block size grows by a factor of about $K$
with each blow-up. \Cref{thm:main-logstar} gives all blocks one size and chooses the servers at random.
This costs only a single logarithmic factor in the number of colors, whatever the number of levels. \Cref{sec:overview}
explains both choices, that of \cref{thm:main-det} on small examples.

\begin{table}[t]
\centering\small
\renewcommand{\arraystretch}{1.25}
\begin{tabular}{@{}l>{\raggedright}p{4.9cm}>{\raggedright\arraybackslash}p{5.8cm}@{}}
\toprule
& deterministic & randomized (oblivious adversary)\\
\midrule
upper, $k=3$ & $O(n^{2/3}\log^{1/3}n)$ \citep{Kie1998} & $O(n^{1/2})$ \citep{Li2026}\\
upper, $k\ge 4$ & $\Ot(n^{1-6/(3k(k-1)-2)})$ \citep{KYY2025} & $O(n^{1-1/(k-1)}(\log n)^{1/(k-1)})$ \citep{Vis1990,Hal1997}\\
\midrule
lower, previous & $\Omega(\log^{k-1}n)$ \citep{Vis1990};\newline $2^{\Omega((\log n)^{1/3}(\log\log n)^{2/3})}$ \citep{Li2026}
  & $\Omega(\log^{k-1}n)$ \citep{Vis1990}; $2^{\Omega((\log n)^{1/3}(\log\log n)^{2/3})}$ \citep[App.~A]{Li2026}\\
lower, this paper & $n^{1/(\lgs n+O(1))}$ (\cref{thm:main-logstar});\newline $n^{1/O((\lgs n)^2)}$, explicit
  (\cref{thm:main-det}) & $n^{1/(\lgs n+O(1))}$ (\cref{thm:main-logstar})\\
\bottomrule
\end{tabular}
\caption{Online coloring of $k$-colorable graphs with $n$ vertices, for fixed $k\ge 3$. Every lower bound for $k=3$ is
also a lower bound for larger $k$. The lower bounds of \cref{thm:main-logstar} hold for every $n$ in the form
$n^{1/(\lgs n+2)}/O(\log n)$. For $k=4$ the deterministic upper bound is $\Ot(n^{14/17})$~\citep[Thm.~4.14]{KYY2025}.}
\label{tab:bounds}
\end{table}

\paragraph{Known graphs.} The bounds of \cref{thm:main-logstar,thm:main-det} also hold for single graphs. The
\emph{online chromatic number} $\chiol(G)$ of a graph $G$ is the least $c$ such that some deterministic online algorithm
uses at most $c$ colors on every presentation order of $G$~\citep[Def.~1]{BV2016}; cf.~\citep{GL1988}. For example,
trees on $N$ vertices have $\chiol\le\log_2N+1$. The reason is that First-Fit, the algorithm that gives each vertex
the least color not used by its earlier neighbors, uses at most this many colors on
them~\citep[Examples~1 and~5]{Kie1998survey}. Bipartite graphs on $N$ vertices have $\chiol=O(\log N)$~\citep[p.~320]{LST1989}. A
lower bound for a class of graphs does not directly give a lower bound for a single graph, because an adversary may
build different graphs against different algorithms. Our adversaries, however, can be made to produce the same graph, up
to isomorphism, against every algorithm. The bounds therefore hold even in Halld\'orsson's known-graph
game~\citep[\S1]{Hal2000}. In this game the algorithm is shown in advance a graph that contains the presented graph, but
it is not told which of its vertices each arriving vertex is. In our case, the graph shown is the final graph itself.

\begin{theorem}[known graphs]\label{thm:main-known}
For every $N\ge 3$ and $L=\lgs N$ there are triangle-free $3$-colorable graphs $G$ and $G'$ on $N$ vertices with
\[
  \chiol(G)\;\ge\;\big\lfloor N^{2/((L+2)(L+3))}\big\rfloor
  \qquad\text{and}\qquad
  \chiol(G')\;\ge\;\frac{N^{1/(L+2)}}{8(1+\ln 4N)} .
\]
Every randomized online algorithm that knows $G'$ in advance has a presentation order of $G'$ on which it uses at least
$N^{1/(L+2)}/(8(1+\ln 4N))$ colors in expectation. For $N\ge 2^{406}$, both bounds for $G'$ hold with $N^{1/(L+3)}$ in
place of $N^{1/(L+2)}/(8(1+\ln 4N))$. For every $k\ge 4$ the same holds for $k$-colorable graphs, with exponents
$2/(L(L+1))$ and $1/L$, and with $N^{1/(L+1)}$ for $N\ge 2^{201}$.
\end{theorem}

\paragraph{What the method can and cannot do.} Two further theorems explain the shape of these bounds. The first,
\cref{thm:main-transfer}, says that deterministic lower bounds for different values of $k\ge 3$ are equivalent up to a
polynomial change in the number of vertices. It is an online counterpart of the transfer of hardness between values of
$k$ in offline coloring~\citep[Lemma~3.3]{WZ2019}.

\begin{theorem}[transfer between values of $k$]\label{thm:main-transfer}
Let $k,k'\ge 3$. There is $j\le\lgs k'+1$ such that for all $N$ and all $K\ge 2$, if $\chidet_{k'}(N)\ge K$
then $\chidet_k(N^{(j+1)(j+2)/2})\ge K$. In particular, if $\chidet_{k'}(n)=\Omega(n^{\varepsilon})$ for a fixed $k'\ge 3$, then $\chidet_3(n)=\Omega(n^{\varepsilon'})$
with $\varepsilon'=2\varepsilon/((\lgs k'+2)(\lgs k'+3))$.
\end{theorem}

Equivalently, $N_k(K)\le N_{k'}(K)^{(j+1)(j+2)/2}$ for every $K\ge 2$, where $N_k(K)$ is the least number of
vertices with which an adversary can force every deterministic algorithm to use $K$ colors on $k$-colorable graphs
(\cref{obs:adv}).

Every hard instance of
\cref{thm:main-logstar,thm:main-det,thm:main-known} is a tower. By the second theorem, \cref{thm:main-ceiling}, First-Fit
uses at most $n^{1/(\lgs n-\lgs k-3)}$ colors on every $k$-colorable tower with at most $n$ vertices, for all large
$n$. Towers therefore cannot improve \cref{thm:main-logstar} beyond the additive constant in the denominator of the
exponent. \Cref{thm:main-ceiling} is stated in terms of the \emph{label posets} $P_j(k)$ defined next, which decide when a full tower is $k$-colorable.
A \emph{down-set}
of a poset is a subset that contains, with each of its elements, every smaller element. For a finite poset $P$, let
$\Jst(P)$ be the poset of its nonempty down-sets, ordered by inclusion. Let $P_0(k)$ be the antichain on
$[k]=\{1,\dots,k\}$, in which no two distinct elements are comparable, and let $P_{j+1}(k):=\Jst(P_j(k))$; thus $P_1(k)$
consists of the nonempty subsets of $[k]$. The sizes $|P_j(3)|=3,7,18,81,8\,569,\dots$ grow like a tower of
exponentials. A full tower with $j$ levels over a clique on $K$ vertices is $k$-colorable exactly when $K\le|P_j(k)|$
(\cref{cor:towers}).

\begin{theorem}[a ceiling for towers]\label{thm:main-ceiling}
Let $k\ge 3$. If First-Fit uses $c\ge 2$ colors on a $k$-colorable $j$-level tower $G$, then $c\le|P_j(k)|$ and
$|V(G)|\ge\binom{c-1}{i+1}$ for every $0\le i\le j$. Consequently there is $n_k$ such that for every $n\ge n_k$, First-Fit
uses at most $n^{1/(\lgs n-\lgs k-3)}$ colors on every $k$-colorable tower with at most $n$ vertices.
\end{theorem}

Two more results complete the picture (\cref{sec:limits}). First, the $2^k-1$ bound of the transparent model extends
from colors to labels. A \emph{labeling} by a poset $P$ gives every vertex $u$ an element $f(u)$ of $P$ such that
$f(u)\not\le f(v)$ for every later neighbor $v$ of $u$; for the antichain on $[k]$ it is a proper $k$-coloring. An
adversary that reveals such a labeling of its graph, each label right after the algorithm has colored the vertex,
forces at most $|\Jst(P)|$ colors (\cref{prop:cap}). A strong adversary must therefore delay its commitments or make
them in a poset with many down-sets: an adversary that reveals labels from $P_j(k)$ in this way forces at most $|P_{j+1}(k)|$
colors. Second, the
clique from which a tower starts can be replaced by an adversary of Kawarabayashi, Yoneda and
Yoneda~\citep[Lemma~4.3]{KYY2026comp}. This gives polynomial lower bounds at slowly growing chromatic number: for every
$n\ge 16$ there is $m\le\log_2\log_2n+\log_2\log_2\log_2n+3$ with $\chidet_m(n)\ge\lfloor(n/16)^{1/3}\rfloor$
(\cref{thm:slow}).

\paragraph{Formal verification.} All numbered results of this paper, including
\cref{thm:main-logstar,thm:main-det,thm:main-known,thm:main-transfer,thm:main-ceiling} exactly as stated, are
formalized in Lean~4 with Mathlib; the proofs are complete and use only Lean's standard axioms. The values found by
computer search are verified by checkers that are proved sound.\footnote{This excludes computed values on which no numbered statement depends, such as the best escape probabilities of \cref{app:oblivious}, for which we did not write a
verified checker.}

\paragraph{Organization.} \Cref{sec:overview} explains the constructions and proofs informally, and
\cref{sec:related} relates them to previous work. \Cref{sec:prelim} fixes
the model and \cref{sec:templates} develops the calculus of labels. \Cref{sec:blowup} proves \cref{thm:main-det},
\cref{sec:known} the part of \cref{thm:main-known} that concerns $G$, \cref{sec:logstar}
proves \cref{thm:main-logstar} and the rest of \cref{thm:main-known}, and \cref{sec:limits} the results on transfer and
limits; \cref{sec:discussion} lists open problems. The appendices verify the thresholds of \cref{thm:main-logstar}
(\cref{app:thresholds}), give smaller explicit instances and exact small cases (\cref{app:refined}) and an oblivious
version of the reservation construction (\cref{sec:rand}), and collect further results (\cref{app:further}).

\section{Technical overview}\label{sec:overview}

Our adversaries resolve the tension between forcing colors and staying $3$-colorable by keeping a clique in the
background and never showing it to the algorithm. They present instead an iterated blow-up of the clique, in which
every vertex is replaced by a block of independent vertices. We first explain the explicit bound of
\cref{thm:main-det}, which is weaker but simpler, and then, in \cref{sec:ov-logstar}, the change that gives
\cref{thm:main-logstar}. The proof of \cref{thm:main-det} rests on four facts.
\begin{enumerate}[label=\arabic*.,nosep]
\item \emph{A blow-up preserves forcing}: an adversary that forces $K$ colors still forces $K$ colors after one
  blow-up, if the servers are chosen after the colors are known (\cref{lem:force}).
\item \emph{A blow-up lowers the label poset by one level}: if the vertices of a graph can be labeled by nonempty
  down-sets of a poset $P$ so that no label is contained in the label of a later neighbor, then its blow-up can be
  labeled in the same way by elements of $P$ (\cref{lem:down}).
\item \emph{The label posets grow like a tower}, so a clique on $K$ vertices becomes $3$-colorable after $\lgs K+1$ blow-ups (\cref{lem:tower}).
\item \emph{Each blow-up multiplies the block size by about $K$}, which gives $K^{O((\lgs K)^2)}$ vertices in total
  (\cref{thm:L}).
\end{enumerate}
Facts~2 and~3 are classical results about arc graphs and shift graphs, and the reservation behind Fact~1 goes back
to Li (\cref{sec:related}). To our knowledge, three things are new: reserving single vertices, which makes each blow-up exactly
one step of the arc-graph calculus; iterating such blow-ups under a clique; and the resulting bound. We explain the four facts on the smallest interesting example,
forcing four colors on a $3$-colorable graph.

\subsection{One blow-up: four colors on a 3-colorable graph}\label{sec:ov-one}

Take the clique $K_4$ on $v_1,\dots,v_4$, in this order. It forces $4$ colors but is not $3$-colorable, so the
algorithm never sees it. Instead each $v_t$ is presented as a block $\sigma(v_t)$ of $7$ independent vertices. While the
algorithm uses at most $3$ colors, some color occurs at least $3$ times in $\sigma(v_1)$, since $7>2+2+2$. The
adversary calls this color the \emph{induced color} $\col(v_1)$, and it reserves three vertices of that color, one for
each later neighbor of $v_1$; we call them the \emph{servers} $x_{1,2}$, $x_{1,3}$ and $x_{1,4}$. Every vertex of
$\sigma(v_2)$ is joined to $x_{1,2}$, so none of them has color $\col(v_1)$, and the induced color $\col(v_2)$ differs from
$\col(v_1)$. In the same way $\sigma(v_3)$ is joined to $x_{1,3}$ and $x_{2,3}$, so $\col(v_3)$ is a third color. The
first vertex of $\sigma(v_4)$ is joined to $x_{1,4}$, $x_{2,4}$ and $x_{3,4}$, which carry three different colors, and
it needs a fourth, as in the complete play with $22$ vertices traced in \cref{fig:blowup}(b).

Two properties of the servers make this work. First, the servers are chosen after their colors are known, so the
adversary can take them from the popular color of each block. The induced
colors are therefore the colors that a deterministic online algorithm would give to the clique, and each of them is a
color of the real algorithm; so any adversary that forces $K$ colors still forces $K$ colors after the blow-up.
Second, each server is joined to a single later block. Had the adversary joined $\sigma(v_t)$ to every vertex of color
$\col(v_s)$ in every earlier block, one vertex of the induced color from each block would span a $K_4$. Private servers
are what let the chromatic number fall, as we show next. The price is the block size: a block whose vertex has $D$
later neighbors needs $D$ vertices of one color, and with at most $K-1$ colors available, $(K-1)(D-1)+1$ vertices
guarantee this. Reserving one copy per later neighbor is the simulation step of Li~\citep[Lemma~2.5]{Li2026}, with
single vertices in place of his bipartite gadgets.

\begin{figure}[t]
\centering
\resizebox{\linewidth}{!}{\figSchematic}\\[4pt]
{\small (a) Two arcs $w\to v$ and $w\to v'$ of the graph one level up, blown up; $K=4$, $D=3$, blocks of $R=7$
vertices.}\\[10pt]
\resizebox{0.74\linewidth}{!}{%
\begin{tikzpicture}[x=1cm,y=1cm]
\input{figures/fig_blowup_one.tex}
\end{tikzpicture}}\\[-1pt]
{\small (b) An exact play: the clique $K_4$ blown up once, with blocks of $7$ vertices.}
\caption{One reservation blow-up. (a) Some color occurs at least $D$ times in $\sigma(w)$; it becomes the induced color
$\col(w)$, and $D$ vertices of that color form the reserve (thick borders). Of the three later neighbors of $w$, the first and the third are drawn. The server $x_{w,v}$ is joined to all of
$\sigma(v)$, so no vertex of $\sigma(v)$ has color $\col(w)$, and $\col(v)\ne\col(w)$. (b) The adversary against the
algorithm that always takes its least-used allowed color among $1,\dots,K-1$. The blocks of $v_1,v_2,v_3$ have induced
colors $1,2,3$, and the first vertex of $\sigma(v_4)$ sees all three and receives color $4$, after $3\cdot 7+1=22$ vertices.}
\label{fig:blowup}
\end{figure}

\begin{figure}[t]
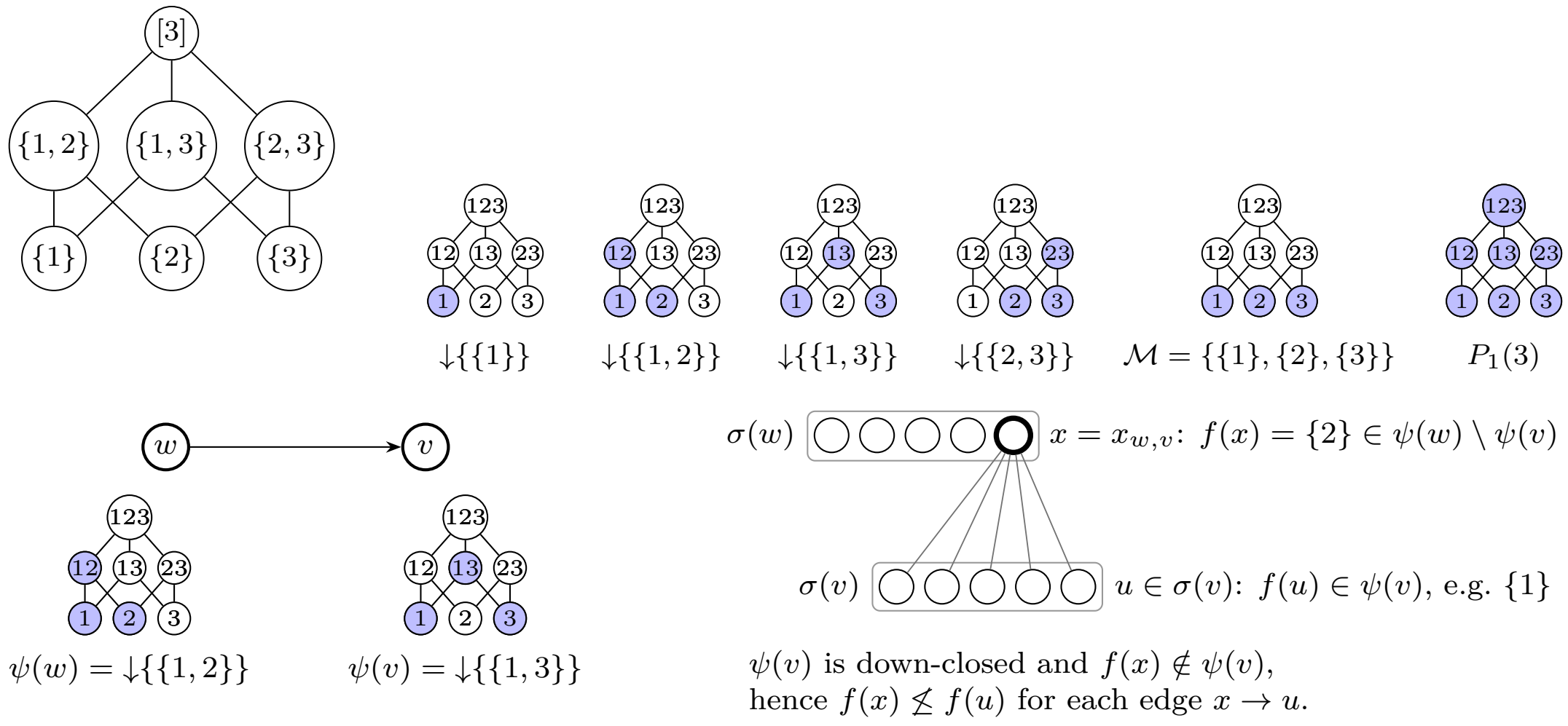

\centering
\resizebox{0.86\linewidth}{!}{\begin{tabular}{@{}c@{\hspace{0.8cm}}c@{}}
\figPosetOne & \figDownsets
\end{tabular}}

\medskip
\resizebox{0.86\linewidth}{!}{\figRule}
\caption{Labels one level up are nonempty down-sets. Top: $P_1(3)$, the nonempty subsets of $[3]$, and six of the
$18$ elements of $P_2(3)$, drawn as down-sets of $P_1(3)$. The four middle ones are pairwise incomparable, so two levels
raise the width from $3$ to $4$, after which the width grows like a tower. Bottom: the rule of \cref{lem:down}, from a labeling $\psi$ one level up to a labeling $f$ of the blow-up.
The server $x$ of the arc $w\to v$ gets a label in $\psi(w)\setminus\psi(v)$, every vertex $u$ of $\sigma(v)$ a label in $\psi(v)$, and since $\psi(v)$ is down-closed, $f(x)\not\le f(u)$.}
\label{fig:templates}
\end{figure}

\paragraph{Why the blow-up is 3-colorable.} Give each vertex of the clique a nonempty set of colors: $\{1,2,3\}$ to
$v_1$, and $\{1,2\}$, $\{1,3\}$, $\{2,3\}$ to $v_2$, $v_3$, $v_4$. No set is contained in a later one. The server
$x_{s,t}$ takes a color that lies in the set of $v_s$ but not in the set of $v_t$, and every other vertex of $\sigma(v_s)$
takes any color from the set of $v_s$. Every edge joins a server $x_{s,t}$ to a vertex of $\sigma(v_t)$, whose color
lies in the set of $v_t$ while the color of $x_{s,t}$ does not, so the coloring is proper. In
\cref{fig:blowup}(b), the servers $x_{1,2}$, $x_{1,3}$ and $x_{1,4}$ all receive online color $1$, while offline they
receive the colors $3$, $2$ and $1$, one outside each of the sets $\{1,2\}$, $\{1,3\}$ and $\{2,3\}$ of the blocks they
serve. The same argument works for
the blow-up of any graph $H$ whose vertices carry nonempty subsets of $\{1,2,3\}$ such that no set is contained in the set
of a later neighbor. For the clique $K_m$ this requires $m$ distinct sets, listed from larger to smaller, so one blow-up
supports a clique of at most $2^3-1=7$ vertices; indeed the blow-up of $K_8$ is not $3$-colorable
(\cref{cor:towers}). \Cref{fig:templates} shows the rule.

\subsection{Iterating: labels of labels, and a tower of exponentials}\label{sec:ov-iterate}

\begin{figure}[t]
\centering
\resizebox{0.9\linewidth}{!}{%
\begin{tikzpicture}[x=1cm,y=1cm]
\input{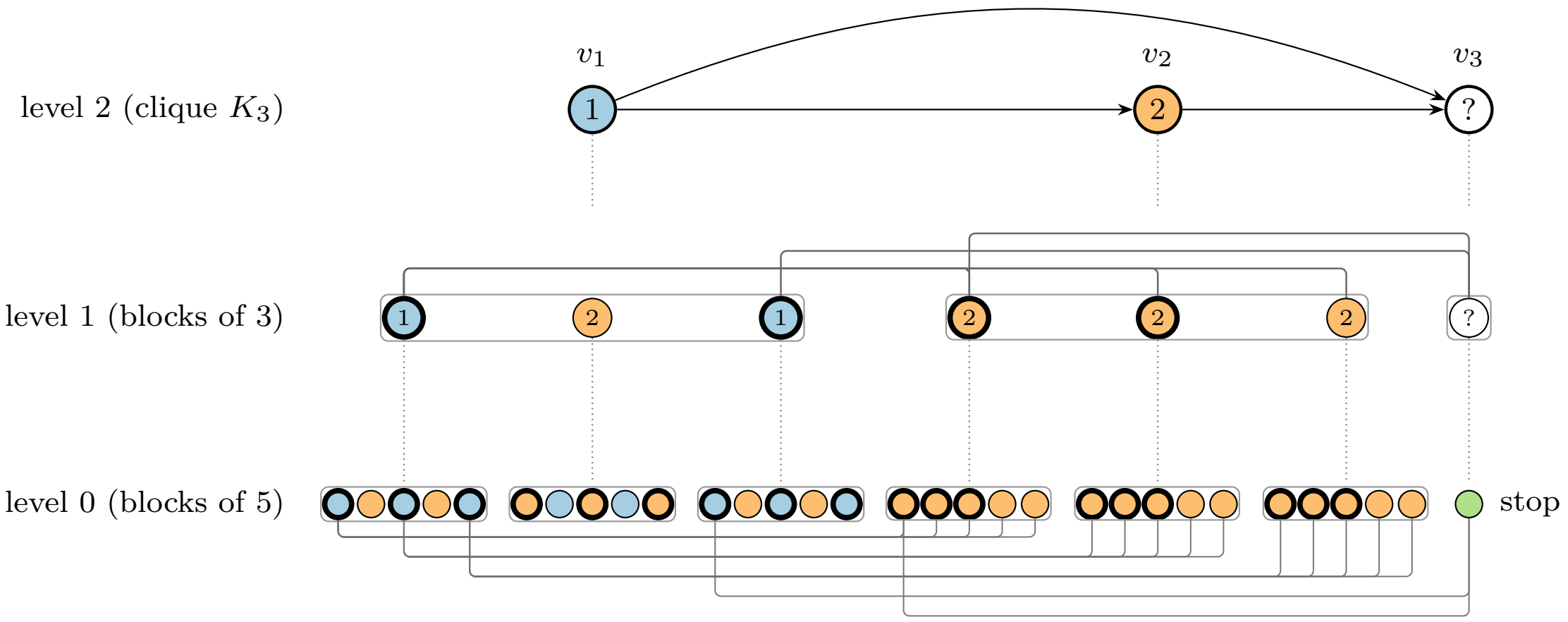}
\end{tikzpicture}}
\caption{A complete play with two blow-ups, $K=3$, against the least-used-color algorithm. The middle row shows the
induced colors of the nodes of level $1$; each lane below a row is one server joined to all of one block. The color of
every server is inherited from the reserve it was taken from, so the first vertex below $v_3$ sees the induced colors
of $v_1$ and $v_2$ and needs a third color.}
\label{fig:blowup2}
\end{figure}

To force more than seven colors, the adversary blows up twice. Each vertex of the clique becomes a block of
\emph{nodes of level $1$}, which the algorithm does not see either, and each node of level $1$ becomes a block of
presented vertices. A node of level $1$ is colored by the induced color of its own block. Such a node serves a whole
block of nodes of level $1$, so its own block of presented vertices must contain one server for each node of that block. For $K=8$ the blocks of the clique
vertices have $7\cdot 6+1=43$ nodes, the blocks of the nodes of level $1$ have $7\cdot 42+1=295$ vertices, and the
adversary forces $8$ colors with at most $88\,796$ vertices. \Cref{fig:blowup2} shows a complete two-level play for
$K=3$.

The colorability argument iterates as well. A node of level $1$ is labeled by a set of colors, as before, and a
vertex of the clique by a nonempty family of such sets that is closed under taking nonempty subsets. A family is not contained in
a later family exactly when it has a set that the later family lacks, and the server of the arc takes that set as its
label. This is Fact~2 for the poset $P_1(3)$ (\cref{lem:down}). Starting from the antichain $P_0(3)$ of three colors, the label posets $P_{j+1}(3)=\Jst(P_j(3))$ have
$3$, $7$, $18$, $81$ and $8\,569$ elements, and the clique $K_m$ survives $j$ full blow-ups exactly when $m\le|P_j(3)|$
(\cref{cor:towers}).

These posets grow like a tower of exponentials. If $P$ has $w$ pairwise incomparable elements, then the down-sets
generated by subsets of them are $2^w-1$ distinct elements of $\Jst(P)$, of which $\binom{w}{\lfloor w/2\rfloor}\ge
2^w/(w+1)$ are pairwise incomparable. The poset $P_2(3)$ already has four pairwise incomparable elements, and from then on
each level raises the width exponentially. Consequently $\lgs K+1$ blow-ups suffice for $K_K$, and at least
$\lgs K-2$ are necessary (\cref{cor:levels}).

This static part is classical. A reservation blow-up of an ordered graph $H$ is the arc graph of $H$ with a private sink
attached to every vertex, except that each sink arc appears as any number of copies, possibly none:
\begin{align*}
  \text{the server }x_{w,v}\;&\leftrightarrow\;\text{the arc }(w,v),\\
  \text{a vertex of }\sigma(v)\text{ that serves no arc}\;&\leftrightarrow\;\text{a copy of the sink arc }(v,s_v),\\
  x_{w,v}\text{ joined to all of }\sigma(v)\;&\leftrightarrow\;\text{the arcs }(w,v)\to(v,z)\text{ of }\delta(H^+),
\end{align*}
where $H^+$ is $H$ with the sinks $s_v$ added and $\delta$ is the arc-graph operator: the arc graph of a digraph has its
arcs as vertices, with an arc $(a,b)\to(b,c)$ for every two consecutive arcs.
Full towers over a clique are, in this sense, the shift graphs of Erd\H{o}s and Hajnal~\citep[proof of Thm.~2]{EH1968}, and
their colorability threshold and the tower growth go back to Harner and Entringer~\citep{HE1972}, Poljak and
R\"odl~\citep{PR1981}, Poljak~\citep{Poljak1991} and Duffus, Lefmann and R\"odl~\citep{DLR}, with recent treatments by
Rorabaugh, Tardif, Wehlau and Zaguia~\citep{RTWZ2018} and Bai et al.~\citep{BDHLW2026}. The nonempty blocks are why our
labels are nonempty down-sets.

\paragraph{Counting.} A node with $D$ later neighbors (post-degree $D$) needs a block of $(K-1)(D-1)+1$ nodes, and each node of that
block may in turn serve a block of this size. The post-degree, and with it the block size, therefore grows by a factor of about
$K$ per blow-up: the blocks of the top vertices (the vertices of the clique) have about $K^2$ nodes, the blocks one level down about $K^3$, and
after $i$ blow-ups about $K^{i+1}$. With $j$ blow-ups the graph has at most
\[
  K\cdot K^2\cdot K^3\cdots K^{j+1}\;=\;K^{(j+1)(j+2)/2}
\]
vertices. With $j=\lgs K+1$ this is $K^{O((\lgs K)^2)}$, and solving for $K$ gives \cref{thm:main-det}.

\subsection{Removing one \texorpdfstring{$\log^*$}{log*}: random servers and unused mass}\label{sec:ov-logstar}

The quadratic exponent comes from the counting: the block size grows by a factor $K$ per blow-up, because a server must
hand a single color to an entire block. \Cref{thm:main-logstar} gives up reserving many vertices of one color. In a
\emph{uniform tower} every arc still has its own server, but all blocks have one size $A=2m$, where $m$ is the number of
top vertices, and the servers are chosen at random. The tower is presented in $m$ phases, one subtree of a top vertex per
phase, and the servers of a phase, its \emph{wiring}, are chosen as follows: every earlier top vertex supplies a uniformly
random child that has not served yet, and below the top the children of a node are matched to the children of its
servers by independent uniformly random bijections.

The analysis follows, for each color, its \emph{unused mass}: the presented vertices of that color
that can still become servers, weighted so that the subtree of each top vertex has mass $1$. Once enough unused vertices
of some color, say red, are available, a new vertex is likely to receive a red server, so the
algorithm has few chances to make red heavier. Yet about three quarters of the total mass is still unused at the end, so
many colors share it. The proof has four steps.
\begin{enumerate}[label=\arabic*.,nosep]
\item \emph{A color with unused mass $u$ is missed with probability at most $e^{-u}$.} Under the random wiring, a
  presented vertex has one earlier neighbor below each earlier top vertex, and that neighbor is uniform over the
  presented vertices below the children that have not served (\cref{lem:miss}).
\item \emph{Heavy colors grow slowly.} Call a color heavy if its unused mass is at least $\ln 4m$. Fewer than $m$
  colors are heavy, so in expectation at most a quarter of each phase misses some heavy color, and only that part can
  add mass to heavy colors.
\item \emph{The final mass is known.} The top vertices contribute mass $m$ and the servers remove $\binom m2/A$, so the
  unused mass at the end is exactly $(3m+1)/4$ (\cref{lem:mass}). A phase adds at most $1$ to any color, so a color
  becomes heavy with mass below $\ln 4m+1$, and afterwards it gains only through the vertices of Step~2.
\item \emph{Counting.} Comparing the final mass with what light and heavy colors can hold gives at least
  $(2m+1)/(4(1+\ln 4m))$ colors in expectation, at every number of levels (\cref{thm:T}).
\end{enumerate}
To our knowledge Steps~1 to~3 are new; the tower calculus supplies colorability. The quantitative core is that no level loses anything: the uniform tower has $n=m(2m)^j$ vertices and is $k$-colorable once $m\le|P_j(k)|$, so $j=\lgs
n+O(1)$ levels suffice, and the bound is $n^{1/(j+1)}/O(\log n)=n^{1/(\lgs n+O(1))}$. The price is the factor $1+\ln 4m$; we prove
$n^{1/(\lgs n+3)}$ for $n\ge 2^{406}$.

\subsection{Variants: deterministic adversaries, randomized algorithms and known graphs}\label{sec:ov-variants}

The random wiring never looks at the colors, so by Yao's principle \cref{thm:main-logstar} holds against randomized
algorithms. The same analysis gives a deterministic adversary that chooses
the wiring of each phase from the colors already used and forces the bound on every play (\cref{thm:T}(ii)).

The wiring changes only the order in which the vertices of a block arrive, not the graph: every wiring gives the same
full tower up to isomorphism (\cref{lem:K1,lem:uniform-graph}). The algorithm may therefore be told the final graph, and the bounds of \cref{thm:main-det,thm:main-logstar} hold for single graphs (\cref{thm:main-known}). For
\cref{thm:main-det} the adversary must also complete every block after winning (\cref{lem:K}).

\subsection{Limits}\label{sec:ov-limits}

First-Fit colors every block of a full tower with the color it gives the corresponding node one level up
(\cref{lem:FF}), so on an initial segment of a full tower it uses exactly as many colors as there are top vertices with a presented
descendant. If it uses $c$ colors, the subtrees of the first $c-1$ top vertices are presented in full. Following
servers down one level maps a walk of $i+1$ top vertices to a walk of $i$ nodes of level $j-1$, injectively, and after
$i$ such steps each of the $\binom{c-1}{i+1}$ increasing sequences of top vertices has its own node of level $j-i$.
The tower therefore has at least $\binom{c-1}{i+1}$ vertices for every $i\le j$. At least $\lgs c-O(1)$ levels are
necessary, so we may take $i=\lgs c-O(1)$, and every tower adversary is capped at $n^{1/(\lgs n-O(1))}$ colors
(\cref{thm:main-ceiling}), an additive constant in the exponent's denominator away from \cref{thm:main-logstar}.

A second limit concerns every adversary that commits to labels, not only towers. Suppose that, right after
each vertex is colored, the adversary reveals its label in a labeling of the graph by a poset $P$. An algorithm that knows the adversary can color each vertex $v$ by the set of labels that $v$
could still receive, which is a nonempty down-set of $P$, and two adjacent vertices always get different sets
(\cref{prop:cap}). For the antichain of $k$ colors this is a variant of the classical $2^k-1$ algorithm for the transparent
model~\citep[\S4.5]{HS1994}. In general, an adversary that commits to labels from $P_j(k)$ forces at most
$|P_{j+1}(k)|$ colors.

Private servers matter as well. In Vishwanathan's levels, one selected set of vertices
serves all later neighbors at once and every gadget contains an edge; such levels raise the clique number by one each,
so they can be stacked at most $k-1$ deep (\cref{prop:S}), which is where the exponent of $\Omega(\log^{k-1}n)$ comes
from. Li's levels reserve a private copy of a gadget for each future neighbor~\citep[Lemma~2.5]{Li2026}, and our blow-ups a private vertex.

Finally, the clique at the top can be replaced by any adversary. With an adversary for $k'$-colorable graphs at the top,
$j$ blow-ups produce an adversary for $k$-colorable graphs (\cref{thm:main-transfer}). With the pairing adversary of
Kawarabayashi, Yoneda and Yoneda~\citep[Lemma~4.3]{KYY2026comp}, against which every color is used at most twice on $2^\ell$ vertices of chromatic number at most $2\ell$, a single blow-up forces $\lfloor(n/16)^{1/3}\rfloor$ colors on graphs of
chromatic number about $\log\log n$ (\cref{thm:slow}).

\section{Relation to previous work}\label{sec:related}

\paragraph{Lower bounds for online coloring.} Previous lower bounds for fixed $k$ that grow with $n$ force many colors on an
independent set and then amplify. Vishwanathan's adversary joins a new gadget of chromatic number one less completely to
an independent set that already carries many colors, so that the gadget's colors must be
new~\citep[\S\S3.1--3.2]{Vis1990}. Kierstead's handbook chapter states the deterministic version~\citep[Thm.~4.4]{KieHandbook1998},
and Kierstead and Kolossa make the construction perfect~\citep[Thm.~0.2]{KK1996}. Li~\citep{Li2026} works with $t$-fold
colorings and a simulation lemma that reserves a separate copy of a gadget for each future neighbor of a simulated
vertex~\citep[Lemma~2.5]{Li2026}; every level of his construction keeps chromatic number $3$, and about $\log K/\log\log
K$ levels, each multiplying the post-degree, give his bound~\citep[\S2.3]{Li2026}. Without reservations, levels of this
kind raise the clique number by one each and can be stacked at most $k-1$ deep (\cref{prop:S}). Our construction keeps
Li's reservation and drops the rest: there are no gadgets and no fold reduction, a clique at the top does all the forcing,
and the blow-ups only lower the chromatic number. Because the reserved copies are single vertices, each blow-up is exactly
one step of the arc-graph calculus, which lowers the chromatic number exponentially~\citep[Thms.~8--9]{HE1972}, so
$\lgs K+O(1)$ blow-ups suffice. Li's randomized lower bound hides the reserved copy in a random
index~\citep[Lemma~A.4]{Li2026}, which is what we do in \cref{sec:rand}. In a different direction, Halld\'orsson and
Szegedy force about $n/\log n$ colors on graphs of chromatic number at most $\log n$~\citep{HS1994}, and Kawarabayashi,
Yoneda and Yoneda force at least $n/2$ colors on $n$-vertex graphs of chromatic number at most $2\lceil\log_2
n\rceil$~\citep[Lemma~4.3 and proof of Thm.~1.1]{KYY2026comp}; \cref{thm:slow} uses the latter construction as a top.
Li, Narayan and Pankratov define ``saturated bins''~\citep[Def.~2.1]{LNP2020}, a notion that is implicit in some earlier adversaries; the
saturated colors of the one-level adversary in \cref{sec:one-level} are a different notion, defined by a flow.

\paragraph{Shift graphs and arc graphs.} A reservation blow-up of $H$ is the arc graph $\delta(H^+)$ of $H$ with a private sink at every vertex, up to the number
of copies of each sink arc (\cref{sec:ov-iterate}). Iterating $\delta$ on a transitive tournament gives the shift
graphs of Erd\H{o}s and Hajnal~\citep[proof of Thm.~2]{EH1968}, a classical family with large chromatic number and no short
odd cycles~\citep[p.~277]{RTWZ2018}. The one-level bounds relating $\chi(\delta(H))$ and $\chi(H)$ are due to Harner and
Entringer~\citep[Thms.~8--9]{HE1972}, the lower one already to Erd\H{o}s and Hajnal~\citep[Thm.~5]{EH1968}, with an
exact formula for symmetric digraphs by Poljak and R\"odl~\citep[Thm.~1.1]{PR1981}. \Cref{sec:templates} gives the exact sources of the label calculus: Poljak's adjunction~\citep[Lemma~1]{Poljak1991} is the
content of \cref{lem:down,lem:up}; the colorability threshold of shift graphs counts all iterated down-sets, with an
equivalent statement about monotone paths by Moshkovitz and Shapira~\citep[Thm.~5 and Eq.~(10)]{MS2014}, while ours,
\cref{cor:towers}, counts nonempty ones; and the tower growth rests on the recursion of Poljak and
R\"odl~\citep[Rem.~3.2]{PR1981} and of Duffus, Lefmann and R\"odl~\citep[Eq.~(1)]{DLR}; see also~\citep[\S4.5]{Kriegel2018}. Offline, the arc graph transfers NP-hardness of
approximate coloring between values of $k$~\citep[Lemma~3.3 and Prop.~2.1]{WZ2019}; \cref{thm:main-transfer} is an
online counterpart, in which the reduction maps adversaries instead of instances and costs a polynomial in the number
of vertices.

\paragraph{Online games on shift graphs.} Shift graphs have appeared in online coloring before, but to our knowledge never to force many colors on graphs of constant chromatic number. The triangle-free presenter of Diwan, Kenkre and Vishwanathan~\citep[Thm.~16]{DKV2013} builds shift graphs
when the algorithm is First-Fit, as observed by Gutowski, Kozik, Micek and Zhu~\citep[p.~6]{GKMZ2014}, who also keep the
vertices of a popular color of a ``fan'' for later use~\citep[proof of Thm.~2]{GKMZ2014}; neither construction bounds the chromatic
number of the presented graph by a constant. In the Builder--Painter game of P\'erez-Gim\'enez, Pra\l{}at and
West~\citep{PGPW2021}, for paths with two edges, Painter colors an induced subgraph of a shift graph online and the analysis maintains iterated
down-set labels~\citep[\S3]{PGPW2021}, but Painter is told which vertex of the shift graph is presented, and Builder
forces only about the static chromatic number. In our setting the identity of a vertex is decided after it is colored,
and this is what makes forcing possible; the same device appears in Gy\'arf\'as and Lehel's tree
construction~\citep[Thm.~2.5]{GL1988}.

\paragraph{Iterated logarithms.} The $\lgs$ behavior of iterated arc graphs is familiar from distributed computing.
Linial's $\Omega(\lgs n)$ lower bound for $3$-coloring a cycle~\citep[Thm.~2.1]{Linial1992} rests on the same operator: his
neighborhood graphs contain shift graphs, and one application of the operator lowers the chromatic number at most
exponentially~\citep[Props.~2.1--2.2]{Linial1992}. The same fact limits the number of levels in our constructions.

\paragraph{Models.} The transparent model~\citep{HS1994}, the random-order model (see~\citep{YY2026random} and the
references there) and the known-graph model~\citep{Hal2000} were discussed in \cref{sec:intro}; in random order,
First-Fit still needs polynomially many colors on some bipartite graphs~\citep{LiLi2026}. In the model of Bartal, Fiat and Leonardi~\citep{BFL2006},
the algorithm knows a host graph and is told which host vertex each arrival is; their polynomial lower bounds concern
graphs of polynomially large chromatic number. B\"ohm and Vesel\'y showed that deciding the online chromatic number of a
given graph is PSPACE-complete~\citep[Thm.~1.1]{BV2016}. Li, Narayan and Pankratov restrict the adversary to graphs with
few connected components at all times~\citep{LNP2020}. Our adversaries start by presenting many isolated vertices, those below the first top
vertex, but presenting one extra vertex first and joining it to all later vertices keeps the graph connected at the
cost of one color. The bound of \cref{thm:main-det} for $k=3$, with one more vertex, therefore also holds for $4$-colorable
graphs that are connected at every step of the presentation.

\paragraph{Offline coloring.} In polynomial time, $3$-colorable graphs can be colored with $O(n^{0.19539})$
colors~\citep{BHL2026}, after a line of work that starts with Wigderson's
$O(\sqrt n)$~\citep{Wig1983,Blum1994,BK1997,KTY2024}, and by a recent result of Fei, Minzer and Wang, coloring them
with any constant number of colors is NP-hard~\citep{FMW2026}. These are statements about polynomial time; online
lower bounds such as ours hold against computationally unbounded algorithms.

\section{Preliminaries}\label{sec:prelim}

\paragraph{Online coloring.} An \emph{ordered graph} is a finite simple graph whose vertices carry a linear order, the
\emph{arrival order}. For a vertex $v$ we write $N^-(v)$ for the set of its earlier neighbors, and we call the number of
its later neighbors its \emph{post-degree}, following Li~\citep{Li2026}. A deterministic online coloring algorithm receives the vertices one at a
time; when $v$ arrives the algorithm learns $N^-(v)$ and nothing else, and it must give $v$ a color (a natural number)
that differs from the colors of $N^-(v)$. Colors are never changed, and the cost of the algorithm on an input is the
number of distinct colors it uses. The inputs are $k$-colorable for a fixed $k\ge 1$, which the algorithm knows; our
lower bounds hold even if it also knows the number $n$ of vertices.

\begin{definition}[online chromatic numbers]\label{def:chi}
The number $\chidet_k(n)$ is the minimum, over deterministic online algorithms, of the maximum cost over all ordered
$k$-colorable graphs on $n$ vertices. Similarly, $\chirand_k(n)$ is the minimum, over randomized online algorithms (probability
distributions over deterministic ones), of the maximum expected cost over all ordered $k$-colorable graphs on $n$
vertices.
\end{definition}

Both quantities are non-decreasing in $n$, since a bad input on $n$ vertices stays bad after an isolated vertex is
appended. For a fixed $n$, a deterministic algorithm has one of finitely many vectors of costs on the ordered graphs with
$n$ vertices, so it suffices to consider distributions with finite support. A worst input for a deterministic algorithm can be built \emph{adaptively}, by looking at the colors already
given, whereas the input for a randomized algorithm is fixed in advance (an \emph{oblivious} adversary). We formalize
adaptive constructions as follows.

\begin{definition}[adversaries]\label{def:adv}
An \emph{adversary} is a deterministic rule that, given the history of a play (the ordered graph presented so far and
the colors given to it), either stops or presents a new vertex together with its set of earlier neighbors. It
\emph{forces $K$ colors} if every play against every deterministic online algorithm is finite and ends with at least
$K$ colors in use. It has \emph{size} at most $N$ and \emph{post-degree} at most $D$ if in every play at most $N$
vertices are presented and every vertex receives at most $D$ later neighbors. It \emph{lands in} a class of ordered
graphs if the final graph of every play belongs to the class.
\end{definition}

Let $N_k(K)$ be the least size of an adversary that forces $K$ colors and lands in the $k$-colorable graphs. If such an
adversary of size at most $n$ exists, then $\chidet_k(n)\ge K$: run the adversary and append isolated vertices until
there are $n$. The converse also holds. Consider the game in which the adversary presents at most $n$ vertices,
stopping when it likes, and wins if the play ends with at least $K$ colors in use and a $k$-colorable graph. Once
colors are renamed in the order of their first use, this is a finite game of perfect information, so by Zermelo's theorem
one of the two players has a winning strategy: an adversary of size at most $n$ that forces $K$ colors and lands in the
$k$-colorable graphs, or an algorithm that uses fewer than $K$ colors on every ordered $k$-colorable graph with at
most $n$ vertices. Only the proof of \cref{thm:main-transfer} uses the converse.

\begin{observation}[determinacy]\label{obs:adv}
We have $\chidet_k(n)\ge K$ if and only if $N_k(K)\le n$. In particular, it suffices to exhibit, for every deterministic
algorithm, an ordered $k$-colorable graph on at most $n$ vertices on which the algorithm uses $K$ colors.
\end{observation}

For randomized algorithms we use the easy direction of Yao's principle~\citep{Yao1977}.

\begin{lemma}[Yao's principle]\label{lem:yao}
Let $\mathcal D$ be a distribution over ordered $k$-colorable graphs on exactly $N$ vertices such that every
deterministic online algorithm uses at least $x$ colors in expectation on $G\sim\mathcal D$. Then
$\chirand_k(n)\ge x$ for every $n\ge N$. In particular, if every deterministic online algorithm uses at least $K$ colors
with probability at least $1-\varepsilon$, then $\chirand_k(n)\ge(1-\varepsilon)K$.
\end{lemma}

\begin{proof}
A randomized algorithm is a distribution over deterministic algorithms that is independent of $G$. By Fubini's theorem
its expected cost on $G\sim\mathcal D$ is at least $x$, so some $G$ in the support forces this expected
cost. Appended isolated vertices do not change the colors of the first $N$ vertices.
\end{proof}

If all graphs in the support of $\mathcal D$ are presentations of one graph $G$, the same argument shows that every
randomized algorithm that knows $G$ in advance has some presentation order of $G$ on which it uses at least $x$
colors in expectation.

\emph{First-Fit} gives each vertex the least color that is not used on its earlier neighbors. It uses at most one more
color than the largest number of earlier neighbors of a vertex.

\paragraph{Posets and labels.} Let $P$ be a finite poset. A set $U\subseteq P$ is a \emph{down-set} if $x\le y\in U$
implies $x\in U$, and ${\down}U$ denotes the down-set generated by $U$. We write $\Jst(P)$ for the poset of
\emph{nonempty} down-sets of $P$, ordered by inclusion, and $\width(P)$ for the largest size of an antichain of $P$. The
\emph{label posets} are $P_0(k)$, the antichain on $[k]$, and $P_{j+1}(k):=\Jst(P_j(k))$ for $j\ge 0$.

\begin{definition}[templates and labelings]\label{def:template}
The \emph{template} $\Tmpl(P)$ is the digraph on $P$ with an arc $a\to b$ whenever $a\not\le b$. A
\emph{$P$-labeling} of an ordered graph $G$ is a map $f:V(G)\to P$ such that $f(u)\not\le f(v)$ for every edge $uv$
with $u$ earlier than $v$; we then say that $G$ \emph{maps to} $\Tmpl(P)$.
\end{definition}

A $P$-labeling restricts to every induced ordered subgraph. In the antichain $P_0(k)$ we have $a\not\le b$ exactly when
$a\ne b$, so a $P_0(k)$-labeling is a proper $k$-coloring. The posets $P_j(k)$ label the levels of our
constructions: a node at level $j$ is labeled by a down-set of labels of level $j-1$, down to the colors at level $0$.

\paragraph{Iterated logarithms.} Logarithms are binary unless written $\ln$. Let $\tow(0)=1$ and $\tow(i+1)=2^{\tow(i)}$, and let
$\lgs x=\min\{i\ge 0:\tow(i)\ge x\}$. For example, $\lgs 2=1$, $\lgs 16=3$, $\lgs 65536=4$, and $\lgs n\le 5$ for every $n\le
2^{65536}$.

\paragraph{Blow-ups.} All our hard instances are iterated blow-ups. The following static definition covers them; the
later sections add the online choices.

\begin{definition}[blow-up]\label{def:blowup}
Let $H$ be an ordered graph. A \emph{blow-up} of $H$ is an ordered graph $G$ together with a partition of $V(G)$ into
nonempty independent \emph{blocks} $\sigma(v)$, $v\in V(H)$, and a partial map $\serve:V(G)\to V(H)$, such that
\begin{enumerate}[label=(\roman*),nosep]
\item the blocks arrive one after the other, in the order of $H$;
\item if $x\in\sigma(w)$ and $\serve(x)=v$, then $v$ is a later neighbor of $w$ in $H$, and we say that $x$
  \emph{serves} the arc $w\to v$;
\item the edges of $G$ are exactly the pairs $xu$ with $\serve(x)=v$ and $u\in\sigma(v)$.
\end{enumerate}
The blow-up is \emph{full} if every arc of $H$ is served by at least one vertex, and it is a \emph{reservation
blow-up} if every arc is served by exactly one vertex, its \emph{server}.
\end{definition}

In other words, every vertex of a blow-up is joined to at most one later block, completely, and has no other later neighbors.

\begin{definition}[towers]\label{def:tower}
A \emph{full $j$-level tower} over the arrival-ordered clique $K_m=(v_1,\dots,v_m)$ is obtained by taking $j$ successive
full blow-ups: $H^{(j)}=K_m$, and $H^{(i)}$ is a full blow-up of $H^{(i+1)}$ for $i=j-1,\dots,0$. The vertices of
$H^{(\ell)}$ are the \emph{nodes of level $\ell$}; the nodes of level $j$ are the \emph{top vertices}, and the nodes of
level $0$ are the vertices of the presented graph $H^{(0)}$. The blocks define a forest whose roots are the top vertices,
and the \emph{subtree} of a node consists of all its descendants. A \emph{$j$-level tower} is an initial segment of a full $j$-level tower, possibly
followed by isolated vertices.
\end{definition}

A full tower is presented \emph{subtree by subtree}: by \cref{def:blowup}(i), applied at every level, the presented
descendants of every node form a contiguous interval of the arrival order, and these intervals come in the order of
the nodes. The block of a node of level $\ell\ge 1$ consists
of nodes of level $\ell-1$. If $x$ serves an arc $w\to v$ between
nodes of level $\ell$, then $x$ is a node of level $\ell-1$ joined to all of $\sigma(v)$ in $H^{(\ell-1)}$.

\section{The calculus of labels}\label{sec:templates}

This section is static: it decides which full towers over a clique are $k$-colorable and how fast the label posets grow.
The later adversaries only have to produce blow-ups; colorability then follows from
\cref{lem:down}, whatever choices were made online. The results are known in the language of arc graphs and
shift graphs, and we give the attribution with each statement. We include the short proofs because the blocks of our
blow-ups are nonempty, which replaces the down-sets of the classical statements by nonempty ones.

\begin{lemma}[cliques and antichains; cf.~{\citep[Lemma~2.4]{BDHLW2026}}]\label{lem:tops}
Let $P$ be a finite poset.
\begin{enumerate}[label=(\roman*),nosep]
\item The arrival-ordered clique $K_m$ maps to $\Tmpl(P)$ if and only if $m\le|P|$.
\item If $P$ has an antichain of size $k'$, then every $k'$-colorable ordered graph maps to $\Tmpl(P)$.
\end{enumerate}
\end{lemma}

\begin{proof}
(i) Every two vertices of a clique are adjacent, and $f(a)\not\le f(a)$ fails, so a labeling of a clique is injective.
Conversely, list $m$ distinct elements $p_1,\dots,p_m$ of $P$ so that larger elements come first. If $s<t$ and $p_s\le
p_t$, then $p_s<p_t$, so $p_t$ would be listed before $p_s$. Hence $p_s\not\le p_t$ for all $s<t$. (ii) Distinct elements
of an antichain are incomparable, so composing a proper $k'$-coloring with an injection of the colors into the
antichain gives a labeling.
\end{proof}

The next two lemmas are the heart of the calculus: labels descend through a blow-up, from $\Jst(P)$ to $P$, and they
ascend through a full blow-up, from $P$ to $\Jst(P)$. In the language of arc graphs this is Poljak's
adjunction~\citep[Lemma~1]{Poljak1991}, in the form of Rorabaugh, Tardif, Wehlau and Zaguia~\citep[Lemmas~4 and~7]{RTWZ2018},
who credit their Lemma~4 to Foniok and Tardif~\citep{FT2015}, and of Bai et al.~\citep[Lemma~2.3]{BDHLW2026}. Their
templates use $\not\ge$ where ours use $\not\le$.

\begin{lemma}[labels descend through blow-ups]\label{lem:down}
If $H$ maps to $\Tmpl(\Jst(P))$, then every blow-up of $H$ maps to $\Tmpl(P)$.
\end{lemma}

\begin{proof}
Let $\psi$ be a $\Jst(P)$-labeling of $H$, and let $G$ be a blow-up of $H$. We choose $f(x)$ for $x\in\sigma(w)$. If
$x$ serves an arc $w\to v$, then $\psi(w)\not\subseteq\psi(v)$, and we choose $f(x)\in\psi(w)\setminus\psi(v)$. Otherwise
we choose any $f(x)\in\psi(w)$, which is possible because $\psi(w)$ is nonempty. Every edge of $G$ joins a vertex $x$ that
serves some arc $w\to v$ to a later vertex $u\in\sigma(v)$. Now $f(u)\in\psi(v)$, and $\psi(v)$ is a down-set that does not
contain $f(x)$, so $f(x)\not\le f(u)$.
\end{proof}

The proof uses only that blocks are independent and that each vertex of a block $\sigma(w)$ is joined to at most one
later block, completely, namely $\sigma(v)$ for a later neighbor $v$ of $w$.
It does not depend on the block sizes, on how many vertices serve an arc, or on how the servers were chosen, and this
separates forcing from colorability in all our constructions.

\begin{lemma}[labels ascend through full blow-ups]\label{lem:up}
If a full blow-up $G$ of $H$ maps to $\Tmpl(P)$ via $f$, then $H$ maps to $\Tmpl(\Jst(P))$ via
$\psi(v):={\down}f(\sigma(v))$.
\end{lemma}

\begin{proof}
Each $\psi(v)$ is a nonempty down-set. Let $w\to v$ be an arc of $H$ and let $x\in\sigma(w)$ serve it. Then $f(x)\not\le
f(u)$ for every $u\in\sigma(v)$, so $f(x)\notin\psi(v)$, while $f(x)\in\psi(w)$. Hence $\psi(w)\not\subseteq\psi(v)$.
\end{proof}

Applying \cref{lem:down,lem:up} once per level and \cref{lem:tops}(i) at the top gives the colorability threshold of
full towers. For shift graphs the threshold counts all iterated down-sets~\citep[\S3]{RTWZ2018}, \citep[Thm.~2.2]{BDHLW2026},
\citep[Thm.~5 and Eq.~(10)]{MS2014}; ours counts nonempty ones.

\begin{corollary}[colorability of iterated blow-ups]\label{cor:towers}
Let $k\ge 2$, $j\ge 0$ and $m\ge 1$. If $m\le|P_j(k)|$, then every $j$-fold blow-up of $K_m$, and every induced ordered
subgraph of one, is $k$-colorable. Conversely, if some $j$-fold blow-up of $K_m$ in which all $j$ blow-ups are full is
$k$-colorable, then $m\le|P_j(k)|$.
\end{corollary}

\begin{proof}
If $m\le|P_j(k)|$, then $K_m$ maps to $\Tmpl(P_j(k))$ by \cref{lem:tops}(i), and $j$ applications of \cref{lem:down} give
a $P_0(k)$-labeling of the $j$-fold blow-up, that is, a proper $k$-coloring; it restricts to every induced subgraph.
Conversely, a proper $k$-coloring is a $P_0(k)$-labeling for every arrival order. If all $j$ blow-ups are full, then
$j$ applications of \cref{lem:up} show that $K_m$ maps to $\Tmpl(P_j(k))$, and \cref{lem:tops}(i) gives $m\le|P_j(k)|$.
\end{proof}

It remains to understand $|P_j(k)|$. Each level can increase the size at most exponentially, and the width grows at least
exponentially once it is at least $4$. For $k=3$ the width is $3$ after one level and $4$ after two.

\begin{lemma}[growth; cf.~{\citep[Obs.~2.5--2.6]{BDHLW2026}}]\label{lem:growth}
Let $P$ be a finite poset.
\begin{enumerate}[label=(\roman*),nosep]
\item If $A\subseteq P$ is an antichain of size $w\ge 1$, then $U\mapsto{\down}U$ embeds the nonempty subsets of $A$,
  ordered by inclusion, into $\Jst(P)$. In particular, $|\Jst(P)|\ge 2^w-1$ and $\width(\Jst(P))\ge\binom{w}{\lfloor w/2\rfloor}$.
\item The poset $\Jst(P)$ has at most $2^{|P|}-1$ elements, and $p\mapsto{\down}\{p\}$ embeds $P$ into $\Jst(P)$.
\item The four nonempty down-sets ${\down}\{\{1,2\}\}$, ${\down}\{\{1,3\}\}$, ${\down}\{\{2,3\}\}$ and
  $\mathcal M=\{\{1\},\{2\},\{3\}\}$ of $P_1(3)$ are pairwise incomparable, so $\width(P_2(3))\ge 4$.
\end{enumerate}
\end{lemma}

\begin{proof}
(i) If ${\down}U\subseteq{\down}U'$, then every $a\in U$ lies below some $a'\in U'$, and $a=a'$ because $A$ is an
antichain; so $U\subseteq U'$. The converse holds because ${\down}$ is monotone. The subsets of $A$ of size $\lceil w/2\rceil$ are nonempty and
form an antichain of size $\binom{w}{\lceil w/2\rceil}=\binom{w}{\lfloor w/2\rfloor}$~\citep{Sperner}. (ii) Down-sets are subsets of $P$. (iii) None of the three
generated down-sets contains another one or $\mathcal M$, since each lacks the other two pairs and one singleton, and
$\mathcal M$ contains none of them, since it has no pair.
\end{proof}

Part (iii) is due to Poljak~\citep[Lemma~3]{Poljak1991}. Iterating part (i) gives a tower through the recursion
$T_0=4$, $T_{i+1}=\binom{T_i}{\lfloor T_i/2\rfloor}$ (that is, $4,6,20,184\,756,\dots$) of Poljak and
R\"odl~\citep[Rem.~3.2]{PR1981} and of Duffus, Lefmann and R\"odl~\citep[Eq.~(1)]{DLR}, who use it to color shift graphs
of tower size~\citep[Lemma~2 and Thms.~4--5]{DLR}; see also~\citep[Lemma~2.1]{PudlakRodl2026}.

\begin{lemma}[tower growth]\label{lem:tower}
For every $j\ge 0$ and $k\ge 4$, $\width(P_j(k))\ge\tow(j+1)$. For every $j\ge 2$, $\width(P_j(3))\ge\tow(j-1)$. For
all $k\ge 2$ and $j\ge 0$, $|P_j(k)|\le\tow(j+\lgs k)$.
\end{lemma}

\begin{proof}
Since $\binom{w}{\lfloor w/2\rfloor}$ is non-decreasing in $w$, \cref{lem:growth}(i) gives $\width(P_i(k))\ge T_i$ for
$k\ge 4$, starting from $\width(P_0(k))=k\ge 4$, and $\width(P_{i+2}(3))\ge T_i$, starting from \cref{lem:growth}(iii).
It remains to show $T_i\ge\tow(i+1)$. For $i\le 2$ this holds by inspection, and $T_3=184\,756\ge 2\cdot 2^{16}=2\tow(4)$.
For $i\ge 3$ we show $T_i\ge 2\tow(i+1)$ by induction. The function $g(w)=2^w/(w+1)$ satisfies
$\binom{w}{\lfloor w/2\rfloor}\ge g(w)$ and is non-decreasing for $w\ge 1$. With $x=\tow(i+1)\ge 2^{16}$,
\[
  T_{i+1}\;\ge\;g(T_i)\;\ge\;g(2x)\;=\;\frac{\tow(i+2)^2}{2x+1}\;\ge\;2\tow(i+2),
\]
where the last inequality is $2^x\ge 4x+2$. The upper bound follows from \cref{lem:growth}(ii) by induction, starting
from $|P_0(k)|=k\le\tow(\lgs k)$.
\end{proof}

\begin{corollary}[number of levels]\label{cor:levels}
Let $K\ge 2$. Then $|P_j(k)|\ge K$ for $j=\max(1,\lgs K-1)$ if $k\ge 4$, and for $j=\lgs K+1$ if $k=3$. Conversely,
$|P_j(k)|\ge K$ implies $j\ge\lgs K-\lgs k$.
\end{corollary}

\begin{proof}
For $k\ge 4$ and $K\ge 3$ we have $j=\lgs K-1\ge 1$ and $|P_j(k)|\ge\tow(\lgs K)\ge K$; for $K=2$, $|P_1(k)|=2^k-1\ge 2$.
For $k=3$ we have $j\ge 2$ and $|P_j(3)|\ge\tow(\lgs K)\ge K$. The converse is the upper bound of \cref{lem:tower}.
\end{proof}

The first values are $|P_j(2)|=j+2$, $|P_j(3)|=3,7,18,81,8\,569$ and $|P_j(4)|=4,15,166$ for $j=0,1,2,\dots$. For $k=2$ the
width stays $2$, matching the logarithmic behavior of bipartite graphs. The counts of all down-sets in
Kriegel~\citep[Table~4.1]{Kriegel2018} and Bai et al.~\citep[p.~4]{BDHLW2026} exceed these values by exactly $j$, and
$|P_2(k)|$ is the Dedekind number of $k$ minus $2$~\citep{OEISA000372}.

\section{Reservation blow-ups: proof of Theorem~\ref{thm:main-det}}\label{sec:blowup}

The adversary of \cref{thm:main-det} is a clique followed by $j$ reservation blow-ups. \Cref{sec:blowup-one} describes
one blow-up as a procedure that runs an inner adversary, and \cref{lem:force} shows that the procedure preserves forcing.
\Cref{sec:blowup-proof} stacks $j$ blow-ups under a clique and counts the vertices.

\subsection{One blow-up}\label{sec:blowup-one}

\begin{construction}[reservation blow-up $B(\cA)$]\label{con:blowup}
Fix $K\ge 2$ and an \emph{inner} adversary $\cA$ with post-degree at most $D\ge 1$, and put
\[
  R:=(K-1)(D-1)+1 .
\]
The adversary $B(\cA)$ plays against a deterministic algorithm $\Alg$ and runs $\cA$ internally. The vertices that $\cA$
presents are \emph{inner vertices}; they are never shown to $\Alg$. For each inner vertex $w$ presented so far, $B(\cA)$
keeps a block $\sigma(w)$ of presented vertices, an \emph{induced color} $\col(w)$, and a \emph{reserve}
$\Res(w)\subseteq\sigma(w)$. When $\cA$ presents an inner vertex $v$ with earlier inner neighbors $W$, the adversary
$B(\cA)$ does the following.
\begin{enumerate}[label=(\arabic*),nosep]
\item For each $w\in W$, it removes one vertex $x_{w,v}$ from $\Res(w)$; this vertex is the server of the arc $w\to v$.
\item It presents $R$ new vertices one at a time, each with earlier neighbors exactly $\{x_{w,v}:w\in W\}$. They form
  the block $\sigma(v)$.
\item It lets $\col(v)$ be a color that $\Alg$ used on at least $D$ vertices of $\sigma(v)$, lets $\Res(v)$ consist of
  $D$ such vertices, and reports $\col(v)$ to $\cA$ as the color of $v$.
\end{enumerate}
The adversary $B(\cA)$ stops when $\cA$ stops, and also as soon as $\Alg$ has used $K$ colors. Ties are broken as
follows: $\col(v)$ is the color, among those that $\Alg$ used on at least $D$ vertices of $\sigma(v)$, that appears first
in $\sigma(v)$; the reserve $\Res(v)$ consists of the first $D$ vertices of $\sigma(v)$ with that color; and each reserve
gives up its vertices in arrival order.
\end{construction}

Three properties have to be checked: steps~1 and~3 are always possible, the colors reported to $\cA$ form a legal
online coloring, and every color reported is a color of $\Alg$. The last property is what makes forcing survive,
because $\cA$ forces $K$ colors on the coloring it is shown.

\begin{lemma}[forcing survives a blow-up]\label{lem:force}
Suppose that $\cA$ forces $K$ colors, has size at most $N$, and has post-degree at most $D$. Then $B(\cA)$ forces $K$
colors, has size at most $RN$, and has post-degree at most $R$. Moreover, there is a deterministic online algorithm
$\Alg'$ for the inner game such that, as long as $\Alg$ has used at most $K-1$ colors, every induced color $\col(v)$ is
the color that $\Alg'$ gives $v$ against $\cA$, and it is a color used by $\Alg$.
\end{lemma}

\begin{proof}
\emph{Step~3 is possible.} If $D=1$ then $R=1$, and the single vertex of $\sigma(v)$ will do. If $D\ge 2$ and every color
occurred on at most $D-1$ vertices of $\sigma(v)$, then the $R>(K-1)(D-1)$ vertices of $\sigma(v)$ would carry at least
$K$ colors, and $B(\cA)$ would already have stopped.

\emph{Step~1 is possible.} The reserve $\Res(w)$ starts with $D$ vertices and loses one for each later inner neighbor
of $w$, of which there are at most $D$, because the inner play is a play of $\cA$ (see below).

\emph{The induced coloring is proper.} The color $\col(v)$ is the color of some $u\in\sigma(v)$, and $u$ is adjacent to
$x_{w,v}$ for every earlier inner neighbor $w$ of $v$. The server $x_{w,v}$ was taken from $\Res(w)$, so it has color
$\col(w)$, and therefore $\col(v)\ne\col(w)$.

\emph{The induced algorithm.} We define $\Alg'$ on every ordered graph, so that the forcing hypothesis on $\cA$ applies
to it. When an inner vertex $v$ arrives, $\Alg'$ carries out steps~1 to~3 against $\Alg$ for the inner vertices up to
$v$, in their order and without the stop rule. If this is possible and gives a color $\col(v)$ that differs from the
colors that $\Alg'$ gave to the earlier neighbors of $v$, then $\Alg'$ gives $v$ the color $\col(v)$. Otherwise it
gives $v$ a color not used on the earlier neighbors of $v$, chosen by a fixed rule. The algorithm $\Alg'$ is a proper
deterministic online algorithm, and its answer depends only on the inner graph up to $v$, because $\Alg$ and all
choices in steps~1 to~3 are deterministic.

\emph{The inner play.} As long as $\Alg$ has used at most $K-1$ colors, the play of $\cA$ that $B(\cA)$ runs coincides
with the play of $\cA$ against $\Alg'$. Indeed, by induction on the inner vertices, the vertices presented up to the end
of $\sigma(v)$ and their colors are exactly what $\Alg'$ computes from the inner graph up to $v$; steps~1 and~3 were
possible, and $\col(v)$ differs from the induced colors of the earlier neighbors of $v$, so $\Alg'$ gives $v$ the color
$\col(v)$ that is reported to $\cA$. Every such $\col(v)$ is the color of a vertex of $\sigma(v)$, hence a color of
$\Alg$.

\emph{Forcing.} Every play of $B(\cA)$ is finite, because every play of $\cA$ is finite and each inner vertex produces
at most $R$ vertices. If $B(\cA)$ stops because $\Alg$ has used $K$ colors, we are done. Otherwise $\cA$ has stopped,
and its play is a complete play against $\Alg'$, so at least $K$ distinct induced colors appear, each of them a color
of $\Alg$.

\emph{Size and post-degree.} Each inner vertex produces at most $R$ vertices. A vertex serves at most one arc, because it
leaves the reserve when it is used, and its later neighbors are then the $R$ vertices of one block. A vertex that never
serves has no later neighbors.
\end{proof}

The graph presented by $B(\cA)$ is a reservation blow-up of the inner graph, restricted to the blocks
that have started, so \cref{lem:down} gives colorability whatever the online choices.

We call $\Alg'$ the \emph{induced algorithm}. When blow-ups are stacked, each level faces the induced algorithm of the level below it.

\begin{lemma}[stacking]\label{rem:stack}
Let $\cA_j$ be an adversary of bounded size and post-degree that forces $K\ge 2$ colors, and for $i=j-1,\dots,0$ let
$\cA_i:=B(\cA_{i+1})$, with $D$ equal to the post-degree bound of $\cA_{i+1}$. Then every $\cA_i$ forces $K$ colors. Moreover, in every play of $\cA_0$ against
a deterministic algorithm $\Alg$, as long as $\Alg$ has used at most $K-1$ colors, the following holds at every level:
the play that $\cA_i$ runs internally is the play of $\cA_{i+1}$ against the induced algorithm of level $i$; step~3 is
possible; and every color seen is a color of $\Alg$. In particular, the server of every arc out of a node $w$ of any
level has the induced color $\col(w)$.
\end{lemma}

\begin{proof}
By \cref{lem:force} and induction, every $\cA_i$ forces $K$ colors with the stated post-degree. For the second
statement, apply the last sentence of \cref{lem:force} to $\cA_0=B(\cA_1)$ and $\Alg$: the play of $\cA_1$ is the play
against the induced algorithm $\Alg'$, and every color of $\Alg'$ in it is a color of $\Alg$. Hence $\Alg'$ has used
at most $K-1$ colors as well, and the same argument applies to $\cA_1=B(\cA_2)$ and $\Alg'$, and so on up the levels.
Step~3 is possible at a level while its algorithm has used at most $K-1$ colors (proof of \cref{lem:force}), and a
server of an arc out of $w$ comes from $\Res(w)$, whose vertices have color $\col(w)$.
\end{proof}

\subsection{Stacking the blow-ups}\label{sec:blowup-proof}

We now stack $j$ blow-ups under a clique. By \cref{lem:force} the post-degree bound of one blow-up is the block size of
the previous one, so the block sizes grow by a factor of about $K$ per level, which makes the exponent of
the size quadratic in the number of levels. Every block of a node of level $\ell$ then has the same size $b_\ell$, unless the play
stops inside it; the blocks of the top vertices have size $b_j$.

\begin{theorem}[the iterated reservation adversary]\label{thm:L}
Let $k\ge 2$, $K\ge 2$ and $j\ge 1$ with $|P_j(k)|\ge K$, and put $b_\ell:=(K-1)^{j-\ell+1}(K-2)+1$ for $1\le\ell\le j$.
There is an adversary that forces every deterministic online algorithm to use $K$ colors, all of whose final graphs are
$k$-colorable, and every play of which has at most
\[
  (K-1)\prod_{\ell=1}^{j}b_\ell+1\;\le\;K^{(j+1)(j+2)/2}\quad\text{vertices}.
\]
\end{theorem}

\begin{proof}
Let $\cA_j$ present the arrival-ordered clique $K_K=(v_1,\dots,v_K)$. It forces $K$ colors, has post-degree $K-1$, and
its final graph maps to $\Tmpl(P_j(k))$ by \cref{lem:tops}(i). For $i=j-1,\dots,0$ let $\cA_i:=B(\cA_{i+1})$, with $D$
equal to the post-degree bound of $\cA_{i+1}$. By \cref{rem:stack}, each $\cA_i$ forces $K$ colors.
At every level, the graph presented is a reservation blow-up of the inner graph restricted to the blocks that have
started (the remark after \cref{lem:force}). Hence every graph presented by $\cA_0$ is a $j$-fold blow-up of an initial
segment $K_m$ of $K_K$, with $m\le K\le|P_j(k)|$, and it is $k$-colorable by \cref{cor:towers}.

\emph{Counting.} The first blow-up has $D=K-1$, so the blocks of the top vertices have $b_j=(K-1)(K-2)+1$ nodes. Each
later blow-up has $D$ equal to the previous block size, so $b_{\ell-1}=(K-1)(b_\ell-1)+1$, which solves to
$b_\ell-1=(K-1)^{j-\ell+1}(K-2)$. In particular $b_\ell\le K^{j-\ell+2}$.

\emph{The play ends no later than the first vertex below $v_K$.} Let $y$ be the first presented vertex below $v_K$, the first vertex
of the first block at every level below $v_K$. Its earlier neighbors are $K-1$ servers, one below each $v_s$ with $s<K$,
and by \cref{rem:stack} the server below $v_s$ has induced color $\col(v_s)$ at every level, as long as fewer than
$K$ colors are in use. The colors $\col(v_1),\dots,\col(v_{K-1})$ are distinct, since the induced coloring of
the clique is proper. Hence, if fewer than $K$ colors are in use when $y$ arrives, these are exactly the $K-1$ colors seen
by $y$, and $y$ receives a $K$-th color. The vertices presented before $y$ lie below $v_1,\dots,v_{K-1}$, so every play
has at most
\[
  (K-1)\prod_{\ell=1}^{j}b_\ell+1\;\le\;K\prod_{\ell=1}^{j}K^{j-\ell+2}\;=\;K^{(j+1)(j+2)/2}\quad\text{vertices.}\qedhere
\]
\end{proof}

\begin{restated}{Theorem~\ref{thm:main-det}}
For every $n\ge 3$ and $L=\lgs n$,
\[
  \chidet_3(n)\;\ge\;\big\lfloor n^{2/((L+2)(L+3))}\big\rfloor\qquad\text{and}\qquad
  \chidet_k(n)\;\ge\;\big\lfloor n^{2/(L(L+1))}\big\rfloor\quad\text{for every }k\ge 4.
\]
In particular $\chidet_k(n)\ge n^{(2-o(1))/(\lgs n)^2}$ for every fixed $k\ge 3$. For all $3\le n\le 2^{65536}$,
$\chidet_3(n)\ge\lfloor n^{1/28}\rfloor$ and $\chidet_k(n)\ge\lfloor n^{1/15}\rfloor$ for every $k\ge 4$.
\end{restated}

\begin{proof}
Let $n\ge 3$, so that $L\ge 2$. For $k=3$ put $e=(L+2)(L+3)/2$ and $K=\lfloor n^{1/e}\rfloor$; we may assume $K\ge 2$.
Since $K\le n$ we have $\lgs K\le L$, so \cref{cor:levels} gives $|P_j(3)|\ge K$ for $j=\lgs K+1\le L+1$. Then
$(j+1)(j+2)/2\le e$, so the adversary of \cref{thm:L} uses at most $K^e\le n$ vertices, and \cref{obs:adv} gives
$\chidet_3(n)\ge K$. For $k\ge 4$ put $e=L(L+1)/2$ and use $j=\max(1,\lgs K-1)\le L-1$. For $n\le 2^{65536}$ we have
$L\le 5$, so $e\le 28$ for $k=3$ and $e\le 15$ for $k\ge 4$. Finally, $\chidet_k(n)\ge\chidet_3(n)$ and
$2/((L+2)(L+3))=(2-o(1))/(\lgs n)^2$.
\end{proof}

\Cref{tab:sizes} lists the size bounds of the construction for small $K$. One blow-up forces $7$ colors on $3$-colorable
graphs with $187$ vertices, and no one-level tower over a clique can force more, since First-Fit uses at most
$|P_1(3)|=7$ colors on a $3$-colorable one-level tower (\cref{thm:main-ceiling}). Two blow-ups force
$8$ colors with $88\,796$ vertices and $18$ colors with $21\,464\,626$ vertices. Both numbers of colors exceed $2^3-1=7$,
the most an adversary can force in the transparent model~\citep[\S4.5]{HS1994}. Our adversary does not reveal its
$3$-coloring vertex by vertex: the final color of a vertex depends on which later block it will
serve. \Cref{app:refined} shrinks these instances by a constant factor.

\begin{table}[t]
\centering\small
\begin{tabular}{@{}rrrrrr@{}}
\toprule
$k$ & $K$ & $j$ & block sizes $b_j,\dots,b_1$ & size bound $(K-1)\prod b_\ell+1$ & $\log(\text{bound})/\log K$\\
\midrule
3 & 7  & 1 & 31 & 187 & 2.7\\
3 & 8  & 2 & 43, 295 & 88\,796 & 5.5\\
3 & 18 & 2 & 273, 4\,625 & 21\,464\,626 & 5.8\\
3 & 81 & 3 & 6\,321, 505\,601, 40\,448\,001 & $1.03\cdot10^{19}$ & 10.0\\
4 & 15 & 1 & 183 & 2\,563 & 2.9\\
4 & 166 & 2 & 27\,061, 4\,464\,901 & $1.99\cdot10^{13}$ & 6.0\\
\bottomrule
\end{tabular}
\caption{Size bounds of the adversary of \cref{thm:L} with the least $j$ such that $|P_j(k)|\ge K$. The exponent
$\log(\text{bound})/\log K$ approaches $(j+1)(j+2)/2$ as $K$ grows.}
\label{tab:sizes}
\end{table}

\begin{remark}[the hard instances]\label{rem:instances}
Every vertex of a blow-up is joined to at most one later block, and that block is independent, so the graphs are
triangle-free. More strongly, a $j$-fold blow-up of $K_K$ is bipartite or has odd girth at least $2j+3$. Indeed, let $C$
be an odd closed walk in a blow-up $G$ of $H$, and map every vertex to the node whose block contains it. This maps $C$ to a
closed walk of the same length in $H$. The earliest vertex $x$ of $C$ has both of its neighbors on the walk later, so both lie
in the block that $x$ serves, and the walk goes back and forth there. Deleting this backtrack leaves an odd closed walk
of length $|C|-2$ in $H$. In a reservation blow-up every arc has one server, so every vertex of our instances has at most $K-1$ earlier
neighbors; hence First-Fit uses at most $K$ colors on them, and it uses exactly $K$ against our adversary. The graphs are hard only because the algorithm does not
know which vertices will become servers.
\end{remark}

\section{Known graphs: proof of Theorem~\ref{thm:main-known}}\label{sec:known}

The adversary of \cref{thm:L} chooses its servers by color, so different algorithms see different graphs. We show that
the isomorphism type of the final graph does not depend on the algorithm once the adversary always completes its
blocks; the algorithm may then be told the final graph in advance. We use Halld\'orsson's known-graph
game~\citep[\S1, p.~2]{Hal2000}. Alice first shows Bob a graph $G$ on $N$ vertices. She then presents vertices
$v_1,v_2,\dots$ with their edges to earlier vertices, such that the graph induced by $v_1,\dots,v_i$ is isomorphic to a
subgraph of $G$ for every $i$, and Bob colors online. Alice never says which vertex of $G$ a presented vertex
corresponds to.

\paragraph{Completing every block.} Let $\cA^{\mathrm{str}}$ be the adversary of \cref{thm:L} with two changes. It never
stops early, so every block of a node of level $\ell$ has exactly $b_\ell$ nodes. In step~3, if no color occurs $D$
times in $\sigma(v)$, it takes for $\col(v)$ a most frequent color of $\sigma(v)$ and fills $\Res(v)$ with $D$
vertices of $\sigma(v)$ chosen by a fixed rule, which is possible because $b_\ell\ge D$. Until the algorithm uses its $K$-th color,
$\cA^{\mathrm{str}}$ makes exactly the moves of the adversary of \cref{thm:L}, so it also forces $K$ colors. After the
$K$-th color, the colorings reported to the inner adversaries may be improper. This does no harm: all these
adversaries are rules defined on every history, and from then on we use only the graph they build (the block sizes,
the reserves of $D$ vertices, and one server per arc).

To compare the final graphs of different plays, we describe a reservation blow-up statically, by the map that says
which vertex serves which arc. For $R\ge 1$ and a digraph $H$ with out-degrees at most $R$, give every vertex $v$ a set
$S(v)$ of $R$ new vertices. A \emph{server assignment} $\rho$ chooses, for every $w$, an injection $v\mapsto\rho(w,v)$ from the set $N^+(w)$ of out-neighbors of $w$ into $S(w)$.
Let $\mathrm{Bl}_\rho(H)$ be the digraph on $\bigsqcup_vS(v)$ with an arc $\rho(w,v)\to u$ for every arc $w\to v$ of $H$
and every $u\in S(v)$, with the sets $S(v)$ ordered as in $H$ when $H$ is an ordered graph oriented by arrival. It is
then a reservation blow-up of $H$ with blocks of size $R$.

\begin{lemma}[the server assignment does not matter]\label{lem:K1}
If $H\cong H'$ are digraphs with out-degrees at most $R$, then $\mathrm{Bl}_\rho(H)\cong\mathrm{Bl}_{\rho'}(H')$ for all
server assignments $\rho$ and $\rho'$, by an isomorphism that maps blocks to blocks.
\end{lemma}

\begin{proof}
Let $\varphi:H\to H'$ be an isomorphism. For every $w$ the sets $\rho(w,N^+(w))\subseteq S(w)$ and
$\rho'(\varphi w,N^+(\varphi w))\subseteq S'(\varphi w)$ have the same size, so some bijection $\pi_w:S(w)\to S'(\varphi w)$
satisfies $\pi_w(\rho(w,v))=\rho'(\varphi w,\varphi v)$ for every out-neighbor $v$ of $w$. The union $\pi$ of these
bijections maps the arc $\rho(w,v)\to u$, $u\in S(v)$, to the arc $\rho'(\varphi w,\varphi v)\to\pi_v(u)$. Hence $\pi$ maps
arcs injectively to arcs, and the two digraphs have the same number of arcs, $R$ times that of $H$.
\end{proof}

Let $H^{(j)}:=K_K$, viewed as a transitive tournament, let $H^{(\ell-1)}:=\mathrm{Bl}(H^{(\ell)})$ with blocks of size
$b_\ell$ for $\ell=j,\dots,1$, and let $H_j(K):=H^{(0)}$. By \cref{lem:K1}, $H_j(K)$ is well defined up to isomorphism.

\begin{lemma}[one final graph]\label{lem:K}
Against every deterministic algorithm, the final graph of $\cA^{\mathrm{str}}$, oriented by arrival, is isomorphic to
$H_j(K)$.
\end{lemma}

\begin{proof}
We induct down the levels. Let $G^{(\ell)}$ be the graph that the adversary produces at level $\ell$, oriented by
arrival; $G^{(j)}=K_K$ in every play. If $G^{(\ell)}\cong H^{(\ell)}$, then the next level presents a block of exactly
$b_\ell$ nodes for every node of $G^{(\ell)}$, and its arcs are exactly $x_{w,v}\to u$ for the arcs $w\to v$ of $G^{(\ell)}$
and $u\in\sigma(v)$. Setting $\rho(w,v):=x_{w,v}$ gives a server assignment, because every vertex leaves a reserve at most
once and reserves never run out. Hence $G^{(\ell-1)}=\mathrm{Bl}_\rho(G^{(\ell)})\cong H^{(\ell-1)}$ by \cref{lem:K1}.
\end{proof}

\begin{restated}{Theorem~\ref{thm:main-known}}
For every $N\ge 3$ and $L=\lgs N$ there are triangle-free $3$-colorable graphs $G$ and $G'$ on $N$ vertices with
\[
  \chiol(G)\;\ge\;\big\lfloor N^{2/((L+2)(L+3))}\big\rfloor
  \qquad\text{and}\qquad
  \chiol(G')\;\ge\;\frac{N^{1/(L+2)}}{8(1+\ln 4N)} .
\]
Every randomized online algorithm that knows $G'$ in advance has a presentation order of $G'$ on which it uses at least
$N^{1/(L+2)}/(8(1+\ln 4N))$ colors in expectation. For $N\ge 2^{406}$, both bounds for $G'$ hold with $N^{1/(L+3)}$ in
place of $N^{1/(L+2)}/(8(1+\ln 4N))$. For every $k\ge 4$ the same holds for $k$-colorable graphs, with exponents
$2/(L(L+1))$ and $1/L$, and with $N^{1/(L+1)}$ for $N\ge 2^{201}$.
\end{restated}

We prove the statements about $G$ here and those about $G'$ at the end of \cref{sec:logstar}.

\begin{proof}[Proof for $G$]
Choose $K$ and $j$ as in the proof of \cref{thm:main-det}, with $N$ in place of $n$. If $K=1$, the edgeless graph on $N$
vertices will do. Otherwise let $G$ consist of $H_j(K)$ and $N-|V(H_j(K))|$ isolated vertices; this is possible because $|V(H_j(K))|=K\prod_\ell b_\ell\le K^{(j+1)(j+2)/2}\le N$.
The graph $G$ is $k$-colorable by \cref{cor:towers} and triangle-free by \cref{rem:instances}.

In the known-graph game, Alice shows $G$, plays $\cA^{\mathrm{str}}$, and then presents the isolated vertices. By
\cref{lem:K} every prefix of the presentation is an initial segment of a graph isomorphic to $G$, so her moves are
legal. Bob knows $G$ in advance, but $G$ is the same in every play, so Bob is a deterministic online algorithm with $G$
built in, and $\cA^{\mathrm{str}}$ forces him to use $K$ colors.

For the online chromatic number, fix a deterministic algorithm $\Alg$, run $\cA^{\mathrm{str}}$ against it, and append
the isolated vertices. By \cref{lem:K} the result is $G$ with some presentation order. On this order $\Alg$ sees exactly
the play, so it uses at least $K$ colors. Since $\Alg$ was arbitrary, $\chiol(G)\ge K$.
\end{proof}

Halld\'orsson also considers a stricter game in which each presented vertex is identified with a vertex of $G$ right
after it is colored~\citep[pp.~2 and~4]{Hal2000}. For fixed $k$, no adversary can force many colors in that game. Bob fixes a proper
$k$-coloring $c$ of $G$ in advance; when $v$ arrives, its earlier neighbors have been identified, and Bob colors $v$
with the set $[k]\setminus c(N^-(v))$. Two adjacent vertices $u$ and $v$, with $u$ earlier, get different sets because
$c(u)$ lies in the set of $u$ but not in the set of $v$, so $2^k-1$ colors suffice. This is the case $P=P_0(k)$ of
\cref{prop:cap}, and it is how Halld\'orsson relates the stricter game to the transparent model.

\section{Uniform towers: proof of Theorem~\ref{thm:main-logstar}}\label{sec:logstar}

The adversary of \cref{thm:L} pays a factor of about $K$ per level. The loss comes from step~3 of \cref{con:blowup}: a
server gives a single color to every vertex of the block it serves, so a block needs $D$ vertices of one color, and
the pigeonhole principle then makes it about $K$ times larger than its post-degree. This section drops the one-color
rule. All blocks have the same size $A=2m$, where $m$ is the size of the clique, and all servers are chosen at random.
Instead of reserving a popular color, we follow, for each color, the \emph{unused mass}: the weight of the earlier
presented vertices of that color that can still become servers. A color with large unused mass is seen by almost
every new vertex, so the algorithm can rarely add mass to it, while the total unused mass is fixed, and
many colors must share it. The argument is a short computation of expectations that covers randomized
algorithms, known graphs and a deterministic adversary at once. \Cref{sec:uniform} defines uniform towers and
the random wiring, \cref{sec:mass} proves the two lemmas about unused mass, \cref{sec:uniform-analysis} assembles them
into \cref{thm:T}, and \cref{sec:logstar-proof} chooses the parameters.

\subsection{Uniform towers with random servers}\label{sec:uniform}

Fix $j\ge 1$ and $m\ge 2$, and put $A:=2m$. The \emph{uniform tower} is a full $j$-level tower over $K_m=(v_1,\dots,v_m)$ in
which every node of positive level has a block of exactly $A$ children. It has $mA^j$ presented vertices. In \emph{phase} $t$ the adversary presents the subtree of $v_t$, in depth-first order.

\paragraph{Servers.} The tower is a $j$-fold reservation blow-up of $K_m$. For a node $u$ of level $\ell\ge 1$ below
$v_t$, including $u=v_t$, and for $s<t$, let $x_s(u)$ be the node of level $\ell-1$ below $v_s$ that serves the arc into
$u$; it is joined to all of $\sigma(u)$. If $\ell\ge 2$, then $x_s(u)$ has exactly $A$ children, and the map $u'\mapsto
x_s(u')$ is a bijection from $\sigma(u)$ onto $\sigma(x_s(u))$. A presented vertex in the block of a node $y$ of level $1$ has
exactly the servers $x_s(y)$, $s<t$, as earlier neighbors: one presented vertex below each earlier top vertex. The \emph{wiring} of phase $t$
consists of the top servers $x_s(v_t)\in\sigma(v_s)$, which must be children of $v_s$ that have not served before, and
the bijections below them. Each top vertex $v_s$ has $m-s<A$ later neighbors, so a child that has not served is
always available.

\begin{construction}[random wiring $\mathcal R_j(m)$]\label{con:R}
In phase $t$, for each $s<t$ independently, the adversary chooses the top server $x_s(v_t)$ uniformly at random among the
children of $v_s$ that have not served, and it chooses every bijection $\sigma(u)\to\sigma(x_s(u))$ uniformly at random,
independently of everything else. It then presents phase $t$. It always presents the whole tower.
\end{construction}

The random choices of phase $t$ are made when phase $t$ starts, so given the history before phase $t$ (the graph
presented and the colors given), they are still drawn as in \cref{con:R}.

\begin{lemma}[one graph]\label{lem:uniform-graph}
For every wiring, the presented graph is a $j$-fold reservation blow-up of $K_m$ with all blocks of size $A$.
It is triangle-free, it is $k$-colorable whenever $m\le|P_j(k)|$, and, as a digraph oriented by arrival, it is
isomorphic to one fixed digraph $G_j(m)$ that does not depend on the wiring.
\end{lemma}

\begin{proof}
Every arc is served by exactly one node, every node below the top serves at most one arc, and every block has $A$
nodes. Colorability is \cref{cor:towers} and triangle-freeness is \cref{rem:instances}. For the isomorphism, apply
\cref{lem:K1} once per level, exactly as in the proof of \cref{lem:K}: the out-degrees are at most $m-1<A$ at the top and
at most $A$ below it, since a node below the top is joined to at most one block of $A$ nodes.
\end{proof}

The wirings differ only in the arrival order inside blocks, so the random wiring is a random presentation order of the
single graph $G_j(m)$.

\subsection{Unused mass}\label{sec:mass}

Give every presented vertex below a top vertex the weight $A^{-j}$, so that the subtree of every top vertex has total weight $1$.
Before phase $t$, a presented vertex below $v_s$, $s<t$, is \emph{eligible} if the child of $v_s$ above it has not served yet. For a
color $c$ let $u_c(t)$ be the total weight of the eligible vertices of color $c$, the \emph{unused mass} of $c$. Let $q_t(c)$ be the
fraction of the vertices of phase $t$ that have no earlier neighbor of color $c$, and let $a_t(c)$ be the fraction of
the vertices of phase $t$ that receive color $c$. These quantities exist only in the analysis; the algorithm does not know
them, and only the adversary of \cref{thm:T}(ii) computes them. Both $u_c(t)$ and $q_t(c)$ are determined by
the history before phase $t$ and the wiring of phase $t$, before the algorithm colors anything in phase~$t$.

\begin{lemma}[bookkeeping]\label{lem:mass}
In every play and for every color $c$ and phase $t$,
\[
  u_c(t+1)\;\le\;u_c(t)+a_t(c),\qquad a_t(c)\le\min\bigl(1,q_t(c)\bigr),\qquad\text{and}\qquad
  \sum_c u_c(m+1)\;=\;\frac{3m+1}{4}.
\]
\end{lemma}

\begin{proof}
In phase $t$ the vertices below $v_t$ become eligible, with total weight $1$ and weight $a_t(c)$ of color $c$, and the vertices below
the top servers of phase $t$ stop being eligible. A vertex with an earlier neighbor of color $c$ cannot receive color
$c$, so $a_t(c)\le q_t(c)$. At the end, all $m$ subtrees have been presented, with total weight
$m$. Each of the $\binom m2$ pairs $s<t$ has used one child of $v_s$, whose subtree has weight $1/A$, and different pairs
use different children. Every presented vertex has some color, so $\sum_cu_c(m+1)=m-\binom m2/A=m-(m-1)/4=(3m+1)/4$.
\end{proof}

\begin{lemma}[colors with large mass are seen]\label{lem:miss}
Fix a deterministic algorithm and play it against $\mathcal R_j(m)$. Condition on the history before phase $t$. Then for
every presented vertex $x$ of phase $t$ and every color $c$,
\[
  \Pr\bigl[\,x\text{ has no earlier neighbor of color }c\,\bigr]\;\le\;e^{-u_c(t)} .
\]
\end{lemma}

\begin{proof}
The history before phase $t$ determines the state of the adversary, namely which children of each $v_s$ have served
(a child has served if and only if its presented descendants have later neighbors), and with it the eligible vertices
and $u_c(t)$. Given the history, the wiring of phase $t$ is drawn as in \cref{con:R}, independently of the earlier
wirings.
Let $v_t=z_j,z_{j-1},\dots,z_1$ be the ancestors of $x$ of levels $j,\dots,1$. The earlier neighbor of $x$ below $v_s$
is $x_s(z_1)$. The top server $x_s(z_j)$ is uniform among the children of $v_s$ that have not served. Given it,
$x_s(z_{j-1})$ is the image of the fixed node $z_{j-1}$ under a uniform bijection onto $\sigma(x_s(z_j))$, so it is
uniform on that block, and so on down to $x_s(z_1)$, which is a presented vertex. All children of a node have subtrees with the same
number of presented vertices, so $x_s(z_1)$ is uniform over the eligible vertices below $v_s$, and these choices are independent for
different $s$. Let $p_s$ be the fraction of the eligible vertices below $v_s$ that have color $c$. Their total weight is at
most $1$, so $p_s$ is at least their weight of color $c$, and $\sum_{s<t}p_s\ge u_c(t)$. Hence the probability is
$\prod_{s<t}(1-p_s)\le\exp(-\sum_sp_s)\le e^{-u_c(t)}$.
\end{proof}

Neighbors of different vertices of the same phase can be dependent; the proof below uses \cref{lem:miss} only through
linearity of expectation.

\subsection{Counting colors}\label{sec:uniform-analysis}

We now assemble the two lemmas. A color whose unused mass
reaches $\ln 4m$, a \emph{heavy} color, holds less than $1+\ln 4m$ at that moment, and by \cref{lem:miss} it gains little
afterwards, so the final mass of \cref{lem:mass} must be spread over many colors.

\begin{theorem}[uniform towers force many colors]\label{thm:T}
Let $j\ge 1$ and $m\ge 2$, and put $X_m:=\dfrac{2m+1}{4(1+\ln 4m)}$.
\begin{enumerate}[label=(\roman*),nosep]
\item Against the random wiring $\mathcal R_j(m)$, every deterministic online algorithm uses at least $X_m$ colors in
  expectation.
\item There is a deterministic adaptive adversary, every play of which is a uniform tower over $K_m$ with $j$ levels,
  that forces every deterministic online algorithm to use at least $\lceil X_m\rceil$ colors.
\end{enumerate}
\end{theorem}

\begin{proof}
Let $b:=\ln 4m$. Call a color $c$ \emph{heavy} in phase $t$ if $u_c(t)\ge b$, and put
\[
  H_t\;:=\sum_{c\text{ heavy in phase }t}q_t(c).
\]
The total unused mass before phase $t$ is at most $t-1<m$, and $b>1$, so fewer
than $m$ colors are heavy in phase $t$.

\emph{A deterministic inequality.} Fix a play, and let $C$ be the number of colors used. Colors that are never used
have no mass. For a color $c$ that is used, let $t_c\le m$ be the last phase at whose start $c$ is not heavy; it exists
because $u_c(1)=0$, and whether $c$ is heavy at the start of phase $m+1$, the end of the play, plays no role. By \cref{lem:mass}, $u_c(t_c+1)\le u_c(t_c)+1<b+1$, and in every later phase, where $c$ is heavy,
the mass of $c$ grows by at most $q_t(c)$. Summing over the colors used and applying \cref{lem:mass} once more,
\begin{equation}\label{eq:mass-count}
  \frac{3m+1}{4}\;=\;\sum_cu_c(m+1)\;\le\;C\,(1+b)\;+\;\sum_{t=1}^mH_t .
\end{equation}

\emph{(i)} The heavy colors of phase $t$ are determined by the history before phase $t$. By \cref{lem:miss} and
linearity of expectation, every heavy color $c$ has $\E[q_t(c)\mid\text{history}]\le e^{-b}=1/(4m)$, so
$\E[H_t\mid\text{history}]\le m\cdot\frac1{4m}=\frac14$. Taking expectations in \eqref{eq:mass-count},
$\frac{3m+1}4\le\E[C]\,(1+b)+\frac m4$, that is, $\E[C]\ge X_m$.

\emph{(ii)} At the start of phase $t$ the adversary knows the history and can compute $H_t$ for every legal wiring of
phase $t$, since $H_t$ depends only on the wiring and on the colors of earlier phases. Given the earlier phases, the random
wiring of phase $t$ is uniform over the legal wirings of phase $t$, and $\E[H_t\mid\text{history}]\le\frac14$ by the
proof of (i), so some legal wiring has $H_t\le\frac14$. The adversary plays
the first such wiring in a fixed order. Then \eqref{eq:mass-count} gives $\frac{3m+1}4\le C(1+b)+\frac m4$ in every
play, so $C\ge X_m$, and $C$ is an integer.
\end{proof}

The proof uses none of the structure of the algorithm and loses nothing from one level to the next; the depth $j$
enters only through colorability, which needs $m\le|P_j(k)|$. In comparison, the reservation adversary of \cref{thm:L}
is guaranteed only about $\sqrt m$ colors with blocks of size $O(m)$, and the one-level adversary of \cref{thm:one-level}
forces $\lfloor m/2\rfloor+1$ colors with even smaller blocks, but only at one level. The adversary of (ii) is explicit, but it searches over all wirings of a phase.

\subsection{Putting things together}\label{sec:logstar-proof}

\begin{restated}{Theorem~\ref{thm:main-logstar}}
For every $n\ge 3$ and $L=\lgs n$,
\[
  \chirand_3(n)\;\ge\;\frac{n^{1/(L+2)}}{8(1+\ln 4n)}\qquad\text{and}\qquad
  \chirand_k(n)\;\ge\;\frac{n^{1/L}}{8(1+\ln 4n)}\quad\text{for every }k\ge 4.
\]
Moreover $\chidet_3(n)\ge\chirand_3(n)\ge n^{1/(\lgs n+3)}$ for all $n\ge 2^{406}$, and $\chidet_k(n)\ge\chirand_k(n)\ge
n^{1/(\lgs n+1)}$ for every $k\ge 4$ and all $n\ge 2^{201}$.
\end{restated}

\begin{proof}
Every deterministic algorithm is a randomized one, so $\chidet_k(n)\ge\chirand_k(n)$. Let $j\ge1$ be the least $j$ with
$|P_j(k)|\ge n$. By \cref{cor:levels} with $K=n$, $j\le L+1$ if $k=3$ and $j\le L-1$ if $k\ge 4$. Put
$M:=(n/2^j)^{1/(j+1)}$ and $m:=\lfloor M\rfloor$. If $m\le 1$, then $n<2^{2j+1}$, so $n^{1/(j+1)}<4$; since $j+1\le L+2$ if $k=3$ and $j+1\le L$ if $k\ge 4$,
the bounds of the first sentence are below $4/8$, and
there is nothing to prove; for the second sentence this case does not occur, because \cref{lem:threshold} gives $M\ge 2$. Otherwise $mA^j=2^jm^{j+1}\le n$ and $m\le n\le|P_j(k)|$. By \cref{lem:uniform-graph}, the
random wiring $\mathcal R_j(m)$, followed by $n-mA^j$ isolated vertices, is a distribution over $k$-colorable ordered
graphs on $n$ vertices, and by \cref{thm:T}(i) every deterministic algorithm uses at least $X_m$ colors on it in
expectation. Yao's principle (\cref{lem:yao}) gives $\chirand_k(n)\ge X_m$. Since $M-1<m\le M$ and $M\ge 2$,
\begin{equation}\label{eq:XM}
  X_m\;\ge\;\frac{2M-1}{4(1+\ln 4M)}\;\ge\;\frac{M}{4(1+\ln 4n)}\;\ge\;\frac{n^{1/(j+1)}}{8(1+\ln 4n)},
\end{equation}
using $2^{j/(j+1)}\le 2$, and $j+1$ is at most $L+2$, respectively $L$. For the second sentence,
\cref{lem:threshold} in \cref{app:thresholds} shows that the second expression in \eqref{eq:XM} is at least $n^{1/(j+2)}$
for $n\ge 2^{406}$ if $k=3$ and for $n\ge 2^{201}$ if $k\ge 4$, and $j+2$ is at most $L+3$ and $L+1$, respectively.
\end{proof}

The thresholds come from the constant factors in \eqref{eq:XM}, not from the method: the first bound of the theorem holds
for every $n$. At small sizes the explicit adversary of \cref{thm:L} is better; for $3$-colorable graphs the
bound of \cref{thm:T} overtakes it at $j=4$ with $m$ of about $600$, that is, for $n$ of about $2^{50}$ (about $2^{54}$
against the smaller instances of \cref{app:refined}).

The bound of \cref{thm:T}(i) is not carried by rare events: the same accounting, with a threshold for heavy colors
that depends on the failure probability, holds with high probability.

\begin{corollary}[forcing with high probability]\label{cor:half}
Let $j\ge1$, $m\ge2$ and $0<\varepsilon<1$. Against the random wiring $\mathcal R_j(m)$, every deterministic online
algorithm uses at least $(2m+1)/(4(1+\ln(4m/\varepsilon)))$ colors with probability at least $1-\varepsilon$.
Consequently, for every $n\ge 3$, every $0<\varepsilon<1$ and $L=\lgs n$ there is a distribution over $3$-colorable
ordered graphs on $n$ vertices on which every deterministic online algorithm uses at least
$n^{1/(L+2)}/(8(1+\ln(4n/\varepsilon)))$ colors with probability at least $1-\varepsilon$; for every $k\ge 4$ the same
holds for $k$-colorable graphs, with $n^{1/L}$ in place of $n^{1/(L+2)}$.
\end{corollary}

\begin{proof}
We run the proof of \cref{thm:T}(i) with $b:=\ln(4m/\varepsilon)$ in place of $\ln 4m$. The derivation of
\eqref{eq:mass-count} holds for every $b>0$, fewer than $m$ colors are heavy because $b>1$, and \cref{lem:miss} now
gives $\E[H_t\mid\text{history}]\le m\,e^{-b}=\varepsilon/4$. Let $H:=\sum_{t=1}^mH_t$. Every $H_t$ is nonnegative and
$\E[H]\le\varepsilon m/4$, so $\Pr[H\ge m/4]\le\varepsilon$ by Markov's inequality. If $H<m/4$, then
\eqref{eq:mass-count} gives $\frac{3m+1}4<C(1+b)+\frac m4$, that is, $C>(2m+1)/(4(1+\ln(4m/\varepsilon)))$. For the
second statement, choose $j$, $M$ and $m$ as in the proof of \cref{thm:main-logstar} and append isolated vertices. If
$m\le 1$ the bound is below $1$, and every algorithm uses at least one color. Otherwise the chain \eqref{eq:XM}, with
$\ln(4m/\varepsilon)\le\ln(4n/\varepsilon)$ in place of $\ln 4m\le\ln 4n$, gives
$(2m+1)/(4(1+\ln(4m/\varepsilon)))\ge n^{1/(j+1)}/(8(1+\ln(4n/\varepsilon)))$.
\end{proof}

{Confidence $1-\varepsilon$ costs only the additive term $\ln(1/\varepsilon)$ in the denominator. With $\varepsilon=1/n$,
for instance, the bound $n^{1/(L+2)}/(8(1+\ln 4n^2))$ holds with probability at least $1-1/n$.}

\begin{proof}[Proof of \cref{thm:main-known} for $G'$]
Choose $j$, $M$ and $m$ as in the proof of \cref{thm:main-logstar}, with $N$ in place of $n$. If $m\le 1$ the bound is
below $1$ and the edgeless graph will do. Otherwise let $G'$ consist of $G_j(m)$ and $N-m(2m)^j$ isolated vertices. It is triangle-free and $k$-colorable by
\cref{lem:uniform-graph}. Every play of the adversary of \cref{thm:T}(ii), followed by the isolated vertices, is a
presentation of $G'$, and so is every outcome of $\mathcal R_j(m)$, by \cref{lem:uniform-graph}. A deterministic
algorithm that knows $G'$ is still a deterministic online algorithm, so \cref{thm:T}(ii) gives $\chiol(G')\ge\lceil
X_m\rceil$, and \cref{thm:T}(i) with the remark after \cref{lem:yao} gives the statement for randomized algorithms.
Finally $X_m\ge N^{1/(j+1)}/(8(1+\ln 4N))$ by \eqref{eq:XM}, and for $N\ge 2^{406}$ (for $k\ge 4$, $N\ge 2^{201}$)
\cref{lem:threshold} gives $X_m\ge N^{1/(j+2)}$, exactly as in the proof of \cref{thm:main-logstar}.
\end{proof}

\section{Transfer, slowly growing chromatic number, and limits}\label{sec:limits}

This section proves \cref{thm:main-transfer}, the bound at chromatic number $\log\log n$ (\cref{thm:slow}), a cap for
adversaries that commit to labels (\cref{prop:cap}) and \cref{thm:main-ceiling}. The first two results replace the clique
at the top of a tower by another adversary; the last two bound what adversaries that commit to labels, and towers, can
force.

\subsection{Other adversaries at the top}

The blow-up machinery of \cref{sec:blowup} never uses that the top is a clique, or that $k$ is fixed. It needs only an
adversary that forces $K$ colors and whose graphs map to $\Tmpl(P_j(k))$.

\begin{restated}{Theorem~\ref{thm:main-transfer}}
Let $k,k'\ge 3$. There is $j\le\lgs k'+1$ such that for all $N$ and all $K\ge 2$, if $\chidet_{k'}(N)\ge K$
then $\chidet_k(N^{(j+1)(j+2)/2})\ge K$. In particular, if $\chidet_{k'}(n)=\Omega(n^{\varepsilon})$ for a fixed $k'\ge 3$, then $\chidet_3(n)=\Omega(n^{\varepsilon'})$
with $\varepsilon'=2\varepsilon/((\lgs k'+2)(\lgs k'+3))$.
\end{restated}

\begin{proof}
Let $j$ be the least $j$ with $\width(P_j(k))\ge k'$. By \cref{lem:tower}, $\width(P_j(k))\ge\tow(j+1)$ for $k\ge 4$ and
$\width(P_j(3))\ge\tow(j-1)$ for $j\ge 2$, so $j\le\lgs k'+1$. Let $K\ge 2$ and $N$ satisfy $\chidet_{k'}(N)\ge K$. By \cref{obs:adv} some
adversary $\cA$ of size at most $N$ forces $K$ colors on $k'$-colorable graphs. Its post-degree is at most $N-1$, and by \cref{lem:tops}(ii) every graph it
presents maps to $\Tmpl(P_j(k))$. Apply $j$ reservation blow-ups (\cref{rem:stack}); by
\cref{lem:down} the final graphs map to $\Tmpl(P_0(k))$, so they are $k$-colorable. The first block size is
$(K-1)(N-2)+1\le KN$, and every later block size is at most $K$ times the previous one, so the size is at most
$N\prod_{i=1}^{j}K^iN=N^{j+1}K^{j(j+1)/2}\le N^{(j+1)(j+2)/2}$, using $K\le N$, and \cref{obs:adv} gives
$\chidet_k(N^{(j+1)(j+2)/2})\ge K$. For the second statement, the hypothesis gives $N_{k'}(K)=O(K^{1/\varepsilon})$,
so by the first statement and \cref{obs:adv}, $N_3(K)=O(K^{(j+1)(j+2)/(2\varepsilon)})$ with $j\le\lgs k'+1$, and
\cref{obs:adv} turns this into $\chidet_3(n)=\Omega(n^{\varepsilon'})$.
\end{proof}

For example $j=2$ suffices for $k=3$ and $k'=4$ by \cref{lem:growth}(iii), so $N_3(K)\le N_4(K)^6$, and $j=1$ suffices
for $k=4$ and $k'\le 6$ because $\width(P_1(4))=\binom42=6$. The case $k=2$ is different: $\width(P_j(2))=2$ for
all $j$, and bipartite graphs need only $O(\log n)$ colors~\citep[p.~320]{LST1989}.

The second replacement uses an adversary of Kawarabayashi, Yoneda and Yoneda~\citep[Lemma~4.3]{KYY2026comp} with few
vertices and logarithmic chromatic number. It delays the choice of its own coloring of a vertex until a second vertex of
the same algorithm color appears.

\begin{construction}[pairing top]\label{con:pairing}
Fix $\ell\ge 1$ and $2\ell$ base colors, grouped into $\ell$ pairs $\{2s-1,2s\}$, $1\le s\le\ell$. The adversary
presents $2^\ell$ vertices, and the $i$-th vertex $v$ gets the $i$-th vector $X_v\in\{0,1\}^\ell$, which chooses the base
color $2s-1+X_v(s)$ from each pair. A presented vertex is \emph{paired} or \emph{unpaired}, and paired vertices carry a
base color $\varphi$. The new vertex $v$ is joined to every earlier paired vertex whose base color is not chosen by
$X_v$, and to no unpaired vertex. If the algorithm gives $v$ the color of an earlier unpaired vertex $u$, then $u$ and $v$
become paired: the adversary chooses the least $s$ with $X_u(s)\ne X_v(s)$ and puts $\varphi(u)=2s-1+X_u(s)$ and
$\varphi(v)=2s-1+X_v(s)$.
\end{construction}

\begin{lemma}[the pairing top; {\citep[Lemma~4.3]{KYY2026comp}}]\label{lem:pairing}
The pairing top presents $2^\ell$ vertices, forces $2^{\ell-1}$ colors, and every graph it presents is
$2\ell$-colorable.
\end{lemma}

\begin{proof}
At most one unpaired vertex has a given color, since a second one would have been paired with it. If $u$ and $v$ are
paired, then $\varphi(u)$ and $\varphi(v)$ are the two base colors of one pair, and every later vertex $w$ chooses one of
them, so $w$ is joined to the member of the pair whose base color it does not choose. That member has the color shared
by $u$ and $v$, so no third vertex receives it. Hence every color is used at most twice, and the $2^\ell$ vertices
receive at least $2^{\ell-1}$ colors. For the coloring, give every unpaired vertex any base color chosen by its own
vector. Then every vertex $w$ has a base color chosen by $X_w$, while every earlier neighbor of $w$ has a base color
that $X_w$ does not choose. A vertex receives later neighbors only after it is paired, and from then on its base color
is fixed, so the base colors form a proper $2\ell$-coloring.
\end{proof}

\begin{theorem}[polynomial bounds at chromatic number $\log\log n$]\label{thm:slow}
Let $K\ge 2$, $\ell=\lceil\log K\rceil+1$ and $c\ge 1$. If $\width(P_{c-1}(m))\ge 2\ell$, then an adversary forces $K$
colors on $m$-colorable graphs with at most $4^cK^{c(c+1)/2}$ vertices. In particular, if $n\ge 16$ and
$\binom{m}{\lfloor m/2\rfloor}\ge\log n$, then $\chidet_m(n)\ge\lfloor(n/16)^{1/3}\rfloor$; this holds for some
$m\le\log\log n+\log\log\log n+3$.
\end{theorem}

\begin{proof}
The pairing top has $N_0=2^\ell\le 4K$ vertices, forces $2^{\ell-1}\ge K$ colors and is $2\ell$-colorable, so by
\cref{lem:tops}(ii) its graphs map to $\Tmpl(P_{c-1}(m))$. Apply $c-1$ reservation blow-ups as in the proof of
\cref{thm:main-transfer}. Their block sizes satisfy $b_{c-1}\le KN_0$ and $b_{\ell'-1}\le Kb_{\ell'}$, so every play has at
most $N_0^cK^{c(c-1)/2}\le 4^cK^{c(c+1)/2}$ vertices, and the final graphs are $m$-colorable by \cref{lem:down}. For the
second statement take $c=2$ and $K=\lfloor(n/16)^{1/3}\rfloor$, so that $16K^3\le n$. If $K=1$ there is nothing to
prove, since every algorithm uses at least one color, so let $K\ge 2$. Then $2\ell\le 2\log K+4\le\log n$,
because $\log K\le(\log n-4)/3$ and $\log n\ge 4$, and $\log n\le\binom{m}{\lfloor m/2\rfloor}=\width(P_1(m))$. Since
$\binom{m}{\lfloor m/2\rfloor}\ge 2^m/(m+1)$, it suffices that $2^m\ge(m+1)\log n$. Put $x=\log\log n\ge 2$ and
$m=\lceil x+\log x\rceil+2$. Then $2^m\ge 4x\log n$ and $m+1\le x+\log x+4\le 4x$, so
$m\le\log\log n+\log\log\log n+3$ suffices.
\end{proof}

For fixed chromatic number our lower bound is $n^{1/(\lgs n+O(1))}$, and at chromatic number $(1+o(1))\log\log n$ it is
already of order $n^{1/3}$. To our knowledge, earlier polynomial lower bounds concern graphs of chromatic number of order at least
$\log n$~\citep{HS1994,Hal2000,KYY2026comp,BFL2006}, or triangle-free graphs whose chromatic number is not bounded~\citep{DKV2013,GKMZ2014}.

\subsection{A cap on adversaries that commit to labels}

Our adversaries choose the final coloring of a vertex long after the algorithm has colored it. The next proposition bounds
what an adversary that commits to labels can force: if it reveals a labeling of its graph by a poset $P$ vertex
by vertex, that is, an online homomorphism into $\Tmpl(P)$, then it forces at most $|\Jst(P)|$ colors. For a
nonempty poset $P$, an adversary \emph{commits to labels from $P$} if, right after the algorithm has colored a vertex $v$, it assigns a label
$\psi(v)\in P$ that is a function of the history so far (the graph presented and the colors given), such that
$\psi(u)\not\le\psi(v)$ for every edge $uv$ with $u$ earlier. For the antichain $P_0(k)$ this is exactly the
transparent model.

\begin{proposition}[a cap for adversaries that commit to labels]\label{prop:cap}
If an adversary forces $K$ colors and commits to labels from a poset $P$, then $K\le|\Jst(P)|$. In particular, an
adversary that forces $K$ colors can commit to labels from $P_j(k)$ only if $j\ge\lgs K-\lgs k-1$.
\end{proposition}

\begin{proof}
We describe a deterministic algorithm that uses at most $|\Jst(P)|$ colors against the adversary; it knows the
adversary and its labeling rule. When $v$ arrives, the algorithm computes $\psi(u)$ for every $u\in N^-(v)$, which is
possible because each such label is a function of the history up to the coloring of $u$. It then puts
\[
  U(v):=\{p\in P:\psi(u)\not\le p\text{ for all }u\in N^-(v)\},
\]
a down-set of $P$, and colors $v$ with $U(v)$, numbering the nonempty down-sets in the order of their first appearance.
The set $U(v)$ is nonempty: whatever color $v$ receives, the adversary then gives it a label $\psi(v)$ with
$\psi(u)\not\le\psi(v)$ for all $u\in N^-(v)$, that is, $\psi(v)\in U(v)$. The coloring is proper: for $u\in N^-(v)$ the
same argument gives $\psi(u)\in U(u)$, while $\psi(u)\notin U(v)$, so $U(u)\ne U(v)$. Hence the algorithm uses at most
$|\Jst(P)|$ colors, and since the adversary forces $K$ colors against it, $K\le|\Jst(P)|$. The second statement follows
from $|P_{j+1}(k)|\le\tow(j+1+\lgs k)$ (\cref{lem:tower}).
\end{proof}

For $P=P_0(k)$ the algorithm of the proof colors each vertex by the set of base colors missing from its earlier
neighborhood, which is a variant of the classical algorithm for the transparent model~\citep[\S4.5]{HS1994},
\citep[Thm.~4.2]{KYY2026comp}. The proposition is about deterministic adversaries, which the algorithm of its proof knows;
the random wiring of \cref{sec:logstar} escapes that algorithm because the wiring's choices are hidden.

\subsection{How far towers can go}

Two facts limit what towers can force. A full $j$-level tower over $K_m$ is $k$-colorable only if
$m\le|P_j(k)|$ (\cref{cor:towers}), so at least $\lgs m-\lgs k$ levels are necessary. The second fact is algorithmic.

\begin{lemma}[First-Fit commutes with blow-ups]\label{lem:FF}
Let $G$ be a full blow-up of $H$, or an initial segment of one. Then First-Fit colors every vertex of $\sigma(v)$ with
the color that First-Fit gives to $v$ on $H$. Consequently, on an initial segment of a full tower over $K_m=(v_1,\dots,v_m)$, First-Fit uses
exactly as many colors as there are top vertices with a presented descendant, and it gives the descendants of $v_t$
the $t$-th smallest color.
\end{lemma}

\begin{proof}
All vertices of $\sigma(v)$ have the same earlier neighbors, namely the servers of the arcs into $v$. By induction on
the arrival order, a server in $\sigma(w)$ has the color that First-Fit gives to $w$ on $H$, so the earlier neighbors of
$\sigma(v)$ carry exactly the First-Fit colors of the earlier neighbors of $v$; here we use fullness. The vertices of
$\sigma(v)$ are pairwise non-adjacent, so each of them gets the least color outside this set, which is the First-Fit
color of $v$. Applying this once per level reduces the second statement to First-Fit on $K_m$.
\end{proof}

\begin{restated}{Theorem~\ref{thm:main-ceiling}}
Let $k\ge 3$. If First-Fit uses $c\ge 2$ colors on a $k$-colorable $j$-level tower $G$, then $c\le|P_j(k)|$ and
$|V(G)|\ge\binom{c-1}{i+1}$ for every $0\le i\le j$. Consequently there is $n_k$ such that for every $n\ge n_k$, First-Fit
uses at most $n^{1/(\lgs n-\lgs k-3)}$ colors on every $k$-colorable tower with at most $n$ vertices.
\end{restated}

\begin{proof}
The graph $G$ is an initial segment $S$ of a full tower, followed by isolated vertices. These receive the smallest
color, which the first vertex of $S$ already has, since $c\ge 2$ forces $S\ne\emptyset$. Hence First-Fit uses $c$ colors
on $S$, and by \cref{lem:FF}, $v_c$ has a presented descendant, so the subtrees of $v_1,\dots,v_{c-1}$ are completely presented.

\emph{The poset bound.} Call a node \emph{alive} if it has a presented descendant. If $w\to v$ is an arc between nodes of
the same level and $v$ is alive, then the subtree of $w$ is completely presented, because $w$ arrives before $v$ and
subtrees are contiguous; so the server of $w\to v$ is alive. Hence the alive nodes of each level form a full blow-up of
the alive nodes of the level above, and the alive top vertices are $v_1,\dots,v_c$. Since $G$ is $k$-colorable, $j$
applications of \cref{lem:up} show that $K_c$ maps to $\Tmpl(P_j(k))$, and \cref{lem:tops}(i) gives $c\le|P_j(k)|$.

\emph{The size bound.} For each arc choose one server $\rho(w,v)$. A walk $u_0\to u_1\to\dots\to u_i$ of completely
presented nodes of one level maps to $(\rho(u_0,u_1),\dots,\rho(u_{i-1},u_i))$, a walk with one arc fewer of completely
presented nodes one level down: $\rho(u_a,u_{a+1})$ is joined to all of $\sigma(u_{a+1})$, which contains
$\rho(u_{a+1},u_{a+2})$. The map is injective, because $\rho(u_a,u_{a+1})$ determines its parent $u_a$, and
$\rho(u_{i-1},u_i)$ determines $u_i$, since a vertex serves at most one arc. The top vertices $v_1,\dots,v_{c-1}$ span a
transitive tournament with $\binom{c-1}{i+1}$ walks of length $i$. Applying the map $i\le j$ times turns them into
distinct nodes of level $j-i$, which have disjoint nonempty sets of presented descendants.

\emph{The asymptotic bound.} We show that $n_k:=\tow(\max(7,\lgs k+3))+1$ will do. Let $n\ge n_k$, so that
$\Lambda:=\lgs n\ge\max(8,\lgs k+4)$, and put $d:=\Lambda-\lgs k-3\ge 1$; the claim is $c\le n^{1/d}$. Since $\log
n>\tow(\Lambda-2)$, we have $d\le\Lambda\le\log\log n$, so $d\log\log n\le(\log\log n)^2\le\log n$, that is, $\log n\le
n^{1/d}$. Let $G$ have $n'\le n$ vertices and put $L=\lgs n'$. If $L\le\Lambda-2$, then $c\le n'\le\tow(\Lambda-2)<\log
n\le n^{1/d}$. Otherwise $L\ge\Lambda-1\ge 7$, so that $L\le\log\log n'$. Let $\lambda\le j$ be the least
$\lambda$ with $|P_\lambda(k)|\ge c$. If $\lgs c\le L-3$, then $c\le\tow(L-3)<\log\log n'\le n^{1/d}$. Otherwise
\cref{cor:levels} gives $\lambda\ge\lgs c-\lgs k\ge L-2-\lgs k$, and $\lambda\le\lgs c+1\le
L+1$. The size bound with $i=\lambda$ gives $n'\ge((c-1)/(\lambda+1))^{\lambda+1}$ if $c>\lambda+1$, and in every case
$c\le1+(\lambda+1)n'^{1/(\lambda+1)}$. The function $y\mapsto y\,n'^{1/y}$ is decreasing for $y\le\ln n'$, and
$y_0:=L-1-\lgs k$ satisfies $2\le y_0\le\lambda+1\le L+2\le\ln n'$, so
\[
  c\;\le\;1+y_0\,n'^{1/y_0}\;\le\;2L\,n'^{1/y_0}\;\le\;n'^{1/(y_0-1)}\;=\;n'^{1/(L-\lgs k-2)}\;\le\;n^{1/d}.
\]
The third inequality is $2L\le n'^{1/(y_0(y_0-1))}$, which holds because $y_0(y_0-1)\log 2L\le L^3\le(\log\log
n')^3\le\log n'$; the last one uses $n'\le n$ and $L-\lgs k-2\ge d$.
\end{proof}

Every play of the adversaries of \cref{thm:L,thm:T} is a tower, so First-Fit is optimal against them up to an additive
constant in the denominator of the exponent. With \cref{thm:main-logstar} this determines the value of the method.

\begin{corollary}[the value of towers]\label{cor:value}
Fix $k\ge 3$, and let $V_k(n)$ be the largest $K$ such that some adversary of size at most $n$ forces $K$ colors and
lands in the $k$-colorable towers. There is $n_k$ such that for every $n\ge n_k$,
\[
  n^{1/(\lgs n+3)}\;\le\;V_k(n)\;\le\;n^{1/(\lgs n-\lgs k-3)} .
\]
Equivalently, $V_k(n)=n^{1/(\lgs n+c)}$ for a real number $c=c(k,n)$ with $-\lgs k-3\le c\le 3$, which may be negative.
\end{corollary}

\begin{proof}
Let $n_k$ be as in \cref{thm:main-ceiling}; it exceeds $2^{406}$. For the lower bound, the adversary of
\cref{thm:T}(ii), with the parameters of the proof of \cref{thm:main-logstar}, has size at most $n$ and forces
$\lceil X_m\rceil\ge n^{1/(\lgs n+3)}$ colors, and by \cref{lem:uniform-graph} each of its plays is a $k$-colorable
tower. For the upper bound, play First-Fit against any adversary of size at most $n$ that lands in the $k$-colorable
towers: by \cref{thm:main-ceiling} the play ends with at most $n^{1/(\lgs n-\lgs k-3)}$ colors.
\end{proof}

A lower bound of the form $n^{\Omega(1)}$, or even $n^{1/o(\lgs n)}$, therefore cannot come from towers over a clique. It
needs a top of bounded chromatic number and polynomial size, which by \cref{thm:main-transfer} is the problem itself, or a
different kind of level.

\section{Discussion and open problems}\label{sec:discussion}

For fixed $k\ge 3$ the online chromatic number of $k$-colorable graphs now lies between $n^{1/(\lgs n+O(1))}$ and
$n^{1-\Theta(1/k^2)}$; for $k=3$ it lies between $n^{1/(\lgs n+3)}$ and $\Ot(n^{2/3})$ for deterministic algorithms, and
between $n^{1/(\lgs n+3)}$ and $O(\sqrt n)$ for randomized ones. The lower bound is still $n^{o(1)}$, and by
\cref{cor:value} no adversary that presents towers can exceed $n^{1/(\lgs n-O(1))}$. We list three directions.

\begin{question}[a polynomial lower bound]\label{q:poly}
Is $\chidet_3(n)\ge n^{\varepsilon}$ for some $\varepsilon>0$? Is $\chirand_3(n)\ge n^{\varepsilon}$?
\end{question}

A positive answer, which is the lower-bound half of Kierstead's challenge~\citep[Problem~5.1]{Kie1998}, must leave towers. For deterministic
algorithms, by \cref{thm:main-transfer} it suffices to find such an adversary for any single fixed $k\ge 3$. Two constraints are known. The
graphs must have bounded odd girth, because graphs of sufficiently large odd girth can be colored online with
$O(n^{\varepsilon})$ colors~\citep[Thm.~1]{YY2026oddgirth}; our $j$-level towers have odd girth at least $2j+3$
(\cref{rem:instances}), which grows with $n$. And by \cref{prop:cap}, an adversary that commits to labels from $P_j(k)$ after every move forces at most
$|P_{j+1}(k)|$ colors. The opposite has also been conjectured: Yoneda and Yoneda conjecture that
for every fixed $k$ some randomized online algorithm uses $n^{o(1)}$ colors~\citep[Conj.~27]{YY2026random}. By
\cref{thm:main-logstar}, such an algorithm would still need $n^{1/(\lgs n+O(1))}$ colors.

\begin{question}[the constant]\label{q:det}
What is the infimum of the constants $c$ such that $\chidet_k(n)\ge n^{1/(\lgs n+c)}$ for all large $n$? Can
the factor $\log n$ in
\cref{thm:main-logstar} be removed?
\end{question}

For adversaries that present towers, \cref{thm:main-ceiling,thm:main-logstar} place the corresponding infimum for $k=3$
between $-5$ and $3$. The
logarithmic factor in \cref{thm:T} comes from the threshold $\ln 4m$ for heavy colors. Even a constant fraction of $m$
at one level, as in \cref{thm:one-level}, does not survive several levels with our bookkeeping: the saturated colors (\cref{sec:one-level}) of different nodes of a subtree need not agree.

\begin{question}[one level, exactly]\label{q:one}
What is the largest number of colors that an adversary can force on reservation blow-ups of $K_m=(v_1,\dots,v_m)$ in which the block of $v_t$ has $\max(1,m-t)$ vertices? \Cref{thm:one-level} gives at least $\lfloor m/2\rfloor+1$, First-Fit gives at most $m$, and exhaustive search gives
$3,4,4,5,6,6,7,7$ for $m=4,\dots,11$ (\cref{app:one-level}).
\end{question}

The answer measures how much a single level must lose; a linear lower bound that survives several levels would remove
the logarithmic factor of \cref{thm:main-logstar}.

Finally, Kierstead and Kolossa color every $k$-colorable perfect graph online with $n^{10k/\log\log n}$
colors~\citep[Thm.~0.1]{KK1996}. By \cref{thm:main-logstar}, general $k$-colorable graphs are therefore asymptotically
harder than perfect ones for every fixed $k\ge 3$, although the separation sets in only at astronomically large $n$.
Whether perfect graphs admit polylogarithmically many colors~\citep[Problem~13]{Kie1998survey} is untouched by our
construction: its graphs are triangle-free, and a triangle-free graph is perfect only if it is bipartite.

\section*{Acknowledgments}
The author is deeply grateful to Allan Borodin for his generous advice and encouragement, and for his invaluable
guidance on the direction of this work.

\paragraph{Use of AI tools.} The author used the AI tools GPT (OpenAI) and Claude (Anthropic) to develop and check
proofs, to draft and edit the text, to write the simulations and computer checks, and to produce a Lean formalization. The author is responsible for all content of this paper.

\bibliographystyle{plainnat}
\bibliography{refs}

\appendix\crefalias{section}{appendix}\crefalias{subsection}{subappendix}

\section{Thresholds for Theorem~\ref{thm:main-logstar}}\label{app:thresholds}

This appendix proves the inequality behind the second sentence of \cref{thm:main-logstar}: with the parameters
of \cref{sec:logstar-proof}, the bound of \cref{thm:T} reaches $n^{1/(j+2)}$ beyond explicit thresholds.

\begin{lemma}[thresholds]\label{lem:threshold}
Let $L=\lgs n$, let $1\le j\le j_0(n)$, where $j_0(n):=L+1$ if $k=3$ and $j_0(n):=\max(1,L-1)$ if $k\ge 4$, and put
$M:=(n/2^j)^{1/(j+1)}$. Then $M\ge 2$ and
\begin{equation}\label{eq:threshold}
  \frac{2M-1}{4(1+\ln 4M)}\;\ge\;n^{1/(j+2)}
\end{equation}
for all $n\ge 2^{406}$ if $k=3$, and for all $n\ge 2^{201}$ if $k\ge 4$.
\end{lemma}

\begin{proof}
Write $X:=\log n$ and let $f_j(X)$ be the base-$2$ logarithm of the left side of \eqref{eq:threshold} minus $X/(j+2)$.
Since $\log M=(X-j)/(j+1)$, we have $\frac{d}{dX}\log(2M-1)\ge\frac1{j+1}$ and
$\frac{d}{dX}\log(1+\ln 4M)=\frac1{(j+1)(1+\ln 4M)}$, so
\[
  f_j'(X)\;\ge\;\frac1{j+1}\Big(\frac1{j+2}-\frac1{1+\ln 4M}\Big),
\]
which is nonnegative once $1+\ln 4M\ge j+2$. In the ranges below, $\log M\ge 39$ (in particular $M\ge 2$) and $j\le 7$, so this holds, and $f_j$ is
non-decreasing in $X$. It therefore suffices to check $f_j\ge 0$ for every admissible $j$ at the left end of each range of
$n$ on which $j_0(n)$ is constant (the values are rounded down):
\begin{center}\small
\begin{tabular}{@{}llrr@{}}
\toprule
$k$ & range of $n$ & $j_0$ & $\min_{1\le j\le j_0}f_j$ at the left end\\
\midrule
$3$ & $2^{406}\le n\le 2^{65536}$ & $6$ & $0.0007$ (at $j=6$)\\
$3$ & $2^{65536}<n\le\tow(6)$ & $7$ & $895.8$ (at $j=7$)\\
$\ge 4$ & $2^{201}\le n\le 2^{65536}$ & $4$ & $0.0077$ (at $j=4$)\\
$\ge 4$ & $2^{65536}<n\le\tow(6)$ & $5$ & $1545.6$ (at $j=5$)\\
\bottomrule
\end{tabular}
\end{center}
For $L\ge 7$ we have $X>\tow(L-2)\ge 2^{65536}$ and $L+3\le\log X$, so $\log M=(X-j)/(j+1)\ge 1$. The left side of \eqref{eq:threshold} is at least
$M/(4(1+\ln 4n))$, as in \eqref{eq:XM}, so $f_j(X)\ge\frac{X-j}{j+1}-\log(4(1+\ln 4n))-\frac X{j+2}\ge
\frac X{(\log X)^2}-1-2\log X$, which is positive.
\end{proof}

The thresholds cannot be halved: inequality \eqref{eq:threshold} fails at $n=2^{405}$ for $j=6$, and at
$n=2^{200}$ for $j=4$. However, the proof of \cref{thm:main-logstar} uses
the least $j$ with $|P_j(k)|\ge n$; for $k=3$ this is $j=5$ whenever $8\,569<n<2^{235}$, since $P_4(3)$ has an antichain of
$235$ elements, and for $j=5$ the left side of \eqref{eq:threshold} is at least $n^{1/8}$ once $\log n\ge 146$. We computed the table and these facts by a program, for every integer $\log n$ up to $65536$.

\section{Smaller explicit instances and exact small cases}\label{app:refined}

The pigeonhole in step~3 of \cref{con:blowup} assumes that a block may use any of $K-1$ colors. Below the top vertex
$v_t$ fewer colors are available, because every presented vertex below $v_t$ sees the $t-1$ distinct induced colors of
$v_1,\dots,v_{t-1}$, and the blocks can shrink accordingly. This changes the size bounds of \cref{tab:sizes} by a constant factor, about $(j+1)(j+2)/2$, but not the exponent (\cref{tab:refined}).

For $1\le t\le K$ put $D_j(t):=K-t$, the post-degree of $v_t$, and for $\ell=j,\dots,1$
\[
  b_\ell(t):=(K-t)\big(D_\ell(t)-1\big)+1,
  \qquad D_{\ell-1}(t):=\max_{t<t'\le K}b_\ell(t'),
\]
with the maximum of the empty set equal to $0$. Since $b_\ell(t)$ is non-increasing in $t$, this is
$D_{\ell-1}(t)=b_\ell(t+1)$ for $t<K$ and $D_{\ell-1}(K)=0$. Let $\cA'$ be the adversary of \cref{thm:L} in which every
block of a node of level $\ell$ below $v_t$ has $b_\ell(t)$ nodes and every reserve of such a node has size
$D_\ell(t)$. Below $v_K$, where the post-degree is $0$, every block is a single node, its reserve is
empty, and $\col$ is the color of that node.

\begin{proposition}[smaller blocks]\label{prop:refined}
If $|P_j(k)|\ge K$, then $\cA'$ forces $K$ colors on $k$-colorable graphs, and every play has at most
$\sum_{t=1}^{K}\prod_{\ell=1}^{j}b_\ell(t)$ vertices.
\end{proposition}

\begin{proof}
Colorability is unchanged, because \cref{lem:down} does not depend on the block sizes. For forcing, the proofs of
\cref{lem:force,rem:stack} apply once steps~1 and~3 of \cref{con:blowup} are possible with the new sizes. We check
step~3 block by block, in the order of completion and while fewer than $K$ colors are in use, assuming it for all
earlier blocks, so that every server taken so far has the induced color of its node. By induction down the levels,
every node below $v_t$ has exactly one server below each $v_s$ with $s<t$ among its earlier neighbors, and that server
has induced color $\col(v_s)$. These $t-1$ colors are distinct and in use, so the induced colors of a block below
$v_t$ lie in a set of at most $K-t$ colors, and $(K-t)(D-1)+1$ nodes contain $D$ nodes of one color. For step~1, a
node of level $\ell-1$ below $v_t$ serves at most one block, which lies below some $v_{t'}$ with $t'>t$ and has $b_\ell(t')\le
D_{\ell-1}(t)$ nodes, so a reserve of $D_{\ell-1}(t)$ nodes suffices; the reserve of $v_t$ itself needs $D_j(t)=K-t$
elements. The vertex bound counts every block completely.
\end{proof}

\begin{table}[h]
\centering\small
\begin{tabular}{@{}rrrrr@{}}
\toprule
$k$ & $K$ & $j$ & \cref{thm:L} & \cref{prop:refined}\\
\midrule
3 & 4 & 1 & 22 & 12\\
3 & 7 & 1 & 187 & 77\\
3 & 8 & 2 & 88\,796 & 14\,484\\
3 & 10 & 2 & 426\,394 & 70\,558\\
3 & 18 & 2 & 21\,464\,626 & 3\,591\,642\\
\bottomrule
\end{tabular}
\caption{Size bounds of the two adversaries, computed exactly from the formulas.}
\label{tab:refined}
\end{table}

\paragraph{A cheaper top.} The flow adversary of \cref{thm:one-level} forces $K$ colors on $K_{2K-2}$ blown up once, with
a block of $\max(1,d)$ vertices for every vertex of post-degree $d$. Used as the top level instead of a clique, it lowers the post-degree that the
first reservation blow-up has to serve from about $K^2$ to $2K$.

\begin{corollary}[a cheaper top]\label{cor:cheap-top}
Let $K\ge 3$ and $j\ge 1$ with $2K-2\le|P_j(k)|$. Running the flow adversary of \cref{thm:one-level} on $K_{2K-2}$ and
then $j-1$ reservation blow-ups below it gives an adversary that forces $K$ colors on $k$-colorable graphs with at most
$2^jK^{(j^2+j+2)/2}$ vertices.
\end{corollary}

\begin{proof}
With $m=2K-2$ and blocks of sizes $\max(1,m-t)$, the flow adversary forces $\lfloor m/2\rfloor+1=K$ colors with
$\binom m2+1\le 2K^2$ vertices. A vertex of $\sigma(v_s)$ serves at most one arc $v_s\to v_t$, and then it is joined to the
$\max(1,m-t)\le 2K-4$ vertices of $\sigma(v_t)$, since $t\ge 2$; so the post-degree is at most $D=2K-4$. Its final graph is
a reservation blow-up of $K_m$, which maps to $\Tmpl(P_{j-1}(k))$ by \cref{lem:tops}(i) and \cref{lem:down}, because
$m\le|P_j(k)|$. Apply $j-1$ reservation blow-ups (\cref{rem:stack}); by \cref{lem:down} the final graphs are
$k$-colorable. The block sizes are $b'_1=(K-1)(D-1)+1\le 2K^2$ and then at most $K$ times the previous one, so the size is
at most $2K^2\prod_{i=1}^{j-1}2K^{i+1}=2^jK^{(j^2+j+2)/2}$.
\end{proof}

At the same $j$ the exponent is smaller by $j$ than the $(j+1)(j+2)/2$ of \cref{thm:L}: for example $8$ colors on
$3$-colorable graphs with at most $92\cdot 78=7\,176$ vertices instead of $88\,796$, and $10$ colors with $20\,944$
instead of $426\,394$. The corollary needs $2K-2\le|P_j(k)|$ rather than $K\le|P_j(k)|$, however, and when this costs
an extra level it is larger: $7$ colors need $j=2$ and $3\,685$ vertices against $187$ with one level, and $18$ colors
need $j=3$ and about $2.7\cdot10^9$ vertices against $21\,464\,626$. Asymptotically the switch costs at most one level, and
the term $(\lgs n)^2$ is unchanged.

\paragraph{Exact small cases.} To see how far these instances are from optimal, recall that $N_k(c)$ is the least
number of vertices with which an adversary forces $c$ colors on $k$-colorable graphs.

\begin{lemma}[two vertices per color]\label{lem:FF-small}
For $c\ge k$, $N_k(c)\ge 2c-k$.
\end{lemma}

\begin{proof}
Let the algorithm play First-Fit. If $\{y_a\}$ and $\{y_b\}$, $a<b$, are color classes that consist of a single vertex,
then when $y_b$ was colored, color $a$ was present on a neighbor of $y_b$, which can only be $y_a$. Hence the singleton
color classes form a clique, and there are at most $k$ of them. Every other color class has at least two vertices, so
$c$ colors need at least $s+2(c-s)\ge 2c-k$ vertices, where $s\le k$ is the number of singletons.
\end{proof}

\begin{proposition}[exact values]\label{prop:small}
For $3\le c\le 9$ we have $N_3(c)=2c-3$, and $N_3(10)\ge 17$. For $k\le c\le c_k$ with $(k,c_k)=(4,10),(5,12),(6,13)$,
$N_k(c)=2c-k$.
\end{proposition}
The lower bounds are \cref{lem:FF-small}. The upper bounds are computer-assisted: adversary strategies (for
$N_3(9)\le15$ a DAG of $2\,863$ positions) found by exhaustive game search and replayed from scratch by an independent
checker, which verifies $k$-colorability after every move and that every list of answers is exactly the set of legal
answers. A second search with a different representation reproduces $N_3(c)=2c-3$ for $c\le 8$, and all thirty
strategies are verified in Lean by a checker that is proved sound and is evaluated in the kernel.

For $k=3$ and $c\le 9$, the best adversaries therefore force colors at the First-Fit rate of two vertices per color, while
the adversary of \cref{thm:L} needs $187$ vertices for $7$ colors, and the refined one of \cref{prop:refined} needs $77$,
against the optimum $11$. The regime $N_3(c)=2c-3$ must end, since Kierstead's algorithm gives $N_3(c)\ge
c^{3/2-o(1)}$~\citep[Thm.~1.6]{Kie1998}.

\section{Oblivious blow-ups}\label{sec:rand}

This appendix shows how the reservations of \cref{sec:blowup} can be made oblivious. The result is a version of
\cref{thm:main-det} that holds with probability $1-\varepsilon$ against randomized algorithms, at the price of a factor
$O(\log n)$; \cref{cor:half} gives a stronger bound with the same guarantee, and the oblivious construction also gives
the small explicit instances of \cref{app:oblivious}. Against a randomized algorithm the input is fixed in advance,
so the adversary cannot reserve vertices of the popular color of a block. It replaces the reserved vertex by a hidden random \emph{index class}: every vertex of a block gets a
uniformly random index, and the vertices with index $a$ serve the $a$-th later neighbor. If the popular color of the
block occurs in every index class, then the block of every later neighbor sees that color, as if it had been reserved. This is Li's
random-index device~\citep[Lemma~A.4]{Li2026}, applied to single vertices instead of bipartite gadgets. Colorability
needs no change, because \cref{lem:down} allows any number of servers per arc. The new requirement is that the popular
color must meet every index class, and this costs a factor $O(\log n)$ in the block sizes.

\subsection{The oblivious blow-up}

\begin{construction}[oblivious blow-up $\OB_R(\mathcal H)$]\label{con:OB}
Let $\mathcal H$ be a distribution over ordered graphs on exactly $N$ vertices $v_1,\dots,v_N$, and let $\delta:[N]\to\N$
depend on the position only, with $d^+_H(v_m)\le\delta(m)$ for every $H$ in the support, where $d^+_H$ is the
post-degree. Sample $H\sim\mathcal H$. For $m=1,\dots,N$, present a block $\sigma(v_m)$ of $R\ge 1$ new vertices. Each
$x\in\sigma(v_m)$ receives a hidden index $\idx(x)$, uniform on $[\delta(m)]$ and independent of everything else (no index
if $\delta(m)=0$). Each vertex of $\sigma(v_m)$ is joined to the whole index class
\[
  I(w,a):=\{x\in\sigma(w):\idx(x)=a\}
\]
for every earlier neighbor $w$ of $v_m$ in $H$, where $v_m$ is the $a$-th later neighbor of $w$.
\end{construction}

Every outcome is a blow-up of $H$ in the sense of \cref{def:blowup}: a vertex $x$ serves the arc to the $\idx(x)$-th later
neighbor of its block's vertex, if that neighbor exists. The input never depends on the colors, so $\OB_R(\mathcal H)$
is a distribution over ordered graphs on exactly $NR$ vertices. A vertex of $\sigma(v_m)$ has post-degree at most $R$ if
$\delta(m)\ge 1$ and $0$ otherwise, again a function of the position.

\begin{lemma}[forcing survives an oblivious blow-up]\label{lem:O}
Let $K\ge 2$ and $\varepsilon\ge 0$, and suppose that
\begin{enumerate}[label=(\alph*),nosep]
\item every $H$ in the support of $\mathcal H$ maps to $\Tmpl(\Jst(P))$;
\item every deterministic online algorithm uses at least $K$ colors on $H\sim\mathcal H$ with probability at least $\rho$;
\item $\sum_{m=1}^{N}\delta(m)\,\big(1-1/\delta(m)\big)^{\lceil R/(K-1)\rceil}\le\varepsilon$, where terms with
  $\delta(m)=0$ are $0$.
\end{enumerate}
Then every graph in the support of $\OB_R(\mathcal H)$ maps to $\Tmpl(P)$, and every deterministic online algorithm
uses at least $K$ colors on $\OB_R(\mathcal H)$ with probability at least $\rho-\varepsilon$.
\end{lemma}

Condition (c) holds trivially if every $\delta(m)$ is $0$, and otherwise in particular if $ND\exp(-R/((K-1)D))\le\varepsilon$
with $D=\max_m\delta(m)$, because $(1-1/x)^c\le
e^{-c/x}$ and $x\,e^{-c/x}$ is increasing in $x$.

\begin{proof}
\emph{Labels.} Every outcome is a blow-up of $H$, so \cref{lem:down} applies.

\emph{Forcing.} The proof compares $\Alg$ with an algorithm for the inner graph that uses the popular colors, and shows
that the comparison fails only if some index class misses its popular color. Fix a deterministic algorithm $\Alg$ for
the presented graph, and let $\iota$ be the family of all indices; $H$ and $\iota$ are independent. For $m\le N$ let
$\pi(v_m)$ be the color that $\Alg$ uses most often on $\sigma(v_m)$ (ties to the least color), and consider the events
\begin{align*}
  C_m&:=\{\text{every class }I(v_m,a),\ a\in[\delta(m)],\text{ contains a vertex of color }\pi(v_m)\},\\
  T_m&:=\{\Alg\text{ has used at least }K\text{ colors by the end of }\sigma(v_m)\},\qquad
  E_m:=\neg T_m\wedge\neg C_m .
\end{align*}

\emph{Step~1: $\Pr(E_m)\le\delta(m)(1-1/\delta(m))^{\lceil R/(K-1)\rceil}$.} Condition on $H$ and on the indices of
$\sigma(v_1),\dots,\sigma(v_{m-1})$. The edges presented up to the end of $\sigma(v_m)$ depend only on these, because the
classes joined to $\sigma(v_m)$ lie in earlier blocks. Hence the coloring of $\sigma(v_m)$, and with it $\pi(v_m)$ and
$T_m$, are fixed. If $T_m$ holds, $E_m$ is impossible. Otherwise $\sigma(v_m)$ carries at most $K-1$ colors, so at least
$c:=\lceil R/(K-1)\rceil$ of its vertices have color $\pi(v_m)$, and their indices are still independent and uniform on
$[\delta(m)]$. All of them miss a fixed index $a$ with probability at most $(1-1/\delta(m))^c$; now sum over $a$.

\emph{Step~2: a simulated inner algorithm.} For each fixed $\iota$ we define a deterministic online algorithm
$\Alg'_\iota$ for the inner graph. It must be defined on every ordered graph, not only on the support of $\mathcal H$; on an input with more than $N$
vertices it gives $v_m$ the color $(1,m)$ defined below for every $m>N$.
When $v_m$ arrives, $\Alg'_\iota$ knows the prefix $H[v_1,\dots,v_m]$, and from it and $\iota$ it rebuilds the presented
graph up to the end of $\sigma(v_m)$, joining $\sigma(v_m)$ to the class $I(w,a)$ even when $a>\delta(l)$ for $w=v_l$, in
which case the class is empty. It runs $\Alg$ on the rebuilt graph and gives $v_m$ a color from one of two disjoint
palettes, encoded injectively in the natural numbers (say $(0,q)\mapsto 2q$ and $(1,m)\mapsto 2m+1$):
\[
  \col(v_m):=(0,\pi(v_m))\ \text{if $m$ is not switched},\qquad \col(v_m):=(1,m)\ \text{if $m$ is switched}.
\]
Here ``$m$ is switched'' means that $T_m$ holds, or that $C_l$ fails for some $l\le m$, or that some $v_l$ with $l<m$
has more than $\delta(l)$ later neighbors in $H[v_1,\dots,v_m]$. The algorithm can decide this from the prefix,
because $\iota$ is a fixed parameter, and switching is monotone in $m$. The coloring is proper on every input. Let
$w=v_l\to v_m$ be an arc, with $v_m$ the $a$-th later neighbor of $w$. If $m$ is switched, $\col(v_m)=(1,m)$, and this
color is given to no other vertex. Otherwise $l$ is not switched either, so $\col(w)=(0,\pi(w))$, the event $C_l$ holds,
and $a\le\delta(l)$. The class $I(w,a)$ then contains a vertex of color $\pi(w)$, every vertex of $\sigma(v_m)$ is
adjacent to it, and $\pi(v_m)$, being the color of a vertex of $\sigma(v_m)$, differs from $\pi(w)$. The tags make the
colors of switched vertices new without any knowledge of the colors that $\Alg$ will use later.

\emph{Step~3: comparison.} Let $H$ be in the support of $\mathcal H$, so that its post-degrees are at most $\delta$. If
$\Alg$ uses at most $K-1$ colors in total and no $E_m$ occurs, then no $T_m$ holds, every $C_m$ holds, nothing is
switched, and every color of $\Alg'_\iota$ has the form $(0,q)$ for a color $q$ of $\Alg$, so $\Alg'_\iota$ uses at
most as many colors as $\Alg$. Hence
\[
  \{\Alg'_\iota\text{ uses at least }K\text{ colors}\}\;\subseteq\;\{\Alg\text{ uses at least }K\text{ colors}\}\cup
  \textstyle\bigcup_mE_m .
\]
By (b), applied to each $\Alg'_\iota$, and by the independence of $H$ and $\iota$, the left-hand event has probability
at least $\rho$. By Step~1 and (c), $\Pr(\bigcup_mE_m)\le\varepsilon$.
\end{proof}

The simulated algorithm uses the indices of $\sigma(v_m)$ only through the switch, as a fixed parameter; the graph it
rebuilds depends only on earlier indices. The real algorithm does learn the classes of earlier blocks once they are used,
but this is harmless, because $C_m$ is decided by the indices of $\sigma(v_m)$, which are independent of everything the
algorithm has seen when it colors $\sigma(v_m)$.

\subsection{Putting things together}

\begin{theorem}[iterated oblivious blow-ups]\label{thm:Lrand}
Let $k\ge 2$, $K\ge 2$, $j\ge 1$ with $|P_j(k)|\ge K$, and $\varepsilon\in(0,1)$. Put $N_j=K$ and $D_j=K-1$, and for
$\ell=j,\dots,1$ choose positive integers $b_\ell$ and put $N_{\ell-1}=N_\ell b_\ell$ and $D_{\ell-1}=b_\ell$. If
\begin{equation}\label{eq:star}
  \sum_{\ell=1}^{j}N_\ell D_\ell\exp\Big(-\frac{b_\ell}{(K-1)D_\ell}\Big)\;\le\;\varepsilon,
\end{equation}
then there is a distribution over ordered $k$-colorable graphs on exactly $N_0$ vertices on which every deterministic
online algorithm uses at least $K$ colors with probability at least $1-\varepsilon$.
\end{theorem}

\begin{proof}
Let $\mathcal H_j$ be the point mass on the arrival-ordered clique $K_K$, and let $\mathcal H_{\ell-1}:=\OB_{b_\ell}(\mathcal
H_\ell)$ with $\delta\equiv D_\ell$. This is a valid bound, because post-degrees are at most $K-1$ in $K_K$ and at most
$b_\ell$ after the blow-up with blocks of size $b_\ell$. The clique maps to $\Tmpl(P_j(k))$ by \cref{lem:tops}(i) and
forces $K$ colors with probability $1$. Apply \cref{lem:O} once per level, with the sufficient condition for (c) stated after it.
By \eqref{eq:star} the losses add up to at most $\varepsilon$, and the graphs of $\mathcal H_0$ are $k$-colorable.
\end{proof}

\begin{corollary}[high probability]\label{thm:main-rand}
Let $n\ge 3$, $\varepsilon\in(0,1)$, $L=\lgs n$, and $\Lambda_n=\lceil\ln((L+1)n^2/\varepsilon)\rceil$. For $k=3$, put
$K:=\lfloor n^{2/((L+2)(L+3))}/\Lambda_n\rfloor$, and for $k\ge 4$, put $K:=\lfloor n^{2/(L(L+1))}/\Lambda_n\rfloor$. There
is a distribution over $k$-colorable ordered graphs on $n$ vertices on which every deterministic online algorithm
uses at least $K$ colors with probability at least $1-\varepsilon$. In particular $\chirand_k(n)\ge(1-\varepsilon)K$.
\end{corollary}

\begin{proof}
Take $j=L+1$ and $e=(L+2)(L+3)/2$ if $k=3$, and $j=L-1$ and $e=L(L+1)/2$ if $k\ge 4$; in both cases $e=(j+1)(j+2)/2$ and
$j\le L+1$. Let $\Lambda=\Lambda_n$ and $K=\lfloor n^{1/e}/\Lambda\rfloor$; we may assume $K\ge 2$, since every algorithm uses at least one color. \Cref{cor:levels}, applied to $n$, gives $|P_j(k)|\ge n\ge K$. Choose
$b_\ell:=(K-1)D_\ell\Lambda$. Then $b_\ell\le MD_\ell$ with $M:=K\Lambda$, so $D_\ell\le M^{j-\ell+1}$ and
\[
  N_0=K\prod_{\ell=1}^{j}b_\ell\;\le\;K\cdot M^{\sum_{\ell=1}^{j}(j-\ell+2)}\;\le\;M^{(j+1)(j+2)/2}\;=\;(K\Lambda)^e\;\le\;n .
\]
Every $N_\ell D_\ell$ is at most $N_0^2\le n^2$, so each term of \eqref{eq:star} is at most $n^2e^{-\Lambda}\le
\varepsilon/(L+1)\le\varepsilon/j$, and \eqref{eq:star} holds. \Cref{thm:Lrand} and padding with isolated vertices give the
distribution, and Yao's principle (\cref{lem:yao}) gives the last claim.
\end{proof}

The loss against \cref{thm:main-det} is a single factor $\Lambda_n=O(\log(n/\varepsilon))$, not one per level, because the exponent
$1/e$ applies to $K$ and $\Lambda$ together. Some loss is unavoidable for this construction. Consider one oblivious
blow-up of $K_4$ with blocks of size $7$, the size used by \cref{thm:L}, and with the indices of each block uniform over
the actual later neighbors of its vertex ($\delta=3,2,1,0$). The adaptive adversary forces $4$ colors with certainty,
but against hidden indices the best deterministic algorithm uses only $3$ colors with probability exactly
$7343/23328\approx 0.31$ (exhaustive dynamic program).

\section{Further results on the method}\label{app:further}

This appendix collects results that complement the main line: why reservations are needed, what is known about the
best possible one-level adversaries, the limits of two other kinds of levels, more about the oblivious construction, and
a comparison with Li's construction. Some of these results are computer-assisted; we say so where they are stated.

\subsection{Reservations are necessary}\label{app:propS}

Vishwanathan's lower bound is built from \emph{levels} in which every vertex $v$ of a simulated graph $H$ is replaced by
a gadget $\Gamma(v)$, a graph that contains an edge, and every vertex of $\Gamma(v)$ is joined to a fixed
\emph{selected} set $X(w)\subseteq V(\Gamma(w))$ for every earlier neighbor $w$ of $v$; the set $X(w)$ is nonempty
whenever $w$ has a later neighbor, and it is the same for all later neighbors of $w$. We call such a level
\emph{reservation-free}. Li's simulation lemma is different: it reserves a private copy of a gadget for every future
neighbor of a simulated vertex~\citep[Lemma~2.5]{Li2026}, so the proposition below does not apply to his levels. Our blow-ups keep Li's reservation, with independent blocks.

\begin{proposition}[reservation-free levels]\label{prop:S}
If the graph one level up contains a clique on $s+1$ vertices, then the graph produced by a reservation-free level
contains a clique on $s+2$ vertices. Consequently, $d$ reservation-free levels over any nonempty graph produce a clique on
$d+1$ vertices, so a chain of $k$ such levels is never $k$-colorable, and a chain of $k-1$ levels is $k$-colorable only
if the graph at the top has no edge.
\end{proposition}

\begin{proof}
Let $q_1<\dots<q_{s+1}$ be a clique one level up. For $r\le s$ the vertex $q_r$ has the later neighbor $q_{s+1}$, so
$X(q_r)$ is nonempty; pick $y_r\in X(q_r)$, and pick an edge $uu'$ inside $\Gamma(q_{s+1})$. For $r<r'\le s$, the vertex
$y_{r'}$ belongs to $\Gamma(q_{r'})$ and $q_r$ is an earlier neighbor of $q_{r'}$, so $y_{r'}$ is joined to $X(q_r)\ni
y_r$. In the same way $u$ and $u'$ are joined to every $y_r$. Hence $\{y_1,\dots,y_s,u,u'\}$ is a clique on $s+2$
vertices. Starting from a single vertex ($s=0$) and applying this $d$ times gives a clique on $d+1$ vertices; starting
from an edge gives one on $d+2$.
\end{proof}

Reservation-free amplification can therefore be stacked at most $k-1$ deep, which is where the exponent $k-1$ of
Vishwanathan's bound $\Omega(\log^{k-1}n)$ comes from: with $k-1$ levels whose amplification factors multiply to $K$, the size
grows like $\exp(O((k-1)K^{1/(k-1)}))$, which gives $K\approx(\log n/O(k))^{k-1}$. The hypothesis that gadgets contain
an edge is needed, since gadgets must force colors by themselves in these constructions; with independent blocks, as in
our blow-ups, levels can be stacked without limit, and private servers keep the chromatic number under control.

\subsection{One level with minimal blocks}\label{sec:one-level}

\Cref{thm:T} loses a logarithmic factor at every number of levels. At a single level and with even smaller blocks, a
deterministic adversary loses only a factor of $2$.

Consider, for $m\ge 1$, reservation blow-ups of $K_m=(v_1,\dots,v_m)$ in which the block of $v_t$ has $b_t\ge\max(1,m-t)$ vertices, as
small as the post-degree of $v_t$ permits. When the block of $v_t$ starts, the adversary picks one vertex of each earlier
block that has not served yet to serve $v_t$; let $\mathrm{Forb}_t$ be the set of their colors. The algorithm must
color all of $\sigma(v_t)$ outside $\mathrm{Forb}_t$. With such small blocks the reservation rule of \cref{con:blowup}
guarantees only about $\sqrt m$ colors, since a block cannot reserve one color for every later neighbor. The
following adversary forces half of the colors.

\begin{theorem}[one level, minimal blocks]\label{thm:one-level}
There is an adversary of this form that forces $\lfloor m/2\rfloor+1$ colors.
\end{theorem}

\emph{The adversary.} The adversary records, for every earlier block, how many of its unused vertices of each color it
intends to hand out in the remaining phases, as an integer flow from blocks to colors. Let $W_t:=m-t+1$ be the number of
phases $t,\dots,m$ in which every earlier block still has to supply a server, and let $a_s(c)$ be the number of unused
vertices of color $c$ in $\sigma(v_s)$. The adversary keeps an integer matrix $Y_t$, with a row for each $s<t$ and a
column for each color, such that $0\le Y_t(s,c)\le\min(a_s(c),W_t)$ and all row and column sums are at most $W_t$. A row
or column is \emph{tight} if its sum is exactly $W_t$, and $\Sat(Y_t)$ is the set of tight columns, the \emph{saturated}
colors. In phase $t$ the adversary does the following.
\begin{enumerate}[label=(\alph*),nosep]
\item It views $Y_t$ as a bipartite multigraph with $|Y_t|$ edges, the sum of its entries, and maximum degree at most $W_t$, which by K\H{o}nig's edge-coloring
  theorem~\citep{Konig1916} splits into $W_t$ matchings, each covering every tight row and column. It plays a matching
  $M_t$ of at most average size, $|M_t|\le|Y_t|/W_t$. A block $s$ matched to color $c$ supplies an unused vertex of color
  $c$; every other block supplies an unused vertex of some color $c$ with $a_s(c)>Y_t(s,c)$.
\item After the algorithm has colored $\sigma(v_t)$, it keeps the rows of $Y_t-M_t$ and adds a row for $t$ that is
  maximal subject to $Y_{t+1}(t,c)\le a_t(c)$ and to all row and column sums being at most $W:=W_{t+1}$.
\end{enumerate}

The potential $\Phi_t:=|Y_t|/W_t+|\Sat(Y_t)|$ counts recorded supply and saturated colors; every phase raises it by one.

\begin{lemma}[potential gain at one level]\label{lem:one-level-gain}
For $1\le t\le m-1$ we have $\Phi_{t+1}\ge\Phi_t+1$.
\end{lemma}

\begin{proof}
\emph{Step (a) is possible.} Before phase $t$ the block $\sigma(v_s)$ has $b_s-(t-1-s)\ge W_t$ unused vertices. An
unmatched row is not tight, so its sum is less than $W_t$, and some color has $a_s(c)>Y_t(s,c)$.

\emph{The matrix $Y':=Y_t-M_t$ satisfies the constraints with $W_t-1$ in place of $W_t$.} A matched pick lowers $Y(s,c)$ and $a_s(c)$ together, and any
other pick lowers only a count $a_s(c)$ that exceeded $Y(s,c)$. Tight rows and columns lose exactly one, all other sums
were already at most $W_t-1$, and an entry equal to $W_t$ is the only nonzero entry of a tight row and is therefore
covered. Moreover $|Y'|\ge|Y_t|(1-1/W_t)$, so $|Y'|/(W_t-1)\ge|Y_t|/W_t$. The tight columns of $Y_t$ stay tight, and every
column of sum $W_t-1$ that $M_t$ does not cover becomes tight; call the set of the latter $N_t$.

\emph{Step (b) gains one.} The old rows do not change, so tight columns stay tight. If the new row $t$ has sum
$W=W_{t+1}$, then $|Y|$ grows by $W$ and $|Y|/W$ grows by one. Otherwise, since $b_t\ge W$, some color $c$ has
$Y_{t+1}(t,c)<a_t(c)$, and by maximality the column of $c$ is tight. The color $c$ occurs in $\sigma(v_t)$, so $c\notin\mathrm{Forb}_t$, and
$\mathrm{Forb}_t$ contains every tight column of $Y_t$ because $M_t$ covers them. Either $c\notin N_t$, and the number of
tight columns grows by $|N_t|+1$, or $c\in N_t$, and it grows by $|N_t|\ge 1$.
\end{proof}

\begin{proof}[Proof of \cref{thm:one-level}]
For $m=1$ there is nothing to prove. Let $m\ge 2$, $C=\lfloor m/2\rfloor$, and suppose that the algorithm uses at most $C$ colors. Column sums are at most $W_t$, so
$|Y_t|/W_t\le C$, and if some color is not tight then $|Y_t|<CW_t$; hence $\Phi_t<2C-1$ unless all $C$ colors are tight.
Since $\Phi_1=0$, \cref{lem:one-level-gain} gives $\Phi_{2C}\ge 2C-1$, so all $C$ colors are tight in phase $2C\le m$.
The matching $M_{2C}$ covers all of them, $\mathrm{Forb}_{2C}$ contains all $C$ colors, and the first vertex of
$\sigma(v_{2C})$ needs a new one.
\end{proof}

\subsection{One level, exactly}\label{app:one-level}

For the minimal one-level game of \cref{sec:one-level} (blocks of sizes $\max(1,m-t)$ over $K_m$, one server per arc),
let $V(m)$ be the largest number of colors an adversary can force. \Cref{thm:one-level} gives $V(m)\ge\lfloor
m/2\rfloor+1$, First-Fit gives $V(m)\le m$, and the reservation adversary with the blocks of \cref{prop:refined} fits into these
blocks, and is thus guaranteed to work, only for $\lfloor(1+\sqrt{4m-7})/2\rfloor+1$ colors (for $m\ge 2$). Exhaustive game search gives the following values.

\begin{center}\small
\begin{tabular}{@{}lcccccccc@{}}
\toprule
$m$ & 4 & 5 & 6 & 7 & 8 & 9 & 10 & 11\\
\midrule
$V(m)$ & 3 & 4 & 4 & 5 & 6 & 6 & 7 & 7\\
$\lfloor m/2\rfloor+1$ (\cref{thm:one-level}) & 3 & 3 & 4 & 4 & 5 & 5 & 6 & 6\\
First-Fit & 4 & 5 & 6 & 7 & 8 & 9 & 10 & 11\\
\bottomrule
\end{tabular}
\end{center}
The upper bounds are certified for every $m$: an algorithm
strategy that survives with $V(m)$ colors is replayed by an independent checker of the unrestricted game. The lower
bounds are certified in the same way for $m\le 9$; for $m=10$ the lower bound was obtained by two independent
exhaustive searches, and $m=11$ follows by monotonicity; the whole table is also verified in Lean. The values rule out the natural guess
$V(m)=\lfloor(2m+2)/3\rfloor$ at $m=11$.

The ratio $V(m)/m$ stays near $2/3$ in this range, above the factor $\frac12$ of \cref{thm:one-level}. Whether some
adversary forces $cm$ colors with small blocks at every level is open (\cref{sec:discussion}).

\subsection{Limits of two other kinds of levels}\label{app:other-levels}

\paragraph{The top block of a one-level construction.} Consider adversaries that present the clique $K_K$ blown up once,
where the adversary may use any number of servers per arc and chooses block sizes online. Let the \emph{budget} $b_s$ be
the supremum, possibly infinite, of the sizes of $\sigma(v_s)$ over all plays. The block of $v_1$ in \cref{thm:L} has $(K-1)(K-2)+1$ vertices, and
this is optimal up to a factor of $2$.

\begin{theorem}[top block]\label{thm:top-block}
Every such adversary that forces $K$ colors has $b_1\ge\binom{K-1}{2}+1$, and more generally
$b_s\ge\binom{K-s}{2}+1$ for $1\le s\le K$.
\end{theorem}
An exhaustive search over all such adversaries confirms the bound for $s=1$ and $K\le 7$, and for all $s$ and $K\le
6$.

\begin{proof}
We prove $b_1\ge\binom{K-1}2+1$. For $2\le s\le K-2$ the algorithm gives each of the first $s-1$ blocks its own color,
which is proper because every edge joins two different blocks, and from block $s$ on it runs the algorithm below with
colors different from these; the same argument then runs with $K-s+1$ in place of $K$. For $s\ge K-1$ the claim is
$b_s\ge 1$: First-Fit colors every block with a single color, because the vertices of a block have the same earlier
neighbors, so an adversary that forces $K$ colors must make all $K$ blocks nonempty. The algorithm knows the
adversary, which is deterministic, so it can simulate it and knows to which block each vertex belongs and which vertices
of $\sigma(v_1)$ have served. It colors $\sigma(v_1)$ in a \emph{staircase}: each new vertex gets the least color
$i\in\{1,\dots,K-2\}$ that so far has fewer than $K-1-i$ vertices. This holds $\binom{K-1}2$ vertices; suppose that
$b_1\le\binom{K-1}2$, so that it colors all of $\sigma(v_1)$. Then for every $r\ge 1$ at most $\max(0,K-1-r)$ colors have at least $r$ vertices. Let $Q$ be the set of colors used on $\sigma(v_1)$. Every
later block $\sigma(v_t)$ has a single set $F_t$ of forbidden colors, the colors of its earlier neighbors, and the
algorithm colors the whole block with one color: with the color of $Q$ that no later block has used yet, that is
allowed, and that has the most vertices of $\sigma(v_1)$ that have not served, if there is one; otherwise with the least
allowed color already in use, if there is one; otherwise with a new color.

Call a color of $Q$ \emph{fresh} if no later block has used it. A fresh color occurs only in $\sigma(v_1)$. Let $g$ be
$K$ minus the number of colors used so far, $\rho$ the number of fresh colors, $m_q$ the number of vertices of color
$q$ in $\sigma(v_1)$ that have not served, and $h(r)$ the number of fresh colors $q$ with $m_q\ge r$. We show that the
invariant
\[
  h(r)\;\le\;\max(0,\;g+\rho-1-r)\qquad\text{for all }r\ge 1
\]
holds throughout. Initially $g+\rho=K$, and the invariant is the staircase property. A round $t\ge 2$ only lowers the
numbers $m_q$ of the colors shown to $\sigma(v_t)$, and every fresh color in $F_t$ is shown by a vertex of
$\sigma(v_1)$, whose $m_q$ therefore drops by at least one. Since $h(r)\ge 1$ forces $g+\rho-1-r\ge 1$, we may ignore
the maximum. A round with an empty block colors nothing and only lowers some $m_q$. For a nonempty block, let
$U_t$ be the set of colors already in use that are not in $F_t$.
\begin{enumerate}[label=(\alph*),nosep]
\item The set $U_t$ is empty. Every fresh color is forbidden, so every $m_q$ drops, and the block takes a new color, so $g$
  drops by one. Hence $h'(r)\le h(r+1)\le g+\rho-2-r=g'+\rho-1-r$.
\item The block takes a fresh color $x\in U_t$, so $\rho$ drops by one; colors of $U_t$ keep $m_q$ and the others
  drop. If $m_x\ge r$, then $x$ leaves the count of $h(r)$ and nothing enters it, so $h'(r)\le h(r)-1\le g+\rho'-1-r$. If
  $m_x<r$, then every fresh color of $U_t$ has $m_q<r$ by the choice of $x$, so $h'(r)\le h(r+1)\le g+\rho'-1-r$.
\item The set $U_t$ is nonempty but contains no fresh color. Every fresh color is forbidden, so $h'(r)\le h(r+1)$, while $g$
  and $\rho$ do not change.
\end{enumerate}
Suppose that the adversary forces $K$ colors. The $K$-th color appears in a round $t^*$ of type (a) with $g=1$. There
are $K-|Q|$ rounds of type (a) among the at most $K-1$ rounds before and including $t^*$, so fewer than $|Q|$ rounds of
type (b) precede $t^*$, and $\rho\ge 1$. The invariant with $g=1$ and $r=1$ gives $h(1)\le\rho-1$, so some fresh color
$q$ has $m_q=0$. But $q\in F_{t^*}$ must be shown by a vertex of $\sigma(v_1)$ that has not served, a contradiction.
\end{proof}

\paragraph{One partial-join level.} In a \emph{partial-join} level the vertices of the block of $v_t$ may have
different earlier neighborhoods, all inside the reserves (monochromatic subsets) of earlier blocks, with at least one
neighbor in each; forcing is then automatic once every play reaches the last block, and the question is how small a $3$-colorable such level can be. One
might hope that a single partial-join level gives a polynomial bound with a large exponent. It cannot give an exponent
above $1/3$.

\begin{theorem}[partial joins]\label{thm:pj}
If a partial-join level over $K_m$ forces $m$ colors, is $3$-colorable in every play against every algorithm and has
at most $N$ vertices, then $N\ge m^3/(3(2\log_2N+1))-m^2$, so $N\ge m^{3-o(1)}$. In particular, a lower bound
$\chidet_3(n)\ge n^c$ obtained from one partial-join level over $K_m$ for arbitrarily large $m$ has $c\le 1/3$.
\end{theorem}
The theorem uses that later vertices have no neighbors outside the reserves of earlier blocks; whether it holds
without this restriction is open.

\begin{proof}
Let $B_1,\dots,B_m$ be the blocks and $R_i\subseteq B_i$ the reserve that the adversary selects after $B_i$; its color is
$\mathrm{col}(i)$. Every vertex of $B_i$ has a neighbor in every earlier reserve, so the algorithm may use only colors of
$[m-1]$, and block $i$ has the \emph{palette} $[m-1]\setminus\{\mathrm{col}(1),\dots,\mathrm{col}(i-1)\}$ of $m-i$
colors. The algorithm RR colors every block $i\le m-1$ \emph{round robin}, giving the vertices of block $i$ the colors of its palette in cyclic order. Since $R_i$ is monochromatic, round robin forces $|B_i|\ge(m-i)(|R_i|-1)+1$.

For $i\le m-1$ let RR$_i$ play RR up to the end of block $i$ and then switch to a \emph{hub} strategy. A later vertex $u$
has a neighbor in $R_i$; let $g(u)$, the \emph{hub} of $u$, be its earliest one, and color $u$ by the pair consisting of $g(u)$ and the color
that the algorithm of Lov\'asz, Saks and Trotter~\citep[Thm.~4.2 and its proof]{KieHandbook1998} gives $u$ inside its group $\{u':g(u')=g(u)\}$. Each group avoids the
color of its hub in every $3$-coloring, so it is bipartite and receives at most $\tau:=2\lfloor\log_2N\rfloor+1$ colors, and there are at
most $|R_i|\tau$ pairs. If $|R_i|\tau\le m-1-i$, then RR$_i$ maps the pairs injectively into the $m-1-i$
colors of $[m-1]\setminus\{\mathrm{col}(1),\dots,\mathrm{col}(i)\}$, so it never uses a color outside $[m-1]$, and the
adversary does not force $m$ colors against it. These colors may also occur on vertices of blocks $1,\dots,i$ outside
the reserves; this is harmless because those vertices have no later neighbors. Hence an adversary that forces $m$
colors has $|R_i|\tau\ge m-i$, and so $|R_i|(2\log_2N+1)\ge m-i$. The adversary is deterministic and RR$_i$ agrees with RR up
to the end of block $i$, so the reserve $R_i$ of the RR-play must satisfy this bound for every $i$ at once. Summing
$|B_i|\ge(m-i)((m-i)/(2\log_2N+1)-1)+1$ over $i$ gives the bound.
\end{proof}

\subsection{The oblivious construction}\label{app:oblivious}

\paragraph{The logarithmic factor is needed.} On one oblivious blow-up of $K_K$, with the indices of $\sigma(v_t)$ uniform
over the $K-t$ later neighbors of $v_t$, First-Fit colors every block with a single color, because its vertices have
the same earlier neighbors. It uses all $K$ colors exactly when every index class is nonempty, so
\begin{align*}
  \Pr[\text{First-Fit uses at most }K-1\text{ colors}]&=1-\prod_{d=1}^{K-1}\mathrm{surj}(R,d),\quad\text{where}\\
  \mathrm{surj}(R,d)&=\sum_{i=0}^{d}(-1)^i\binom di\Big(1-\frac id\Big)^{R},
\end{align*}
the probability that $R$ uniform balls hit all $d$ bins. For fixed $\varepsilon$ and $K\to\infty$, forcing with
probability $1-\varepsilon$ therefore needs $R\ge(1-o(1))(K-1)\ln((K-1)/\varepsilon)$ at the top level. This is a factor
$\ln((K-1)/\varepsilon)$ more than the $K-1$ vertices that the top block needs to give each later neighbor one server,
and for large $K$ still less than the $(K-1)(K-2)+1$ vertices of the adaptive top block; but hidden index classes,
unlike reservations, cost a logarithmic factor per server.

The best deterministic algorithm escapes this oblivious blow-up with the following probabilities:
\begin{center}\small
\begin{tabular}{@{}lllllll@{}}
\toprule
$K=4$, block size $R$ & 1--3 & 4 & 5 & 6 & 7 & 8\\
\midrule
best escape probability & $1$ & $263/324$ & $1211/1944$ & $1735/3888$ & $7343/23328$ & $90857/419904$\\
First-Fit & $1,1,5/6$ & $11/18$ & $91/216$ & $61/216$ & $485/2592$ & $2881/23328$\\
\bottomrule
\end{tabular}
\end{center}
These values are exact: a dynamic program over all deterministic algorithms in rational arithmetic, implemented
twice, and a raw game-tree search for small $R$.

\paragraph{Explicit instances.} {One oblivious blow-up of $K_7$ with blocks of $150$ vertices, with indices uniform over the
actual later neighbors of each vertex, gives a distribution on $3$-colorable graphs
with $1\,050$ vertices on which every deterministic algorithm uses $7$ colors with probability at least $0.915$. Two
oblivious blow-ups of $K_8$, indexed in the same way, with blocks of $245$ and then $27\,293$ vertices, give $53\,494\,280$ vertices and
$8$ colors with probability at least $0.906$. Since every algorithm uses at least one
color, every randomized algorithm then uses at least $8-7\cdot 0.094>7.25$ colors in expectation, more than the
transparent limit $7$. Both probabilities are computed from condition~(c) of \cref{lem:O}.}

\paragraph{Transfer for randomized algorithms.} The proof of \cref{thm:main-transfer} also works for oblivious
adversaries, with \cref{lem:O} in place of \cref{lem:force}. If a distribution on $k'$-colorable graphs with $N$
vertices and post-degree at most $D$ forces $K\ge 2$ colors with probability $\rho>0$ (so $N,D\ge 1$), $0<\varepsilon<1$,
and the $j$ of the proof of \cref{thm:main-transfer} is at least $1$ (otherwise there is nothing to do), then $j$
oblivious blow-ups with
$R_1=\lceil(K-1)D\ln(jND/\varepsilon)\rceil$ and $R_{i+1}=\lceil(K-1)R_i\ln(jNR_1\cdots R_i\cdot R_i/\varepsilon)\rceil$
give a distribution on $k$-colorable graphs that forces $K$ colors with probability $\rho-\varepsilon$. Lower bounds
$n^c$ or $2^{(\log n)^c}$ that hold with constant probability against every deterministic algorithm
also transfer from any $k'\ge 3$ to $k=3$, with a smaller $c$. This does not cover $\chirand_{k'}$: an expected cost of $x$ gives $x/2$ colors with probability
at least $x/(2N)$, but in general not with constant probability.

\subsection{Comparison with Li's construction}\label{app:li}

\begin{figure}[ht]
\centering
\begin{tikzpicture}[x=1.05cm,y=0.20cm]
  \draw[->] (0,0) -- (7.0,0) node[right,font=\scriptsize] {$K$};
  \draw[->] (0,0) -- (0,17.5) node[above,font=\scriptsize] {vertices};
  \foreach \a/\l in {1/2,2/4,3/8,4/16,5/32,6/64} {\draw (\a,0) -- (\a,-0.3) node[below,font=\scriptsize] {$2^{\l}$};}
  \draw (0,0) -- (0,-0.3) node[below,font=\scriptsize] {$2$};
  \foreach \b in {0,4,8,12,16} {\draw (0,\b) -- (-0.08,\b) node[left,font=\scriptsize] {$2^{2^{\b}}$};}
  \input{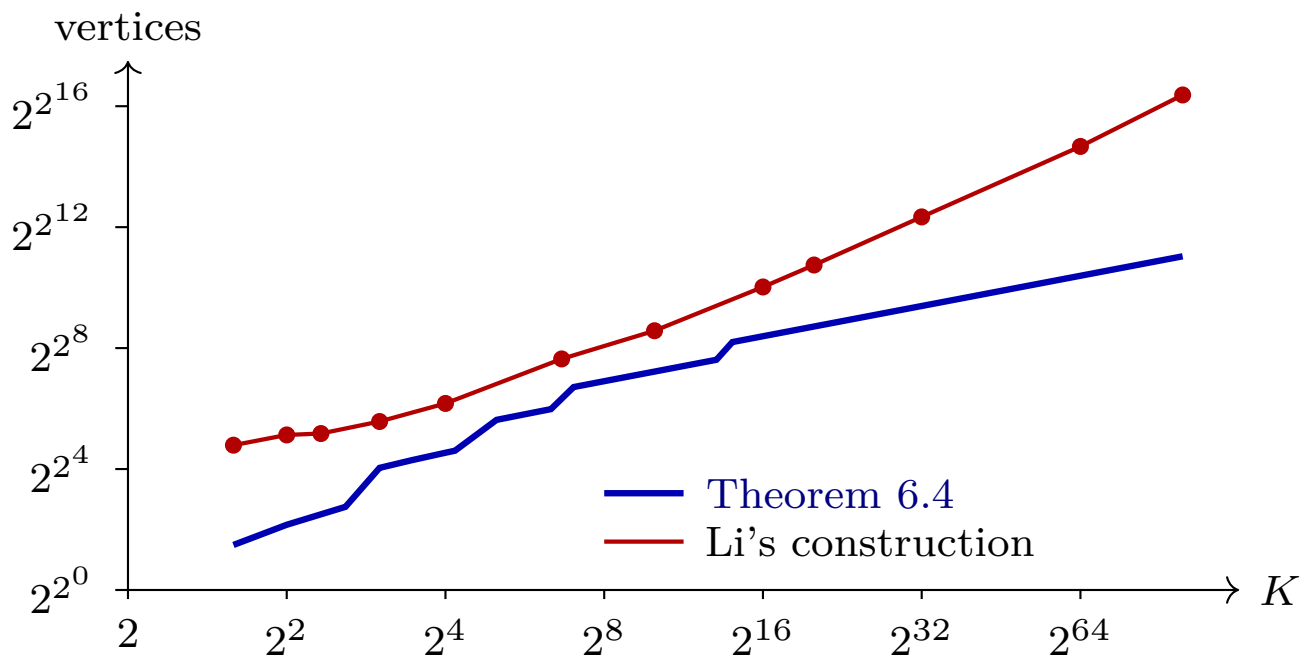}
  \draw[ours] (3.0,3.2) -- (3.5,3.2) node[right,black,font=\scriptsize,align=left] {\cref{thm:L}};
  \draw[li] (3.0,1.6) -- (3.5,1.6) node[right,black,font=\scriptsize,align=left] {Li's construction};
\end{tikzpicture}
\caption{Vertices needed to force $K$ colors on $3$-colorable graphs, on doubly logarithmic axes: the adversary of
\cref{thm:L} with the least sufficient number of levels, and the recursion above for Li's construction.}
\label{fig:li}
\end{figure}

\Cref{fig:li} compares the number of vertices with which the adversary of \cref{thm:L} forces $K$ colors on
$3$-colorable graphs with the corresponding number for Li's construction~\citep{Li2026}. Li states his bounds
asymptotically and without the constants of his recursion, so his curve is an evaluation of one parameter choice that
we derived from his lemmas~\citep[\S\S2.1--2.3]{Li2026}. For $K\ge 3$, with $\beta:=\lceil\log_2K\rceil$${}\ge 2$, $h$ the least integer
with $\beta^h\ge K$,
and thresholds $t_i:=\beta^i$ for $i<h$ and $t_h:=K$, it starts from $(N,D)=(1,1)$ and applies, for $i=h-1,\dots,0$ with
$t=t_i$ and $q=t_{i+1}$, the step
\[
  j:=\Big\lceil\frac{2q-t}{\lceil t/2\rceil}\Big\rceil,\quad B:=K(2^{j+1}-1),\quad R:=KqD,\quad
  N\gets NRB,\quad D\gets RB+B-1,
\]
where $B$ bounds the size of a gadget copy and $R$ the number of copies per simulated vertex. The figure plots the
final $N$; other constants change the curve but not its growth rate
$2^{\Theta((\log K)^3/(\log\log K)^2)}$.

\end{document}